# Proxy Variable in OECD Database: Application of Parametric Quantile Regression and Median Based Unit Rayleigh Distribution

Iman M. Attia *

Imanattiathesis1972@gmail.com

*Department of Mathematical Statistics, Faculty of Graduate Studies for Statistical Research, Cairo University, Egypt

**Abstract:** This study delves deeply into the captivating world of parametric quantile regression, unveiling the innovative Median Based Unit Rayleigh (MBUR) distribution. This remarkable one-parameter model offers a fresh perspective for statistical analysis, unlocking new pathways for exploration. The estimation process is intricately crafted through a cleverly re-parameterized maximum likelihood function, brought to life with a compelling real-world dataset that animates the theoretical concepts. Moreover, the author takes a comprehensive journey into the realms of inference and goodness of fit, weaving together a rich tapestry of insights that underscore the robust capabilities and adaptability of the MBUR distribution. Recognized for its transformative potential, the MBUR distribution is positioned to reshape analytical practices, garnering widespread acceptance and enthusiasm among statisticians and researchers alike. The author undertook a thorough analysis of real-world data characterized by proportions, which unveiled significant deviations from the assumptions of normality and homoscedasticity. This intricate dataset revealed the presence of outliers, rendering traditional regression methods and generalized linear models unsuitable for effective analysis. In contrast, parametric quantile regression emerged as a robust alternative, gracefully handling the challenges posed by outliers while negating the requirement for normality and accommodating heteroscedasticity. This study, by delving into the complexities of proportionate data, highlights the promising potential of both parametric quantile regression and the median-based unit Rayleigh for insightful and accurate analysis.

## Introduction

Quantile regression is a powerful tool for examining the relationships between covariates and response variables, particularly when dealing with highly skewed distributions. This method becomes especially valuable when the distribution has a well-defined, closed-form quantile function, allowing for reparameterization of the probability density function (PDF) and the log-likelihood function. In these scenarios, parametric quantile regression presents a compelling alternative to traditional least squares regression models, particularly when the underlying assumptions—such as normality and homoscedasticity—may not hold true.

Moreover, quantile regression demonstrates robustness in the presence of outliers, which can distort the estimation of regression coefficients and affect the inference process. The extensive literature on quantile regression showcases its varied applications across many disciplines.

Notably, using the median as a conditioning point is often more resilient to outliers and skewed data compared to the mean. This makes it a preferable choice in regression models influenced by these characteristics, although any quantile can be employed for analysis.

The revolutionary transmuted Unit Rayleigh distribution, unveiled by Korkmaz et al. in 2021 [1], stands out as an intriguing and innovative alternative to the well-established beta and Kumaraswamy regression models. In the same remarkable year, Korkmaz and Chesneau introduced the unit Burr-XII distribution, along with a sophisticated quantile regression model that expands the horizons of statistical analysis [2].

Adding to this wealth of knowledge, Korkmaz et al.[3] also presented the novel Arcsecant hyperbolic normal distribution, accompanied by a set of powerful quantile regression models. Their contributions in 2021 were marked by a dedication to advancing the field. In 2023, Korkmaz et al. [4] further enriched the discourse by introducing the Arcsecant hyperbolic Weibull distribution, demonstrating its practical applications in quantile regression tailored for proportionate data sourced from the OECD platform. This body of work highlights a vibrant progression in statistical methodologies, showcasing creativity and rigor in tackling complex data challenges.

Korkmaz et al. [5] have also conducted a significant investigation into the unit Chen distribution and its quantile regression model, shedding light on their potential applications and benefits. Their research offers valuable insights that can enhance our understanding and utilization of these statistical techniques. In a fascinating turn, Kumar et al. [6] employed the Unit-Gompertz model to adeptly characterize inter-record times, showcasing its versatility

Mazucheli et. al., have significantly advanced the field of quantile regression models, particularly with his pioneering work on unit distributions. Their contributions include the refined unit Weibull regression models [7], the innovative unit Birnbaum-Saunders model [8], and the sophisticated Vesicek quantile and mean regression models designed for proportional data [9]. Further showcasing thier expertise, they introduced a one-parameter unit Lindley application in 2019 [10], marking yet another milestone in their impactful careers.

Mazucheli et al. contributions [11] in the field of quantile regression are noteworthy, but he is certainly not the only scholar to have significantly advanced this area of study . For instance, Noufaily and Jones [12] have provided an in-depth exploration of the generalized gamma distribution, including its application within the framework of parametric quantile regression modeling. This work originated from Noufaily's Ph.D. thesis in 2011 and has been further disseminated through published research since 2013.

Leiva and his team have generously shared their expertise on Birnbaum-Saunders distributions, providing detailed descriptions and analyses that have evolved since their initial work in 2016 [13]. Their insights are further supported by a range of contributions from multiple authors, including Sánchez, Leiva, and Galea, who have collectively documented the characteristics of this distribution in several studies [14]&[15]. Gracia-Papani et al. have also provided a detailed explanation of this distribution [16]. Sanchez et al. introduced the quantile regression model for the Wiebull distribution [17].

Exploring the realm of quantile regression models reveals a rich array of diverse distributions that enhance the analytical capabilities available to researchers. The insights gained from these models, introduced by Marchant et al. [18] , Leão et al. [19] , and Leiva et al. [20], are invaluable and offering robust methodologies to improve data interpretation.

To facilitate these analyses, various R packages have been developed that implement quantile regression techniques. In 2020, Mazucheli collaborated with Alves to create the Vasicekreg package, designed to handle specific types of regression tasks. The following year, they introduced the Ugomquantreg package, which further extends capabilities in this domain. Meanwhile, in the same year of 2020, Mazucheli partnered with Menezes to develop the unitBSQuantReg package, which focuses on unit-based distributions. Menezes also contributed to the toolkit by creating the UWquantreg package, designed for weighted quantile regression. Lastly, in 2021, the esteemed statistician Roger Koenker unveiled the quantreg package, a foundational tool widely utilized in the field of quantile regression analysis.

Numerous authors have delved into the intricacies of parametric quantile regression, a powerful statistical tool used to analyze proportional data. This approach allows researchers to uncover and illuminate the complex relationships between variables and their predictors, offering a deeper understanding of how these elements interact within various contexts. To mention some of those authors: [21], [22], [23], [24], [25], [26], [27], [28], [29], [30], [31].

This paper presents an in-depth exploration of the innovative Median-based unit Rayleigh (MBUR) distribution, previously introduced by the author. This new approach is specifically designed for conducting quantile regression analysis, enabling researchers to gain valuable insights into real-world data applications. The author effectively demonstrates the feasible advantage of the MBUR distribution, highlighting its potential to connect advanced statistical theory with meaningful results in data analysis. The author utilized OECD data in employing the parametric MBUR quantile regression using the response variables which are distributed as MBUR. In section one, the author revises the parametric MBUR quantile regression and the link functions. In section two, the author discusses the Goodness of Fit criteria (GoF) and the model specification diagnostic tests. In section three, the author exposes the analysis results and discussion of using the OECD datasets. In section four, the author comprehends the conclusions and the future work.

## Section One: Methodology

### 1.1. Parametric MBUR Quantile Regression Model

The response variable adheres to the Median Based Unit Rayleigh (MBUR) distribution, which adds complexity to the analysis. To uncover a causal relationship between this variable and its influencing covariates, it is essential to define both the parametric function and the link function meticulously. Given that the response variable may exhibit significant skewness and potentially breach the assumptions of normality and homoscedasticity, employing parametric quantile regression could provide a resilient alternative. Nevertheless, it is vital to delve into a myriad of other methodologies to ensure the most accurate and reliable estimation outcomes. The MBUR distribution was discussed by Attia [32]. It has the following PDF, CDF, and quantile function as respectively expressed in equations (1-3). The parametric quantile regression (PQR) depends on the quantile function.

$$f(y) = \frac{6}{\alpha^2}\left[1 - y^{\frac{1}{\alpha^2}}\right]y^{\left(\frac{2}{\alpha^2}-1\right)} \ , 0 < y < 1 \ , \alpha > 0 \qquad (1)$$

$$F(y) = 3y^{\frac{2}{\alpha^2}} - 2y^{\frac{3}{\alpha^2}} \ , \ 0 < y < 1 \ , \alpha > 0 \qquad (2)$$

$$y = F^{-1}(y) = \left\{-.5\left(cos\left[\frac{cos^{-1}(1-2u)}{3}\right] - \sqrt{3} \ sin\left[\frac{cos^{-1}(1-2u)}{3}\right]\right) + .5\right\}^{\alpha^2} \qquad (3)$$

Re-parameterize the PDF and CDF of MBUR using the quantile function, $= c^{\alpha^2}$, where c is $c = -.5\left(cos\left[\frac{cos^{-1}(1-2u)}{3}\right] - \sqrt{3}\,sin\left[\frac{cos^{-1}(1-2u)}{3}\right]\right) + .5$ , $u \in (0,1)$ . U represents the chosen percentile, if it is the median, so u=0.5, if it is the 25th percentile so u=0.25 so $\alpha^2 = \frac{ln(y)}{ln(c)}$.When replacing u=0.5 in c, this gives $ln(.5) = -.6931$ , $\alpha^2 = \frac{ln(y)}{ln(c)} = \frac{ln(median)}{ln(c)}$ . As y is the median corresponding to u=0.5.

## 1.2. The Link Function

The linear predictor, $\varphi = X'B = \beta_0 + \beta_1 x_{i1} + \cdots + \beta_k x_{ik}$ ; $i = 1,\ldots,n$ & $k = 1,\ldots,k$ , can be expressed using different link functions as the logit, clog-log, or the log-log link function. This linear predictor represents the median that should be estimated, where n is the number of cases or observations and k is the number of variables. These link functions of the median are discussed in details by Attia [33].

### 1.2.1. Logit Link Function

Using the logit function of the median, which will be called $\mu$

$$Logit\ (median) = logit(\mu) = \varphi = X'B = \beta_0 + \beta_1 x_{i1} + \cdots + \beta_k x_{ik}\ ;\ i = 1,\ldots,n\ \ \&\ \ k = 1,\ldots,k$$

where n is the number of cases or observations and k is the number of variables. Logit median is the linear combination of variables.

So; $\mu = \frac{e^{X'B}}{1+e^{X'B}} = \frac{e^\varphi}{1+e^\varphi}$ & $\alpha^2 = \frac{ln(y)}{ln(c)} = \frac{ln(median)}{ln(c)} = \frac{ln\left(\frac{e^{X'B}}{1+e^{X'B}}\right)}{ln(c)} = \frac{ln\left(\frac{e^\varphi}{1+e^\varphi}\right)}{ln(c)}$ , $0 < y < 1$ , $\alpha > 0$

### 1.2.2. Complementary Log-Log Link Function

$\log\{-\log(1 - median)\} = X'B = \varphi = \beta_0 + \beta_1 x_{i1} + \cdots + \beta_k x_{ik}$ ; $i = 1,\ldots,n$ & $k = 1,\ldots,k$

$$median = 1 - e^{-e^{X'\beta}} = 1 - exp\left(-exp(X'\beta)\right)\ \ \&\ \ \alpha^2 = \frac{ln(median)}{ln(c)} = \frac{ln\left(1 - e^{-e^\varphi}\right)}{ln(c)}$$

### 1.2.3. Log-Log Median Link Function

A link function other than the logit can be used. The author used the log-log function

$\log\{-\log(median)\} = X'B = \varphi = \beta_0 + \beta_1 x_{i1} + \cdots + \beta_k x_{ik}$ , $i = 1,2,\ldots.,n$ & $k = 1,2,\ldots.,k$

$$median = e^{-e^{X'\beta}} = exp\left(-exp(X'\beta)\right)\ \ \&\ \ \alpha^2 = \frac{ln(median)}{ln(c)} = \alpha^2 = \frac{ln\left(e^{-e^{X'\beta}}\right)}{ln(c)} = \frac{-e^{X'\beta}}{ln(c)}$$

## Section Two: Goodness of Fit criteria

## 2.1. Diagnostic Tests for Model Specification

Model adequacy, as regards the appropriate response variable distribution and the used link function, can be appraised using two residual-based diagnostics: randomized quantile residuals (RQ)[34] and Cox-Snell residuals (CS)[35]. The RQ residuals follow approximately a standard normal distribution, and the (CS) residuals follow a standard exponential distribution , [36] , [37].

### 2.1.1. Randomized quantile (RQ) residuals: $\quad r_i^{RQ} = \Phi^{-1}\left(F(y_i, \widehat{\beta})\right)$

$\Phi$ is the standard normal CDF. F is the re-parameterized MBUR CDF. $y_i$'s are the observations and $\hat{\beta}$ are the estimated regression coefficients. These residuals are approximately distributed as standard normal. When the model is correctly specified, these residuals approximately follow the standard normal distribution.

### 2.1.2. Cox-Snell (CS) residuals: $\quad r_i^{CS} = -log\left(1 - F(y_i, \widehat{\beta})\right)$

Cox-Snell residuals are approximately distributed as a standard exponential distribution with a scale parameter one. The negative logarithm mentioned above represents the cumulative distribution function (CDF) of the standard exponential distribution with this scale parameter. When the model is correctly specified, these residuals will approximately follow a standard exponential distribution.

## 2.2. Model Selection Criteria:

Model selection is conducted by identifying the options with the minimum values of AIC (Akaike Information Criterion), BIC (Bayesian Information Criterion), and Corrected AIC, ensuring that the chosen models are optimal for analysis. $AIC = -2l(\hat{\theta}) + 2\,p$, $CAIC = AIC + \frac{2\,p(p+1)}{n-p-1}$ , $BIC = -2l(\hat{\theta}) + p\log(n)$ , where p is the number of parameters and n is the sample size.

In the study conducted by Sánchez, Leiva, Saulo, and colleagues in 2021 [14], they applied a specific methodological approach, $R_M^2$ , that corresponds directly to established practices, usual $R^2$ , in mean regression. This approach is instrumental in yielding reliable estimates and is meticulously defined within the context of regression analysis. Through these criteria and methodologies, researchers can effectively compare and assess the suitability of various models, facilitating improved decision-making in statistical modeling, $R_M^2 = 1 - exp\left(\frac{2}{n}\left(l(\breve{\theta}) - l(\hat{\theta})\right)\right)$, where $l(\breve{\theta})\ and\ l(\hat{\theta})$ are the maximum log-likelihood for the model without covariates (null model) and the model with all covariates (Full model).

## Section Three: OECD Data Analysis, Results and Discussion

Appendix A discloses the dataset attained from the OECD database, which stands for Organization for Economic Co-operation and Development. The author used the Nealder Mead optimizer in MATLAB to estimate the parameters and used the finite central difference method to estimate the variance. The data is available at https://stats.oecd.org/index.aspx?DataSetCode=BLI

In OECD platform, many predictors are recorded. The author conduct distributional fit for the indicators after transforming these indicators into ratio defined on the unit interval. The different unit distributions tested was mainly Beta, Kumaraswamy and MBUR distribution. The variables fitting the MBUR distribution were used as response variable and the other that do not fit this particular distribution were used as predictors. The author tries to find the relationship between the response variables and the possible predictors. The response variables were: educational attainment, water quality, quality of support network and feeling safe walking alone. The predictors used in this analysis were the employment rate, air pollution, life satisfaction, life expectancy, homicide rate. The median based unit Rayleigh distribution (MBUR) fits the response variables. Because these variables exhibit various shapes as they are skewed and violating the normality assumption, the author used parametric quantile regression and specifically the parametric median regression as median can be considered a good candidate in cases where normality assumption is violated and in presence of outlier. Table 1 shows the descriptive analysis of the predictors and Table 2 depicts the descriptive statistics of the response variables.

Table 1: Summary of descriptive statistics for each predictor

|  | Air pollution | Homicide rate | Life satisfaction | Life expectancy | Employment rate |
|---|---|---|---|---|---|
| Mean | 13.5098 | 3.2902 | 6.6 | 80.261 | 67.6829 |
| Standard deviation | 6.3942 | 6.2518 | 0.7253 | 3.8452 | 8.8188 |
| Skewness | 0.8055 | 2.7476 | -0.4365 | -2.1144 | -1.2031 |
| Kurtosis | 2.8203 | 9.9481 | 2.7183 | 9.4199 | 4.6194 |
| Min. | 5.5 | 0.2 | 4.9 | 64.2 | 39 |
| Max. | 28.5 | 26.8 | 7.9 | 84.4 | 80 |
| 25th percentile. | 8.175 | 0.5 | 6.1 | 78.45 | 64.25 |
| Median | 12 | 0.9 | 6.5 | 81.7 | 70 |
| 75th percentile | 17.125 | 2.025 | 7.225 | 82.9 | 74 |

These variables when transformed into ratio they fit either the Beta distribution or the Kumaraswamy distribution. They did not fit the MBUR. These variables were subjected to the following transformation: divided by 100 and then taking the log of the result. This transformation was used to make the responses and predictors in the same scale so that the regression can be applied.

Table 2 : Summary  of descriptive statistics for each response variable.

|  | Water quality | Educational attainment | Quality of support network | Feeling safe walking alone |
|---|---|---|---|---|
| Mean | 0.8332 | 0.7810 | 0.9078 | 0.7207 |
| St.d. | 0.0972 | 0.1568 | 0.0538 | 0.143 |
| Skewness | -0.6059 | -1.1480 | -1.1760 | -0.9486 |
| Kurtosis | 2.9144 | 3.2077 | 3.5406 | 3.0353 |
| Min | 0.62 | 0.42 | 0.77 | 0.4 |
| Max | 0.98 | 0.95 | 0.98 | 0.93 |
| 25th percentile | 0.7775 | 0.705 | 0.89 | 0.655 |
| Median | 0.83 | 0.83 | 0.93 | 0.76 |
| 75th percentile | 0.91 | 0.895 | 0.95 | 0.8225 |

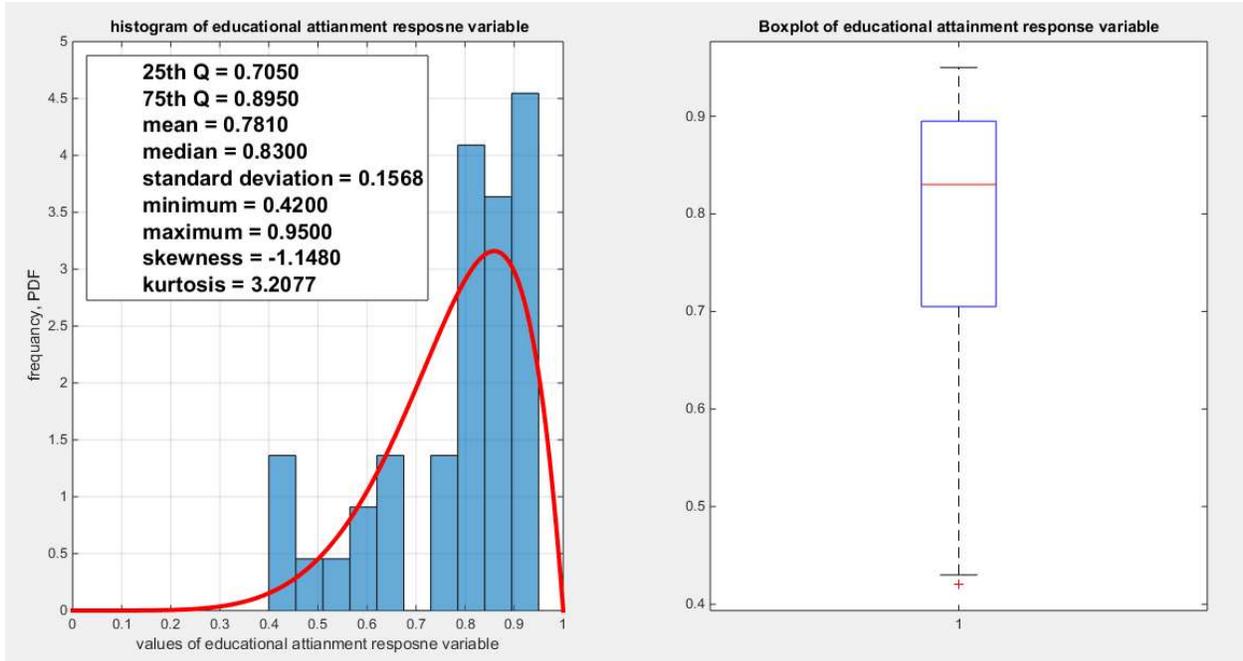

Fig. 1 shows the histogram and the boxplot of the educational attainment variable

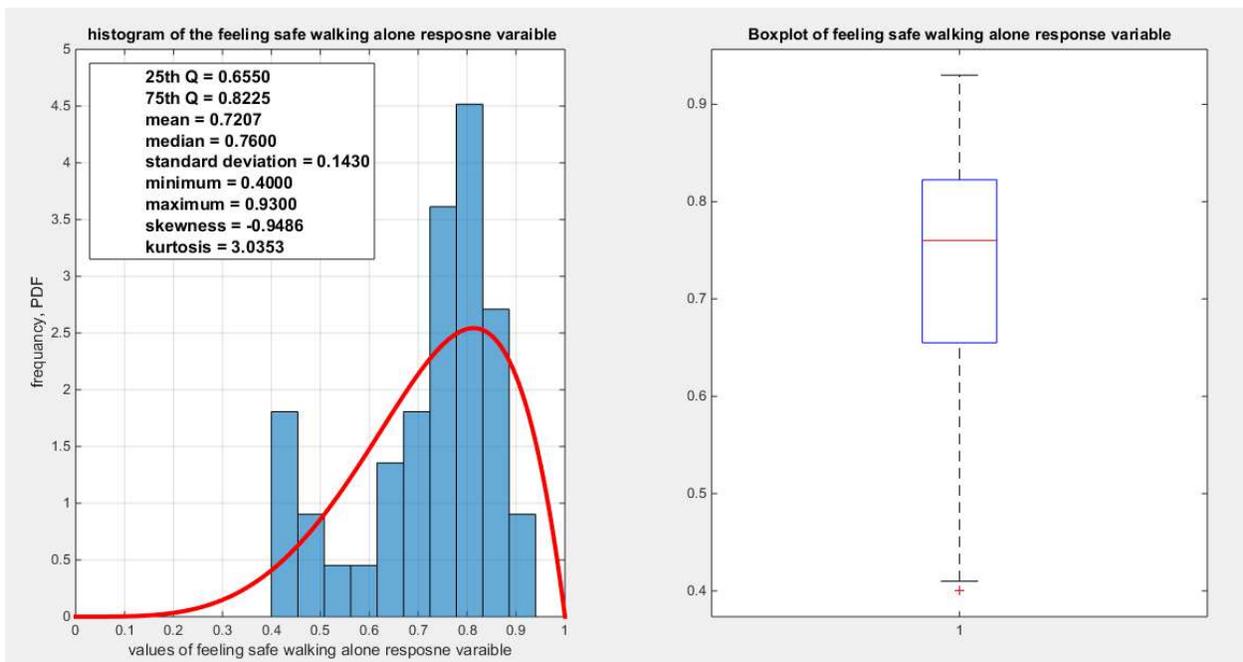

Fig. 2 shows the histogram and the boxplot of the feeling safe walking alone variable

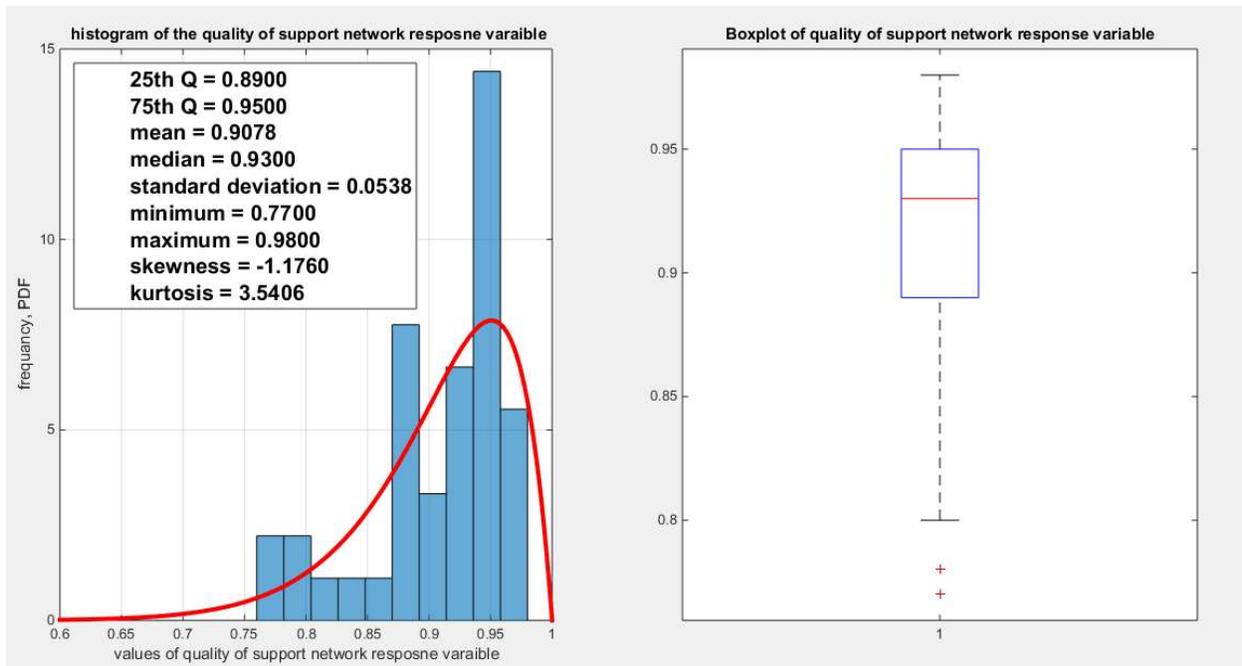

Fig. 3 shows the histogram and the boxplot of the quality of support network variable

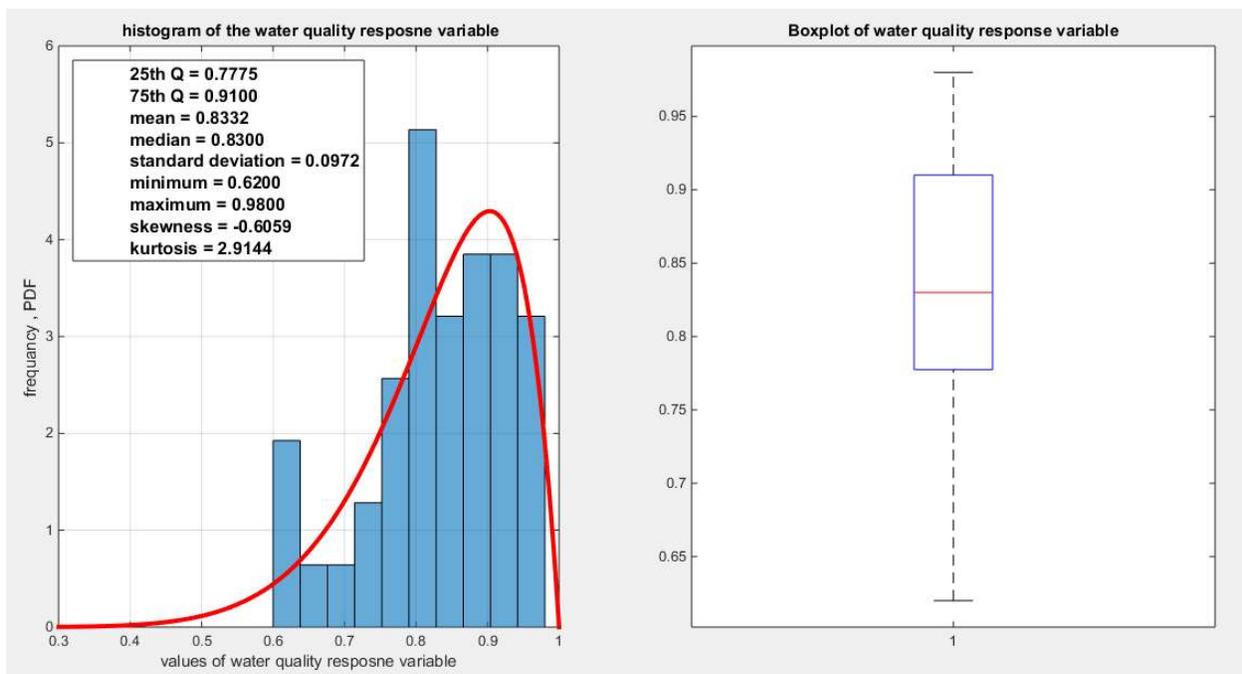

Fig. 4 shows the histogram and the boxplot of the water quality variable

Table(1-2) and Figures (1-4) show that the response variables exhibit variable degree of left skewness (nearly more or less normal kurtosis, mesokurtic) with the presence of one or two extreme values in the lower tail. The MBUR fits all these variables comparable to beta and Kumaraswamy even outperform them like fitting the water quality and educational attainment as evident from the better AIC, CAIC, BIC and HQIC and higher value for log likelihood. The database

contains 41 observations and this is the highest number of observations for a variable in this database. But the education attainment indicator is 40 observations because no recorded value is there from Japan as shown in the table of Appendix A. Other indicators have less number of observations. For this reason, these indicators were chosen among the other.

For each response variable three different link functions were used and the results were compared. The first link function is the logit, the second one is the log-log complementary link function or the clog-log link function, and the third is the log-log link function. The aspects of comparison are the LL, AIC, CAIC, BIC, HQIC, LRT, and pseudo R squared. The estimation of regression coefficient is done using the Nelder Mead optimizer and the variance covariance matrix was gained by Finite Central difference formula. For each analysis, figures depicting the estimated curve, QQ plot for residuals and plotting residuals against predictors.

## First Response Variable: The Educational Attainment

Regressing educational attainment on four predictors; one at a time then all in one full model gives the following results. Figure 5 shows the scatter plot detecting the relationship between the response variable and each predictor (transformed). The author presented figures for the logit model, figures that illustrate the estimated curve, plot between the predictor and the residuals, and the QQ plot for the empirical residuals against the theoretical residuals.

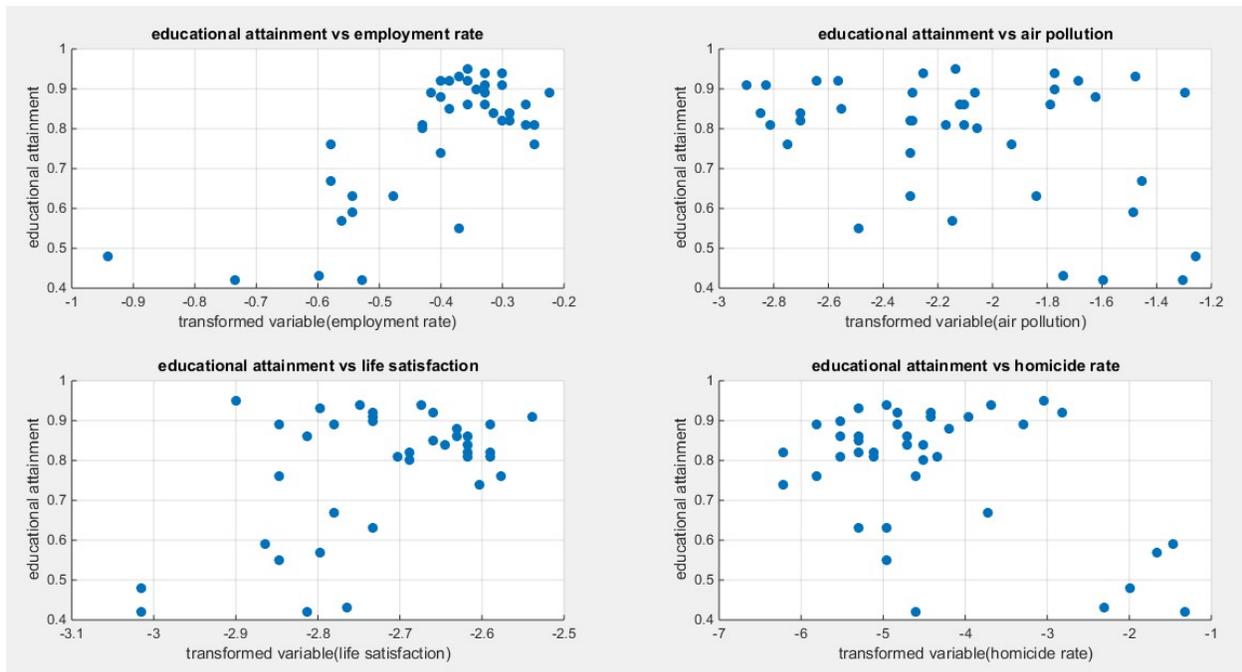

Fig. 5 shows the scatter plot of the response variable and each of the predictor. The relationship is nonlinear.

Figure 5 shows that the relationship between the educational attainment and each of the predictors are nonlinear. Table 3-6 show the results obtained from regressing the educational attainment on each predictor using different link functions and comparing the statistical indices as regards the estimated coefficients , the Likelihood Ratio Test (LRT ) and its p value, AIC, CAIC, BIC, HQIC , and the Log Likelihood (LL).

Table 3: regressing education attainment on employment rate

| | Logit link function | | Log-log complementary | | Log-log median | |
|---|---|---|---|---|---|---|
| B0 | 3.2292 | | 1.3407 | | -3.0983 | |
| B1 | 4.2400 | | 2.0353 | | -3.6239 | |
| LL | 37.9883 | | 37.6605 | | 37.8513 | |
| Wald stat. of b0 | 7.9837(p<0.025) | | 7.4141(p<0.025) | | 8.8996(p<0.025) | |
| Wald stat. of b1 | 4.2511(p<0.025) | | 4.1825(p<0.025) | | 4.4422(p<0.025) | |
| AIC | -71.9766 | | -71.3209 | | -71.7025 | |
| CAIC | -71.6523 | | -70.9966 | | -71.3782 | |
| BIC | -68.5988 | | -67.9432 | | -68.3248 | |
| HQIC | -70.7553 | | -70.0997 | | -70.4813 | |
| LRT | 23.9192 (p=1.0046e-6) | | 23.2636 (p=1.4125e-6) | | 23.6451 ( p=1.158e-6) | |
| R-squared | 0.4501 | | 0.4410 | | 0.4463 | |
| P-value for randomized quantile residuals | 0.4557 | | 0.5458 | | 0.447 | |
| p-value for Cox-Snell residuals | 0.4557 | | 0.5458 | | 0.447 | |
| Variance-covariance matrix | 0.1636 | 0.3831 | 0.0327 | 0.0833 | 0.1212 | 0.2684 |
| | 0.3831 | 0.9948 | 0.0833 | 0.2368 | 0.2684 | 0.6655 |
| QR vs. predictor(tau,p) | -0.017 ,0.8886 | | -0.0156, 0.8978 | | -0.0156,0.8978 | |
| CS vs. predictor(tau,p) | -0.017 ,0.8886 | | -0.0156, 0.8978 | | -0.0156,0.8978 | |

Table (3) shows that the predictor is significant as likelihood ratio test (LRT) is highly significant. The R squared is also high for this predictor. The AIC, CAIC, BIC, HQIC and LL are more or less equal across the different models. The residuals plotted against the predictors show no specific trend and they are randomly scattered. The QQ plot of the randomized quantile residuals shows perfect alignment with the diagonal all through its course in contrast with the Cox Snell residuals that show this perfect alignment at the lower tail. The estimated curve between the estimated median and the transformed predictor is increasing reflecting that the more the employment rate is, the more the percentage attaining the education is. The figure for the clog-log shows the same pattern. The log-log figure has the same pattern. The difference is mainly manifested in the slope of the estimated curve. To assess the assumption of constant variance in the median parametric regression model, residual-based diagnostic tests were conducted using both randomized quantile (RQ) and Cox-Snell (CS) residuals. For each type of residual, an auxiliary regression of the squared residuals on the corresponding predictor was estimated by ordinary least squares, and the null hypothesis of homoscedasticity ($H_0$: constant variance) was tested. The results indicted no significant relationship between the squared residuals and the predictor variable (CS: p=0.702, R-squared=0.00389; RQ: p=0.681, R-squared=0.00449), suggesting that the variance of the residuals remained approximately constant across the range of the predictor. Furthermore, the magnitude of the CS residuals were within a reasonable range (one value 2.3231), which supports the absence of heteroscedasticity. These findings provide evidence that the fitted median regression model satisfies the homoscedasticity assumption. These results are from the logit model. Figures 6-12 show the previous results.

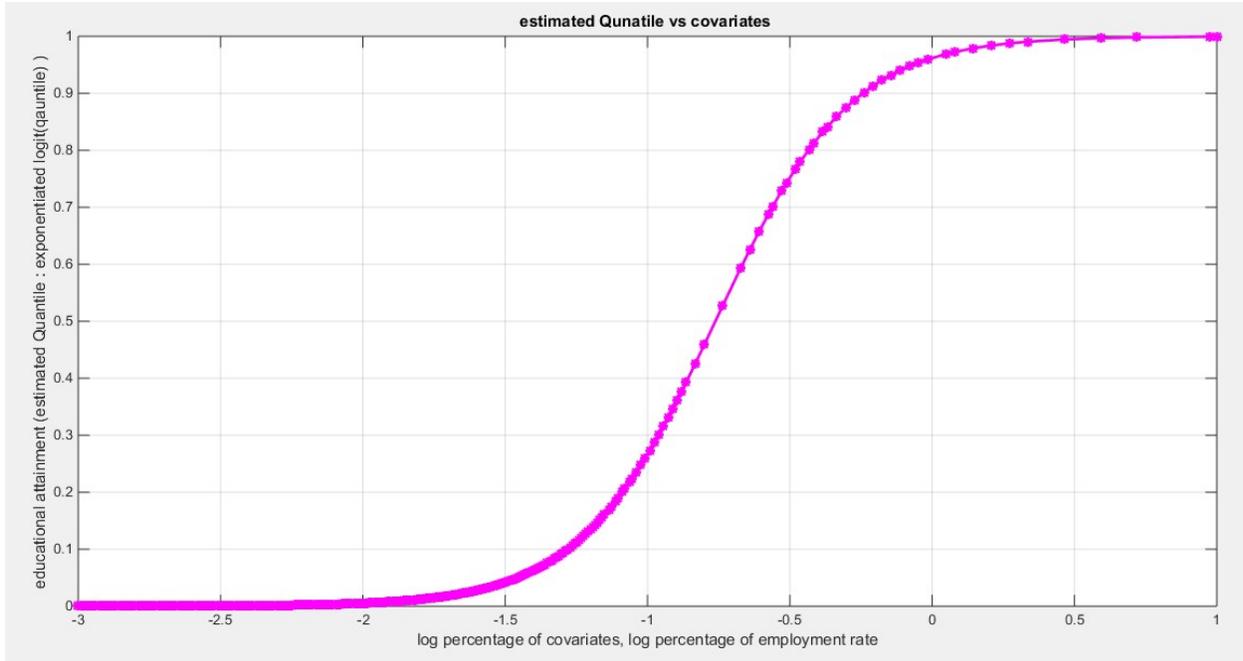

Fig.6 shows the estimated curve plotting the transformed predictor against the estimated median for the logit link.

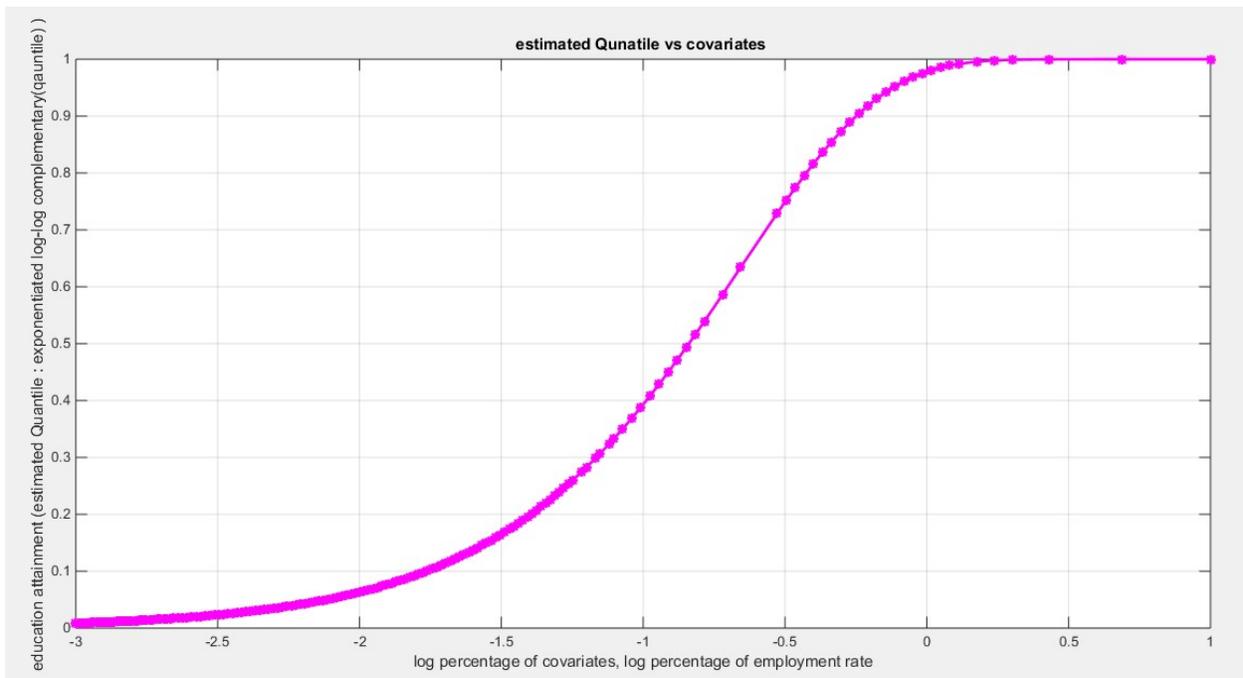

Fig. 7 shows the estimated curve plotting the transformed predictor against the estimated median for clog-log link.

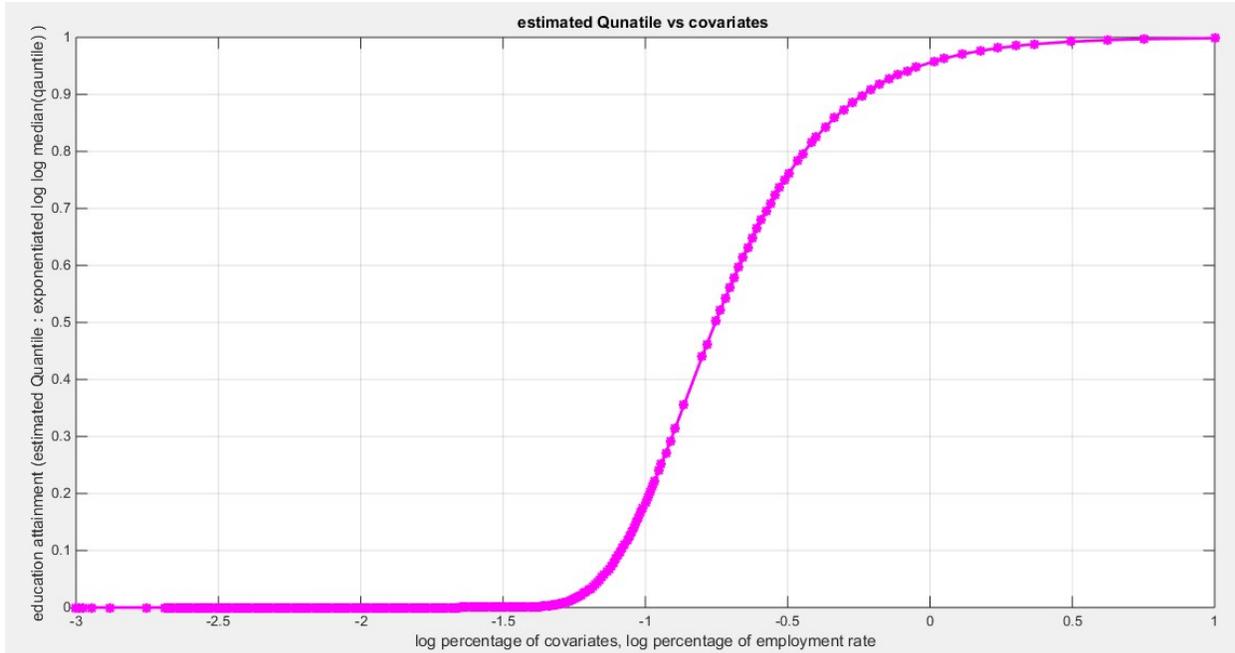

Fig. 8 shows the estimated curve plotting the transformed predictor against the estimated median for the log log link.

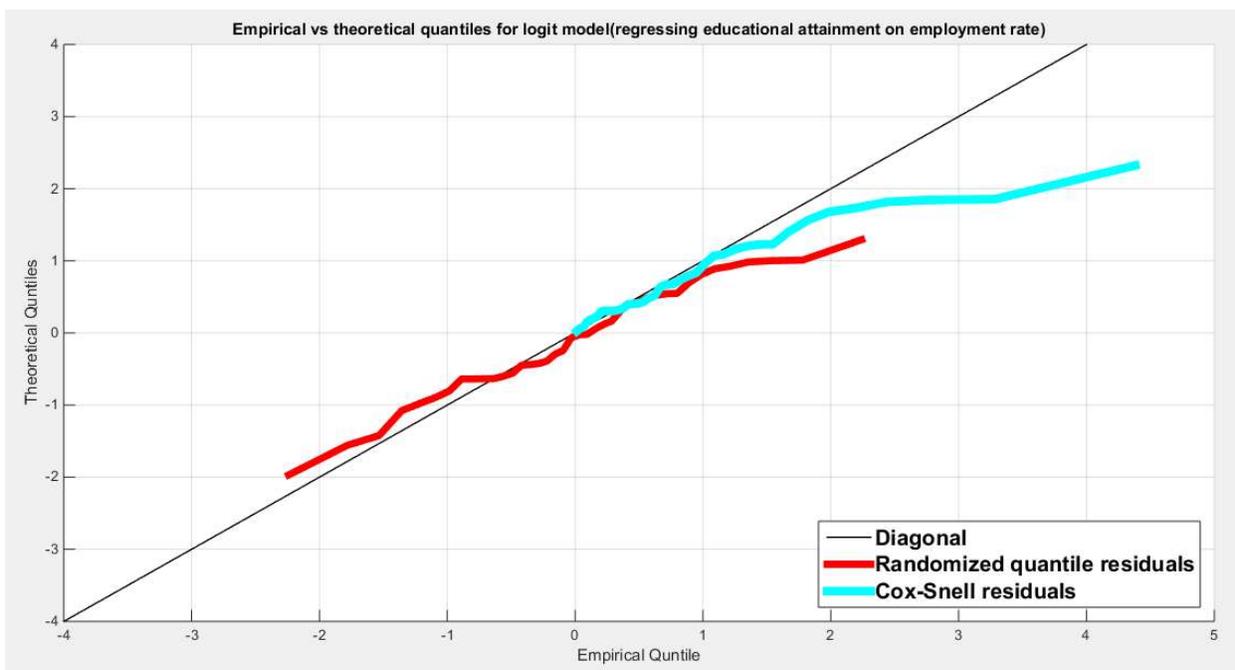

Fig. 9 shows the QQ plot of the empirical quantiles and the theoretical qunatiles for both types of residuals.

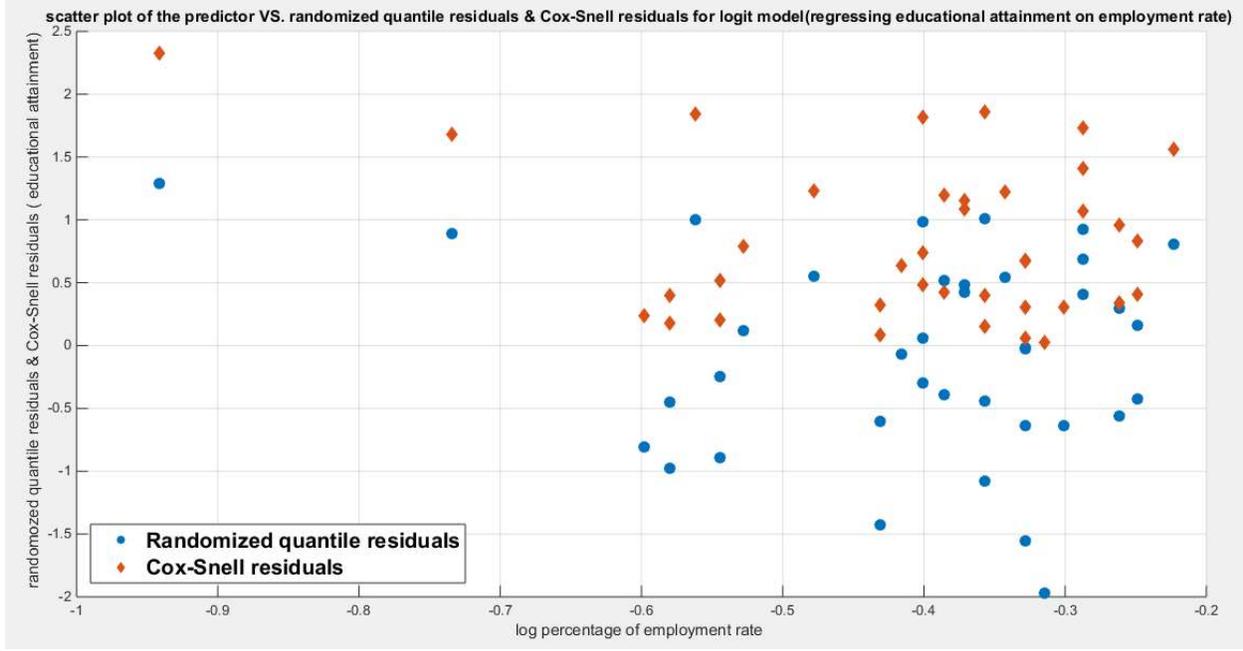

Fig. 10 shows the scatter plot of residuals of both types against the transformed predictors.

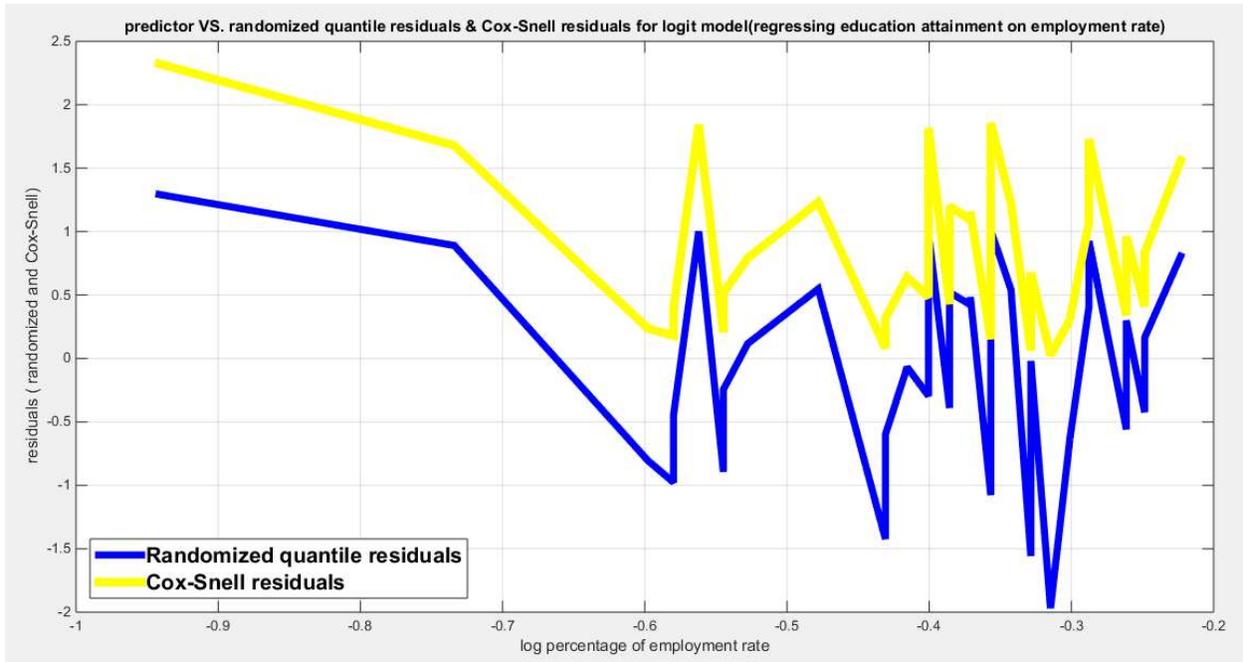

Fig. 11 shows the plot of residuals of both types against the transformed predictors.

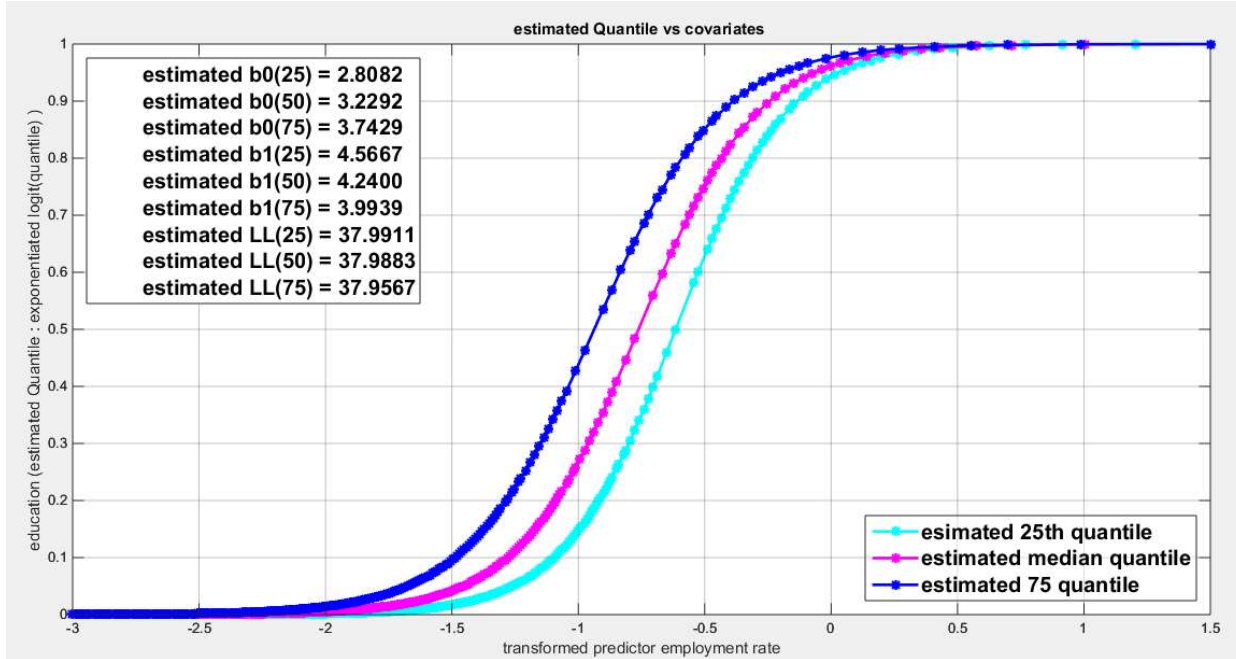

Fig. 12 shows parallel quantile curves across 25th , 50th ( median), 75th percentiles, suggesting that the predictor exerts a uniform influence on the response consistent with homoscedasticity. ( for the logit link)

Table 4: regressing education attainment on air pollution

| | Logit link function | | Log-log complementary | | Log-log median | |
|---|---|---|---|---|---|---|
| B0 | -0.1476 | | -0.2714 | | -0.1201 | |
| B1 | -0.7461 | | -0.3594 | | 0.6736 | |
| LL | 30.1307 | | 30.0430 | | 30.1626 | |
| Wald stat. of b0 | 0.2555(p>0.025) | | 0.9217(p>0.025) | | 0.2358(p>0.025) | |
| Wald stat. of b1 | 2.8383(p<0.025) | | 2.7979(p<0.025) | | 2.8670(p<0.025) | |
| AIC | -56.2613 | | -56.086 | | -56.3251 | |
| CAIC | -55.9370 | | -55.7616 | | -59.0008 | |
| BIC | -52.8836 | | -52.7082 | | -52.9474 | |
| HQIC | -55.0400 | | -54.8647 | | -55.1038 | |
| LRT | 8.2039 (p=0.0042) | | 8.0286 (p=0.0046) | | 8.2677 (p=0.004) | |
| R-squared | 0.1854 | | 0.1819 | | 0.1867 | |
| P-value for randomized quantile residuals | 0.9949 | | 0.9980 | | 0.9917 | |
| p-value for Cox-Snell residuals | 0.9949 | | 0.9980 | | 0.9917 | |
| Variance-covariance matrix | 0.3338 | 0.1481 | 0.0867 | 0.0370 | 0.2594 | 0.1166 |
| | 0.1481 | 0.0691 | 0.0370 | 0.0165 | 0.1166 | 0.0552 |
| QR vs. predictor(tau,p) | 0.0219, 0.8520 | | 0.0219, 0.8520 | | 0.0219, 0.8520 | |
| CS vs. predictor(tau,p) | 0.0219, 0.8520 | | 0.0219, 0.8520 | | 0.0219, 0.8520 | |

Table 4 shows that the predictor is significant as likelihood ratio test (LRT) is highly significant; the R squared is also high for this predictor but less than that of the employment rate. The AIC, CAIC, BIC, HQIC and LL are more or less equal across the different models. The LL is less than that of the employment rate. The residuals plotted against the predictors show no specific trend and they are randomly scattered. The QQ plot of the randomized quantile residuals shows perfect alignment with the diagonal all through its course in contrast with the Cox Snell residuals that show this perfect alignment at the lower tail and the center. The estimated curve between the estimated median and the transformed predictor is decreasing reflecting that the more the pollution of the air is, the less the percentage attaining the education is. The figure for the clog-log shows the same pattern. The log log figure has the same pattern. The difference is mainly manifested in the slope of the estimated curve. To assess the assumption of constant variance in the median parametric regression model, residual-based diagnostic tests were conducted using both randomized quantile (RQ) and Cox-Snell (CS) residuals. For each type of residual, an auxiliary regression of the squared residuals on the corresponding predictor was estimated by ordinary least squares, and the null hypothesis of homoscedasticity ($H_0$: constant variance) was tested. The results indicted no significant relationship between the squared residuals and the predictor variable (CS: p=0.0692, R-squared=0.0843; RQ: p=0.0394, R-squared=0.107), suggesting that the variance of the residuals remained approximately constant across the range of the predictor. Furthermore, the magnitude of the CS residuals were within a reasonable range (six values between 2.2755 and 2.8886), which supports the absence of heteroscedasticity. These findings provide evidence that the fitted median regression model satisfies the homoscedasticity assumption. These results are from the logit model. Figures 13-19 show the previous results.

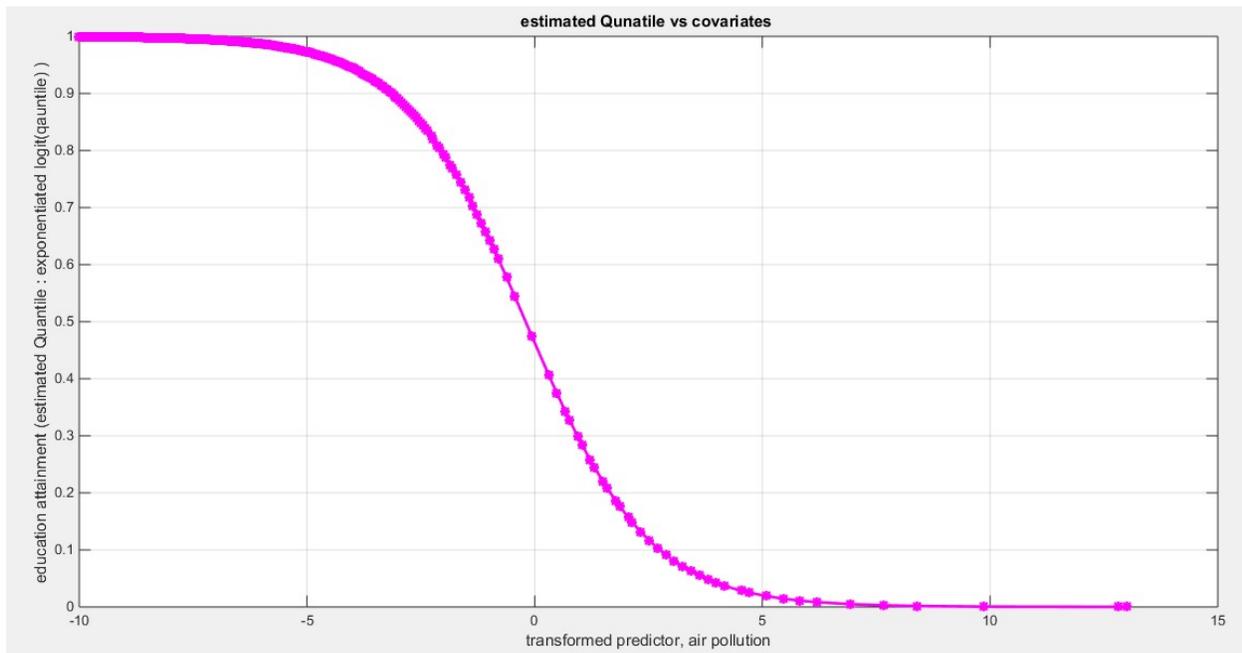

Fig .13 shows the estimated curve plotting the transformed predictor against the estimated median (for the logit link).

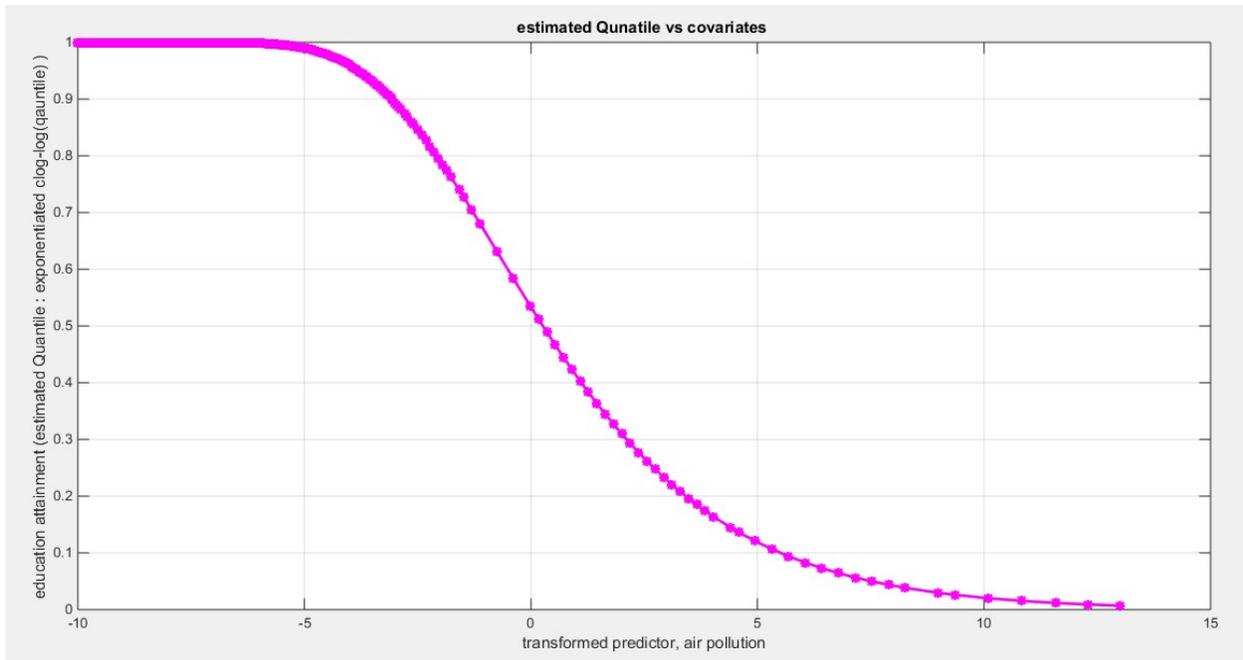

Fig.14 shows the estimated curve plotting the transformed predictor against the estimated median (for the clog-log link).

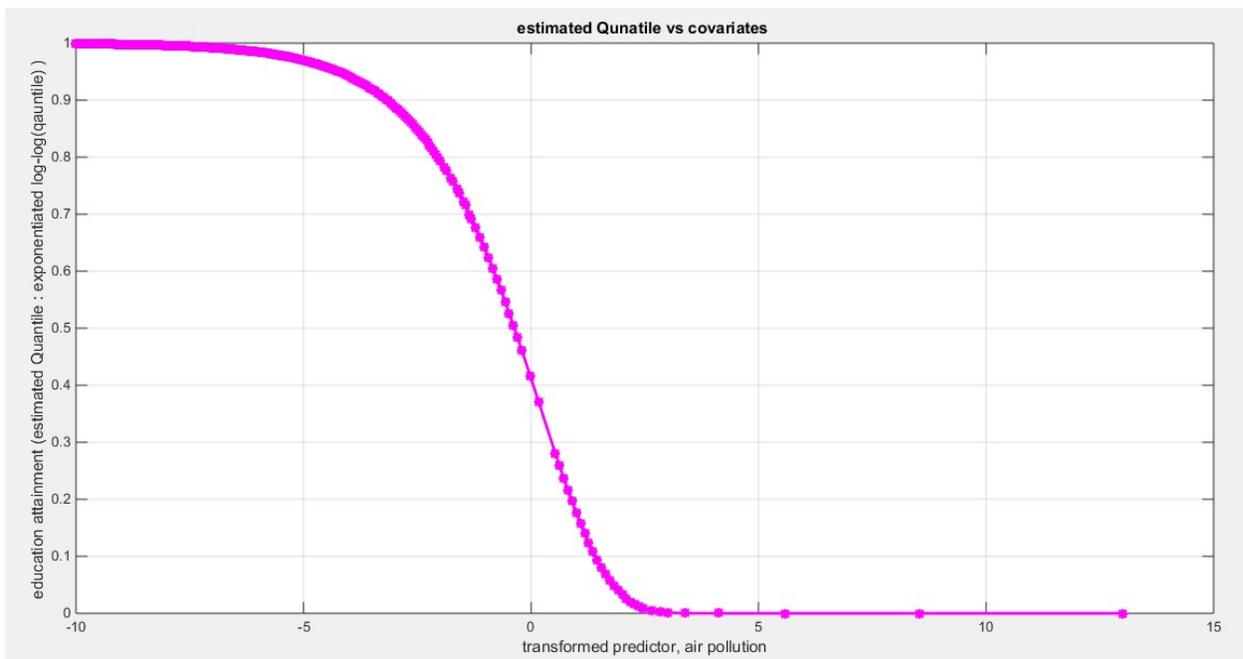

Fig. 15 shows the estimated curve plotting the transformed predictor against the estimated median (for the log-log link).

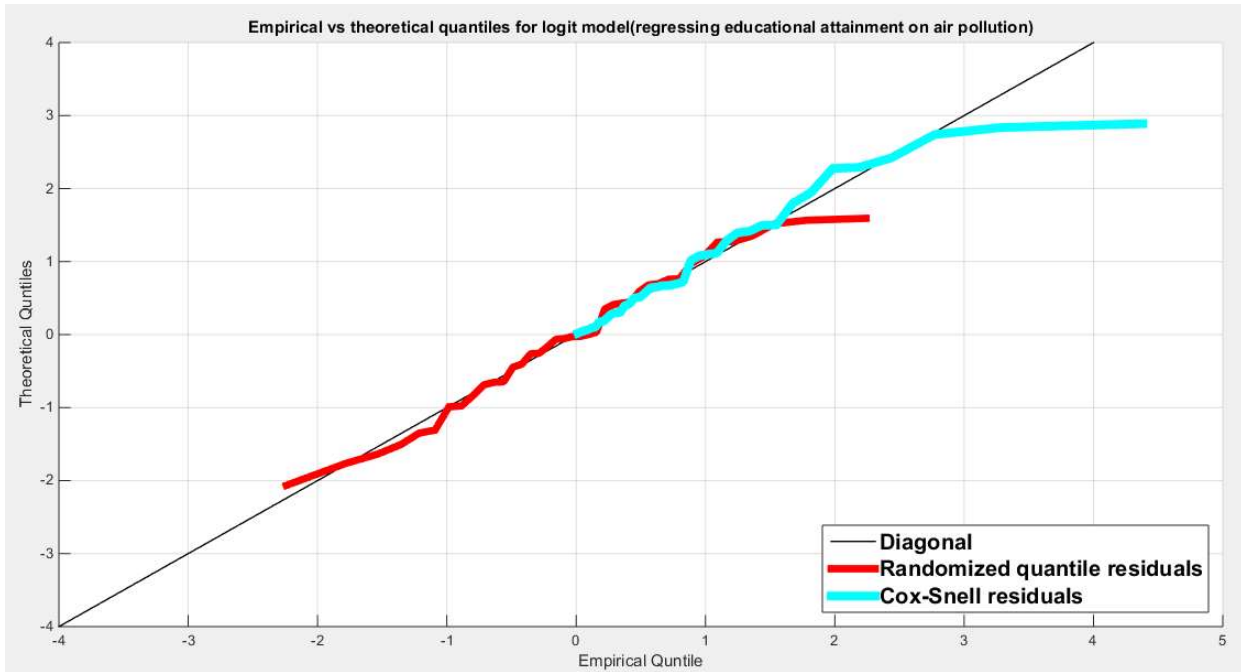

Fig. 16 shows the QQ plot of the empirical quantiles and the theoretical qunatiles for both types of residuals.

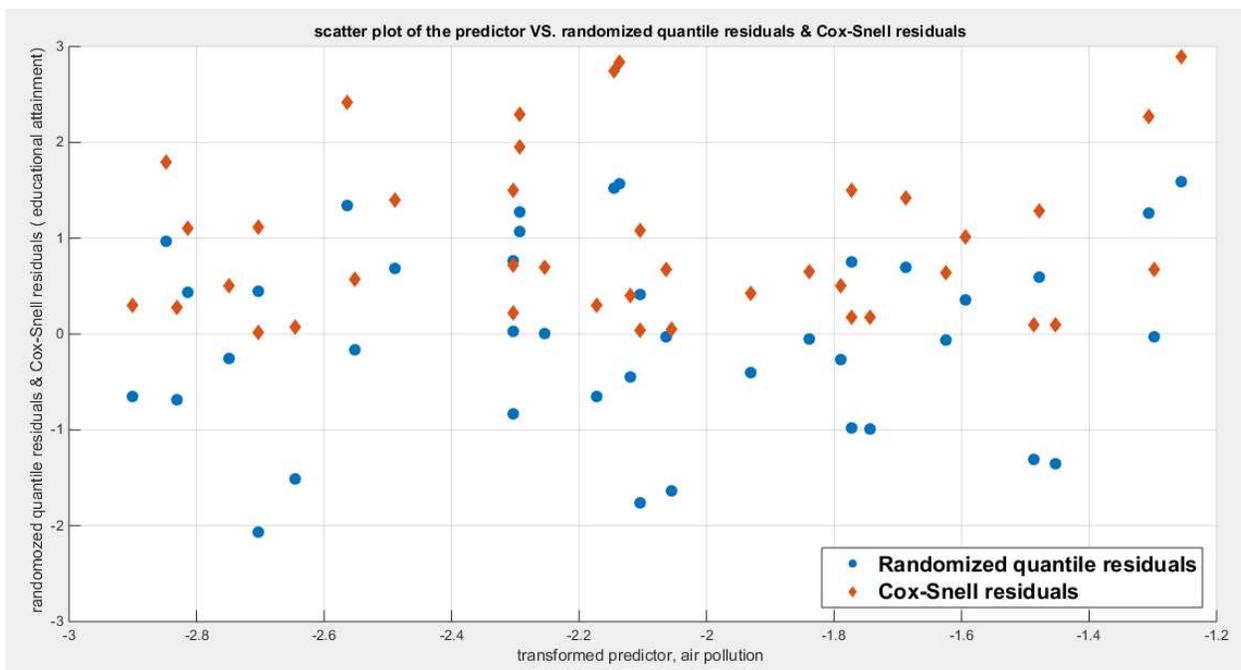

Fig. 17 shows the scatter plot of residuals of both types against transformed predictors.

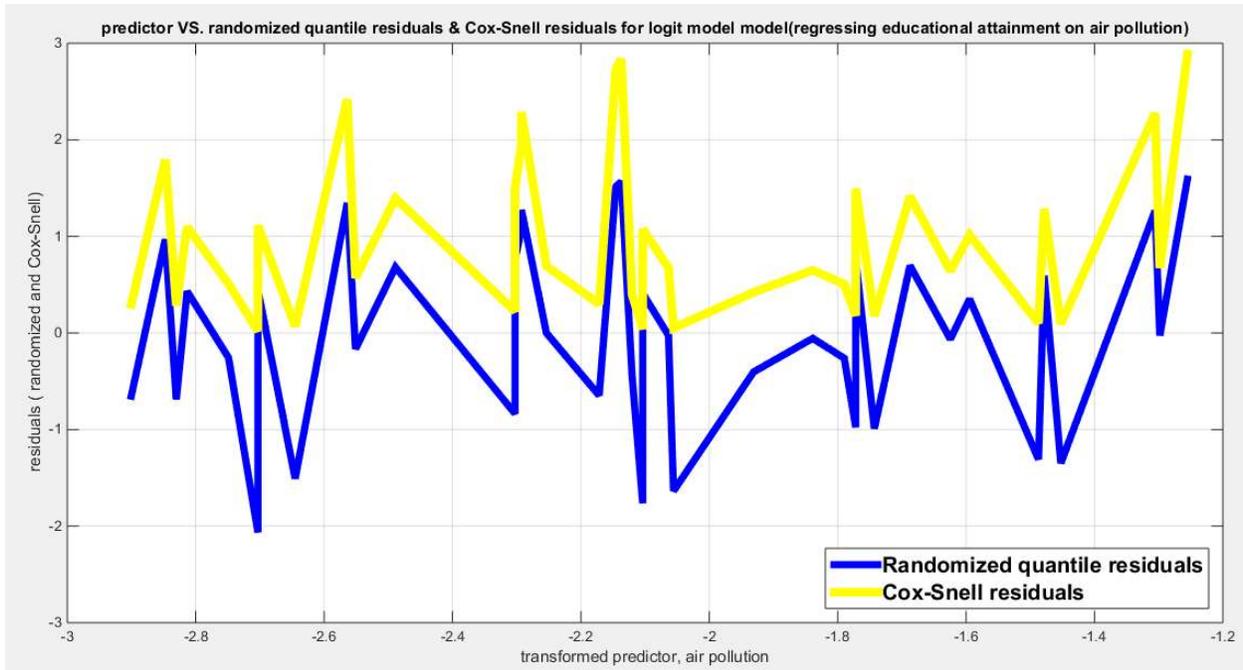

Fig. 18 shows the plot of residuals of both types against transformed predictors.

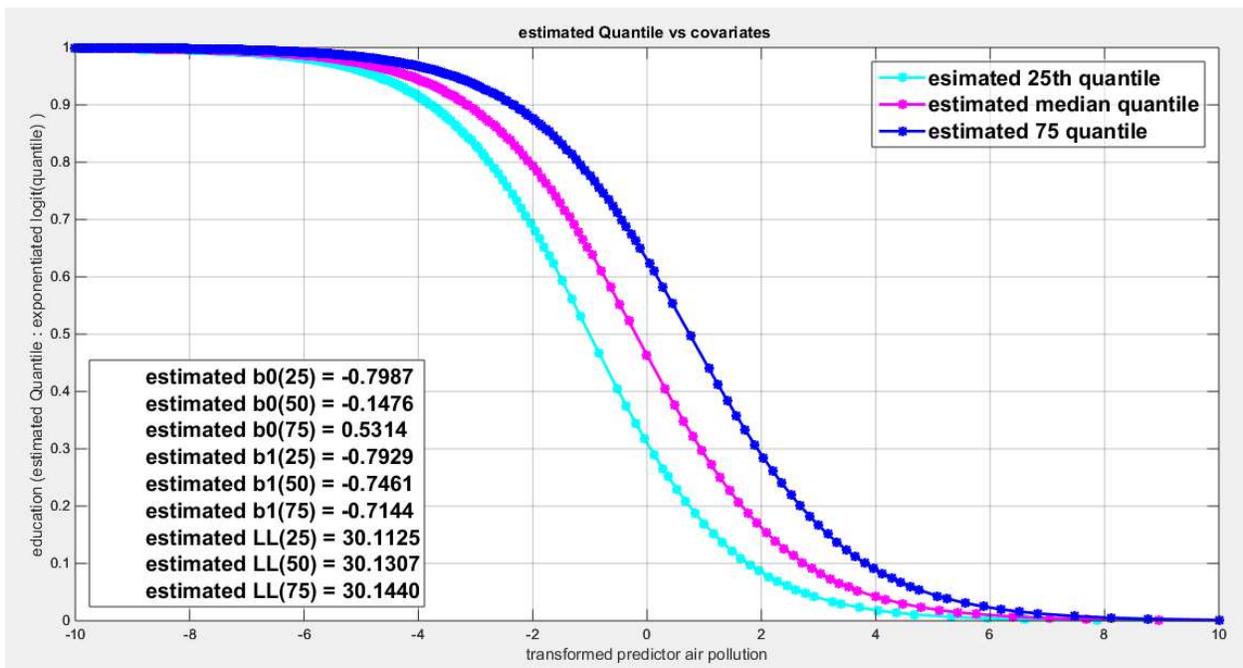

Fig. 19 shows parallel quantile curves across 25th , 50th ( median), 75th percentiles, suggesting that the predictor exerts a uniform influence on the response consistent with homoscedasticity. ( for the logit model)

Table 5: regressing education attainment on life satisfaction

| | Logit link function | | Log-log complementary | | Log-log median | |
|---|---|---|---|---|---|---|
| B0 | 11.9162 | | 5.5801 | | -10.8925 | |
| B1 | 3.8412 | | 1.8667 | | -3.4231 | |
| LL | 32.6712 | | 32.4988 | | 32.7175 | |
| Wald stat. of b0 | 3.897(p<0.025) | | 3.4421(p<0.025) | | 4.086(p<0.025) | |
| Wald stat. of b1 | 3.4101(p<0.025) | | 3.0962(p<0.025) | | 3.4988(p<0.025) | |
| AIC | -61.3425 | | -60.9975 | | -61.435 | |
| CAIC | -61.0182 | | -60.6732 | | -61.1107 | |
| BIC | -57.9647 | | -57.6198 | | -58.0573 | |
| HQIC | -60.1212 | | -59.7762 | | -60.2137 | |
| LRT | 13.2851 (p=2.6752e-4) | | 12.94019 p=3.2162e-4) | | 13.3776 (p= 2.5465e-4) | |
| R-squared | 0.2826 | | 0.2764 | | 0.2843 | |
| P-value for randomized quantile residuals | 0.6323 | | 0.6865 | | 0.6334 | |
| p-value for Cox-Snell residuals | 0.6323 | | 0.6865 | | 0.6334 | |
| Variance-covariance matrix | 9.3500 | 3.4414 | 2.6280 | 0.9767 | 7.1061 | 2.6057 |
| | 3.4414 | 1.2688 | 0.9767 | 0.3635 | 2.6057 | 0.9572 |
| QR vs. predictor(tau,p) | -0.1467 , 0.1946 | | -0.1467, 0.1946 | | -0.1467 , 0.1946 | |
| CS vs. predictor(tau,p) | -0.1467 , 0.1946 | | -0.1467, 0.1946 | | -0.1467 , 0.1946 | |

Table 5 shows that the predictor is significant as likelihood ratio test (LRT) is highly significant, it is more or less around 13 lesser than that of the employment rate whose LRT is around 23; the R squared is also high for this predictor; it is around 0.28 but less than that of the employment rate whose R square is around 0.44. The AIC, CAIC, BIC, HQIC and LL are more or less equal across the different models. The LL is around 32 which is less than that of the employment rate whose LL is around 37. The residuals plotted against the predictors show no specific trend and they are randomly scattered. The QQ plot of the randomized quantile residuals shows perfect alignment with the diagonal all through its course in contrast with the Cox Snell residuals that show this perfect alignment at the lower tail and the center. The estimated curve between the estimated median and the transformed predictor is increasing reflecting that the more the life satisfaction is, the more the percentage attaining the education is. The figure for the clog-log shows the same pattern. The log-log figure has the same pattern. The difference is mainly manifested in the slope of the estimated curve. To assess the assumption of constant variance in the median parametric regression model, residual-based diagnostic tests were conducted using both randomized quantile (RQ) and Cox-Snell (CS) residuals. For each type of residual, an auxiliary regression of the squared residuals on the corresponding predictor was estimated by ordinary least squares, and the null hypothesis of homoscedasticity ($H_0$: constant variance) was tested. The results indicted no significant relationship between the squared residuals and the predictor variable (CS: p=0.0908, R-squared=0.0734; RQ: p=0.0436, R-squared=0.103), suggesting that the variance of the residuals remained approximately constant across the range of the predictor. Furthermore, the magnitude of the CS residuals were within a reasonable range (five values between 2.0575 and 3.7885), which supports the absence of heteroscedasticity. These findings provide evidence that the fitted median regression model satisfies the homoscedasticity assumption. These results are from the logit model. Figures 20-26 show the previous results.

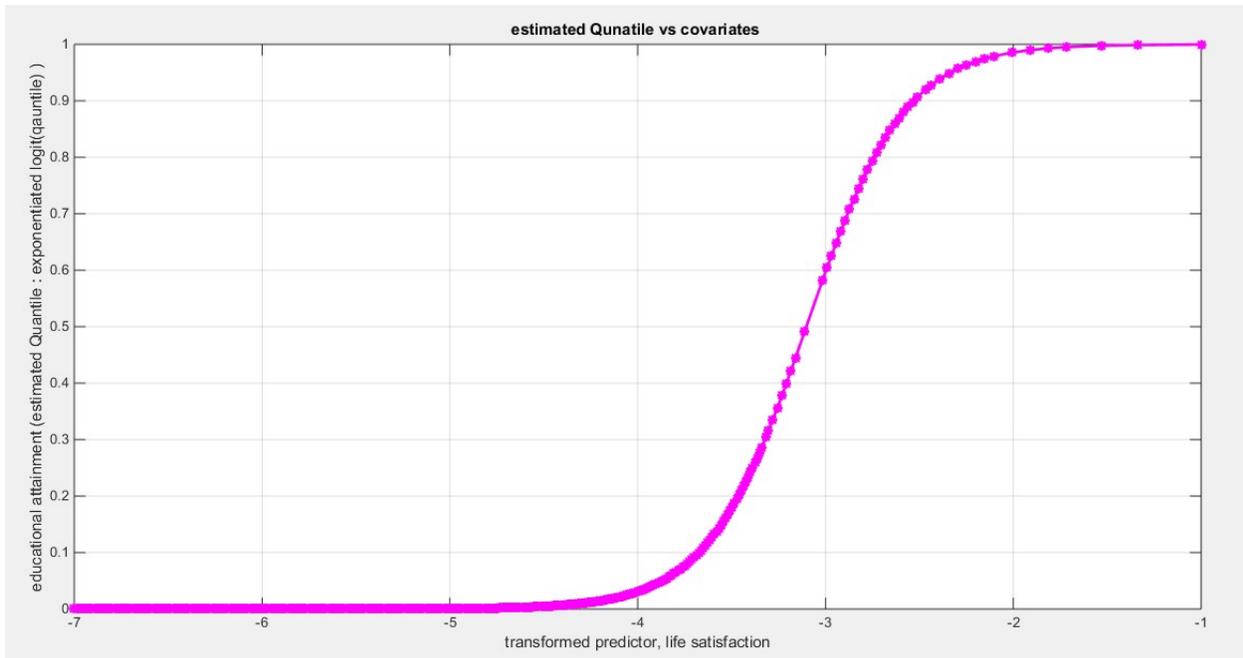

Fig. 20 shows the estimated curve plotting the transformed predictor against the estimated median (for the logit link).

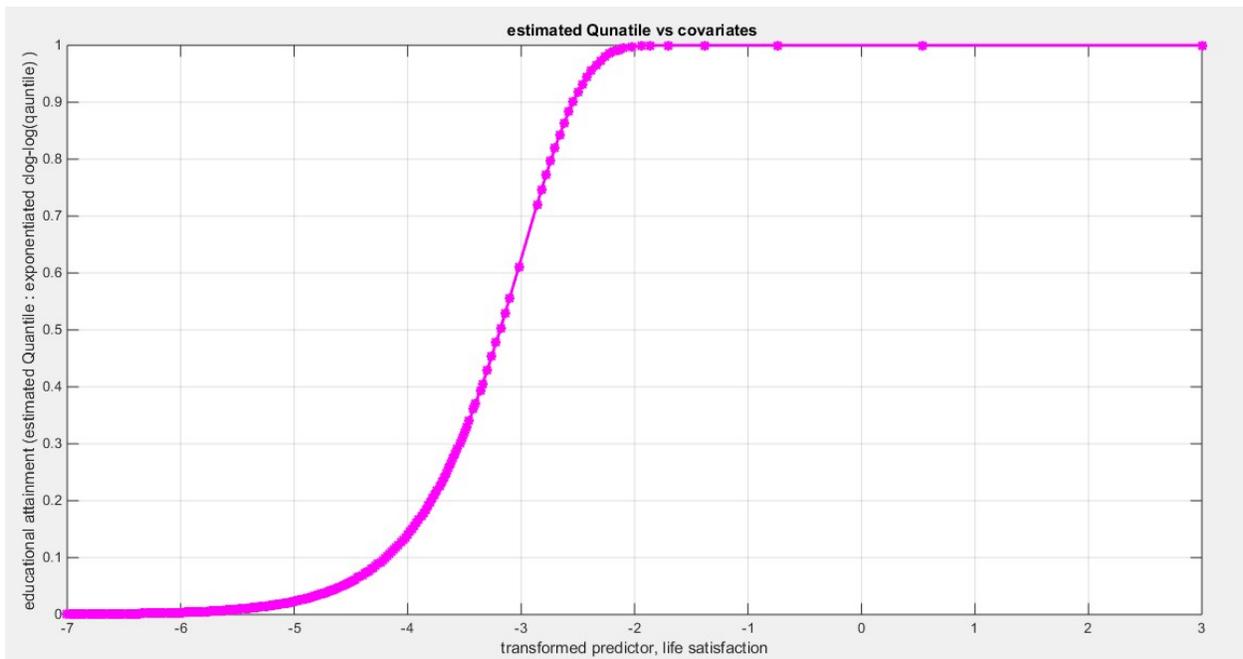

Fig. 21 shows the estimated curve plotting the transformed predictor against the estimated median (for the clog-log link).

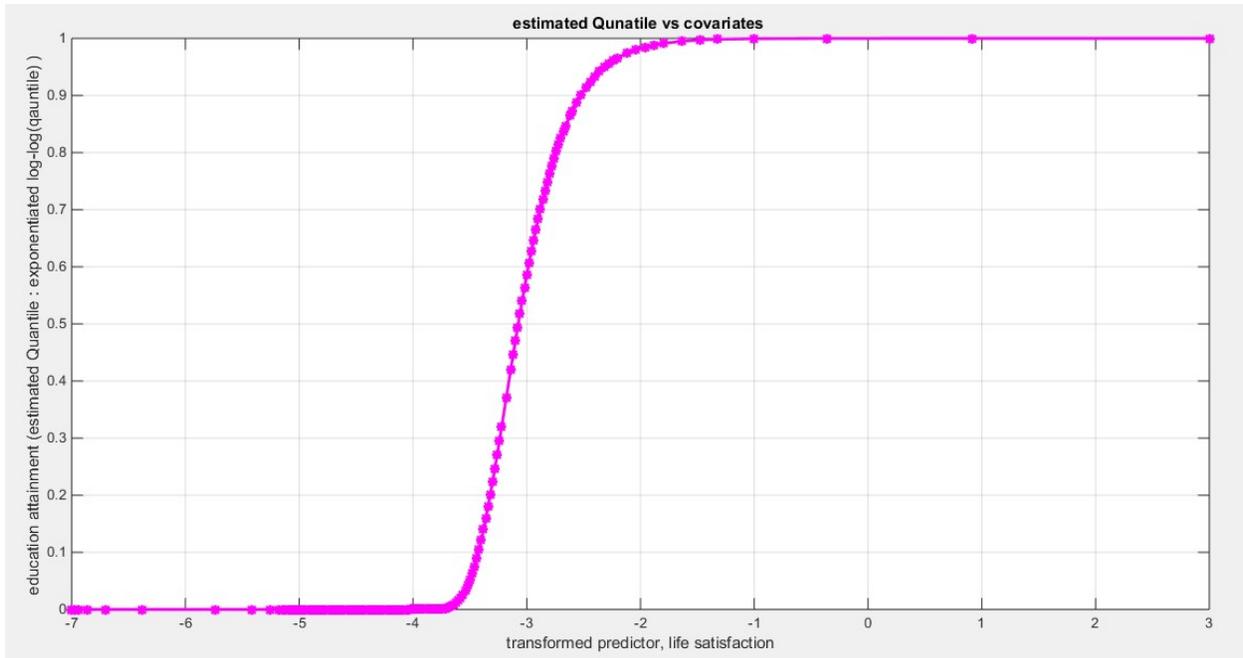

Fig. 22 shows the estimated curve plotting the transformed predictor against the estimated median (for the log-log link).

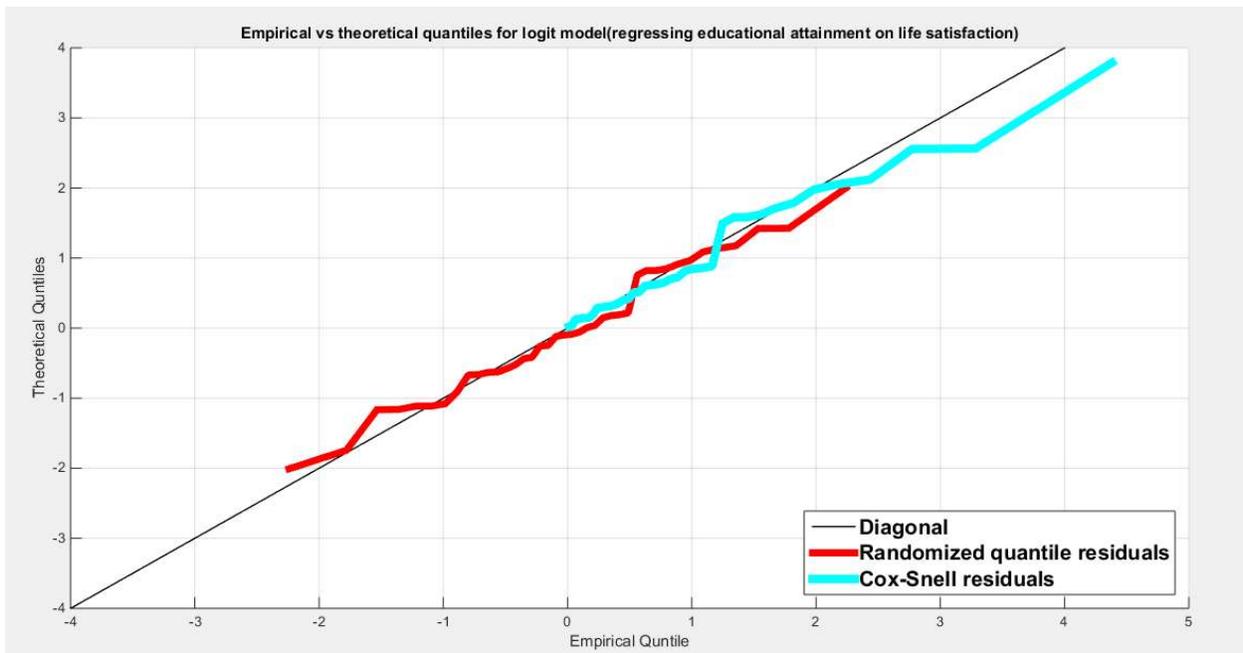

Fig. 23 shows the QQ plot of the empirical quantiles and the theoretical qunatiles for both types of residuals.

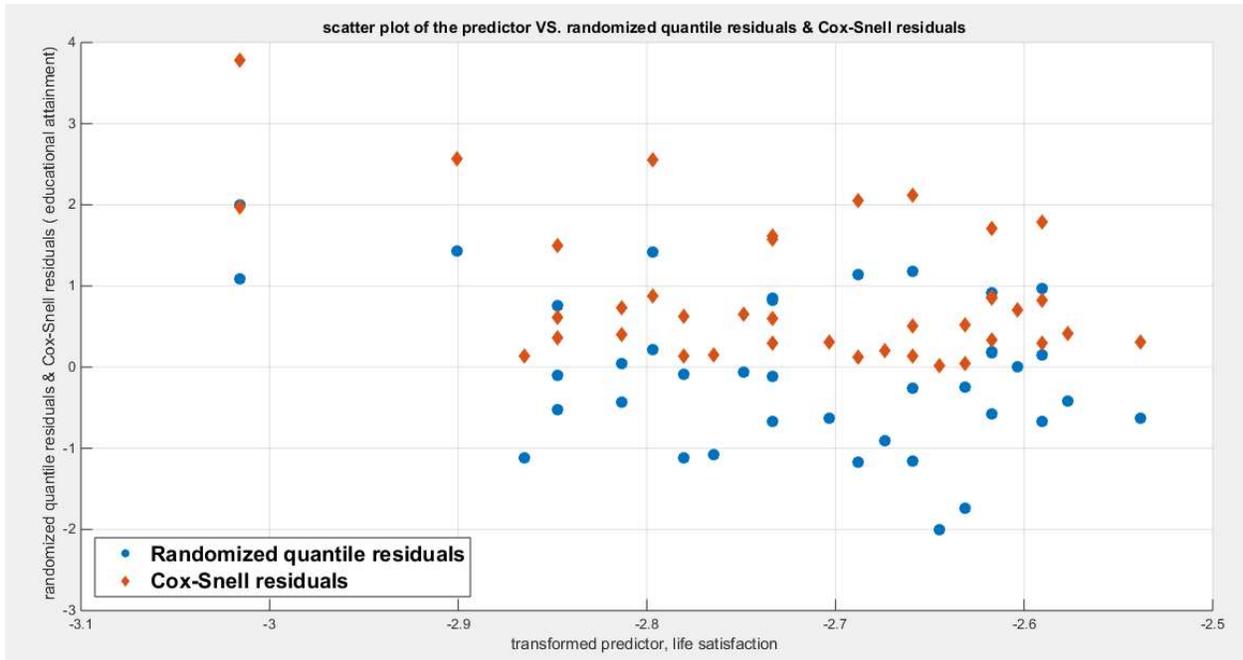

Fig. 24 shows the scatter plot of residuals of both types against transformed predictors.

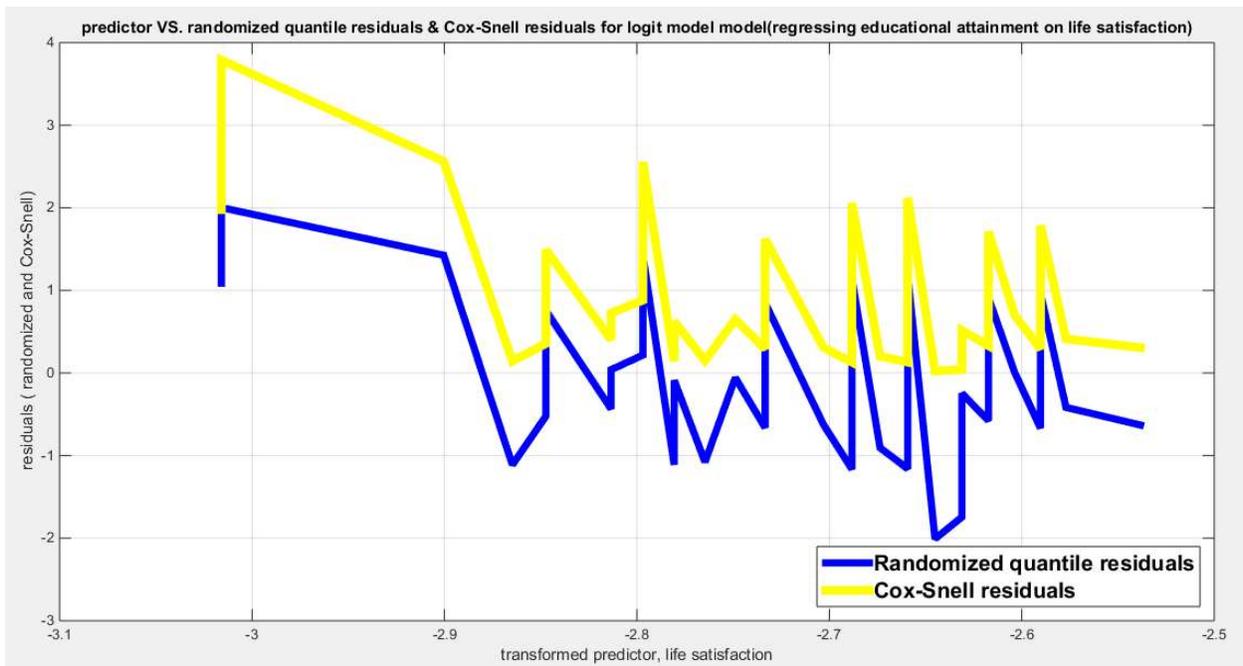

Fig. 25 shows the plot of residuals of both types against transformed predictors.

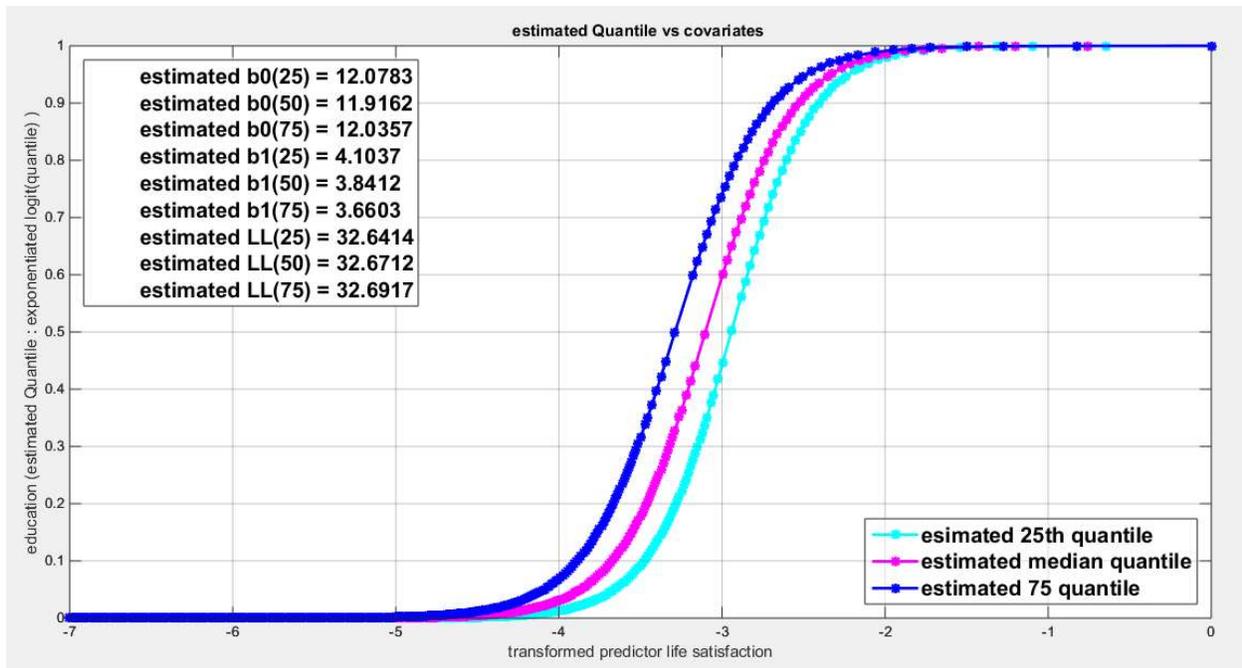

Fig. 26 shows parallel quantile curves across 25th, 50th ( median), 75th percentiles, suggesting that the predictor exerts a uniform influence on the response consistent with homoscedasticity. ( for the logit link)

Table 6: regressing education attainment on homicide rate

| | Logit link function | | Log-log complementary | | Log-log median | |
|---|---|---|---|---|---|---|
| B0 | 0.2813 | | -0.0635 | | -0.5062 | |
| B1 | -0.2619 | | -0.1260 | | 0.2367 | |
| LL | 30.5249 | | 30.2753 | | 30.6194 | |
| Wald stat. of b0 | 0.6419 (p>0.025) | | 0.2708 (p>0.025) | | 1.3386 (p>0.025) | |
| Wald stat. of b1 | 2.8079 (p<0.025) | | 2.6273(p<0.025) | | 2.8918(p<0.025) | |
| AIC | -57.0498 | | -56.5505 | | -57.2389 | |
| CAIC | -56.7255 | | -56.2262 | | -56.9146 | |
| BIC | -53.9721 | | -53.1728 | | -53.8611 | |
| HQIC | -55.8285 | | -49.4468 | | -56.0176 | |
| LRT | 8.9924 | | 8.4931 (p=0.0036) | | 9.1815 (p=0.0024) | |
| R-squared | 0.2013 | | 0.1913 | | 0.2051 | |
| P-value for randomized quantile residuals | 0.9571 | | 0.9360 | | 0.9628 | |
| p-value for Cox-Snell residuals | 0.9571 | | 0.9360 | | 0.9628 | |
| Variance-covariance matrix | 0.192 | 0.0391 | 0.055 | 0.0109 | 0.143 | 0.0295 |
| | 0.0391 | 0.0087 | 0.0109 | 0.0023 | 0.0295 | 0.0067 |
| QR vs. predictor(tau,p) | 0.0078, 0.9534 | | 0.0078,0.9534 | | 0.0078 , 0.9534 | |
| CS vs. predictor(tau,p) | 0.0078, 0.9534 | | 0.0078,0.9534 | | 0.0078 , 0.9534 | |

Table 6 shows that the predictor is significant as likelihood ratio test (LRT) is highly significant, it is more or less around 8.5 lesser than that of the air pollution whose LRT is around 8; the R squared is also high for this predictor; it is around 0.2 but more than that of the air pollution whose R square is around 0.18. The AIC, CAIC, BIC, HQIC and LL are more or less equal across the different models. The LL is around 30 which is less than that of the employment rate whose LL is around 37. The residuals plotted against the predictors show no specific trend and they are randomly scattered. The QQ plot of the randomized quantile residuals shows perfect alignment with the diagonal all through its course in contrast with the Cox Snell residuals that show this perfect alignment at the lower tail and the center. The estimated curve between the estimated median and the transformed predictor is decreasing reflecting that the more the homicide rate is, the more the percentage attaining the education is. The figure for the clog-log shows the same pattern. The log-log figure has also the same pattern. The difference is mainly manifested in the slope of the estimated curve. To assess the assumption of constant variance in the median parametric regression model, residual-based diagnostic tests were conducted using both randomized quantile (RQ) and Cox-Snell (CS) residuals. For each type of residual, an auxiliary regression of the squared residuals on the corresponding predictor was estimated by ordinary least squares, and the null hypothesis of homoscedasticity ($H_0$: constant variance) was tested. The results indicted no significant relationship between the squared residuals and the predictor variable (CS: p=0.237, R-squared=0.0366; RQ: p=0.379, R-squared=0.0205), suggesting that the variance of the residuals remained approximately constant across the range of the predictor. Furthermore, the magnitude of the CS residuals were within a reasonable range (four values between 2.2194 and 3.3203), which supports the absence of heteroscedasticity. These findings provide evidence that the fitted median regression model satisfies the homoscedasticity assumption. These results are from the logit model. Figures 27-33 show the previous results.

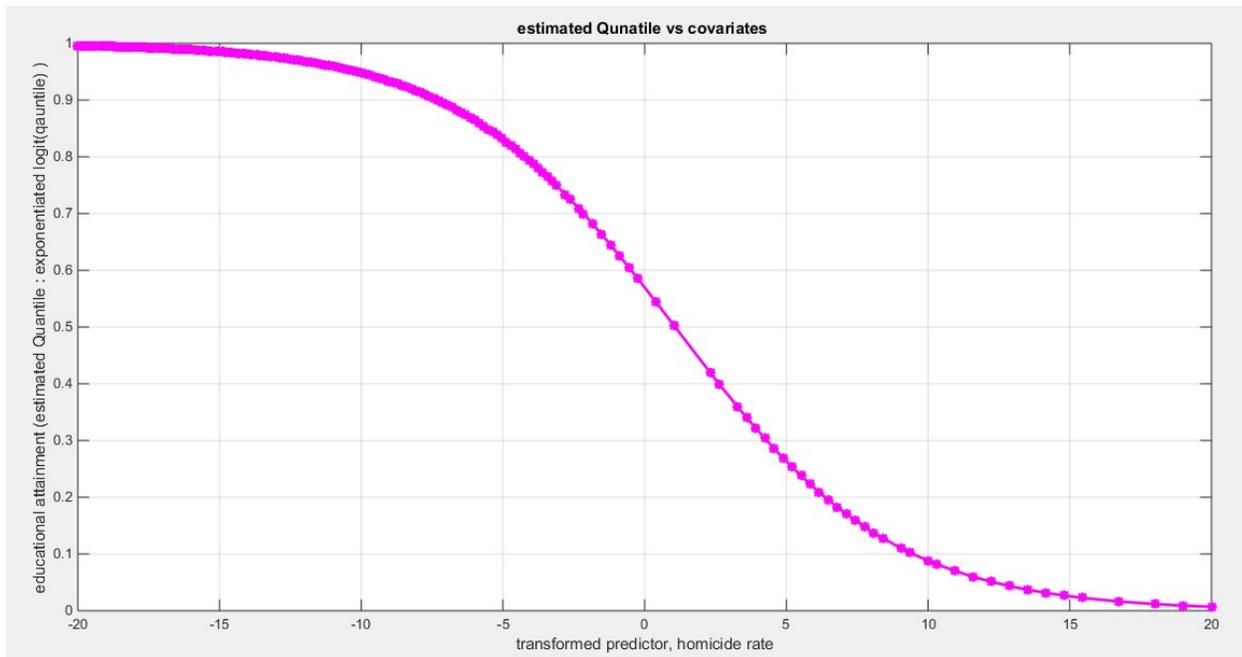

Fig. 27 shows the estimated curve plotting the transformed predictor against the estimated median (for the logit link).

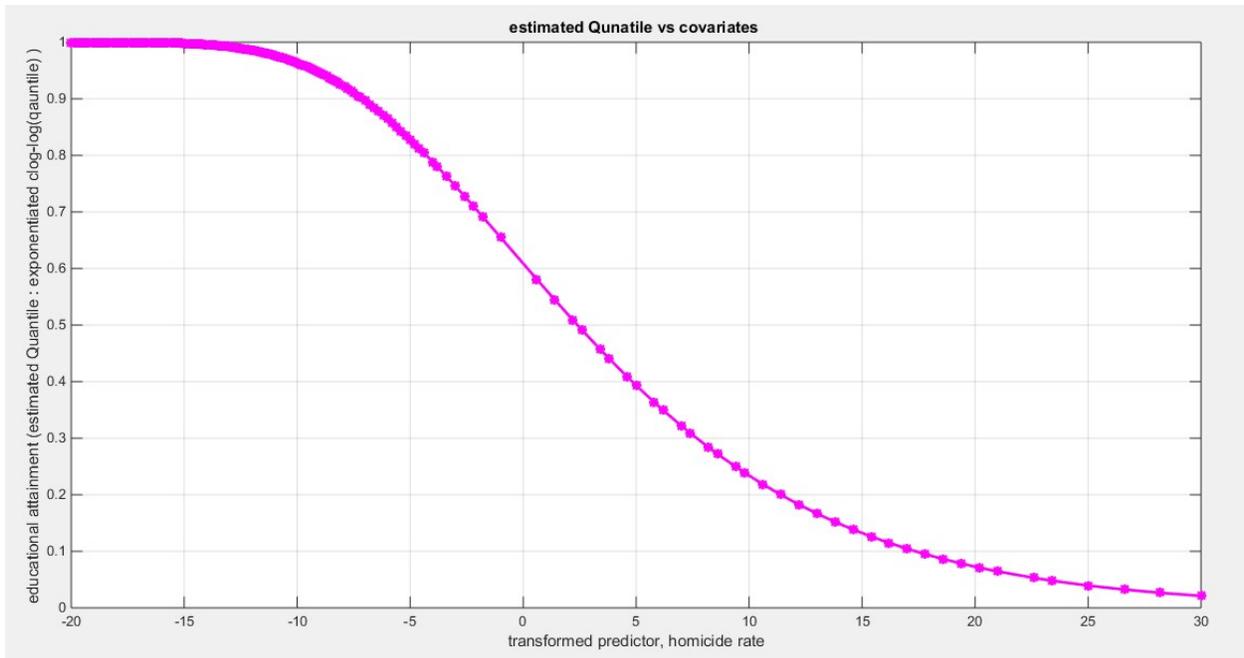

Fig. 28 shows the estimated curve plotting the transformed predictor against the estimated median (for the clog-log link).

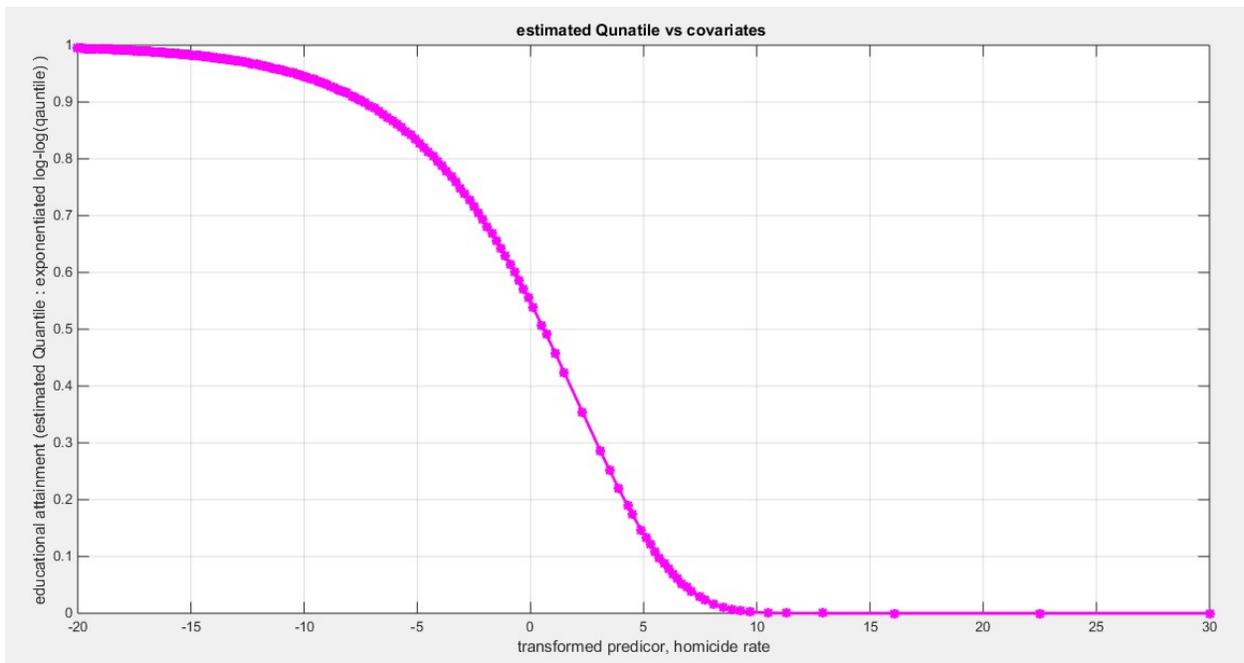

Fig. 29 shows the estimated curve plotting the transformed predictor against the estimated median (for the log-log link).

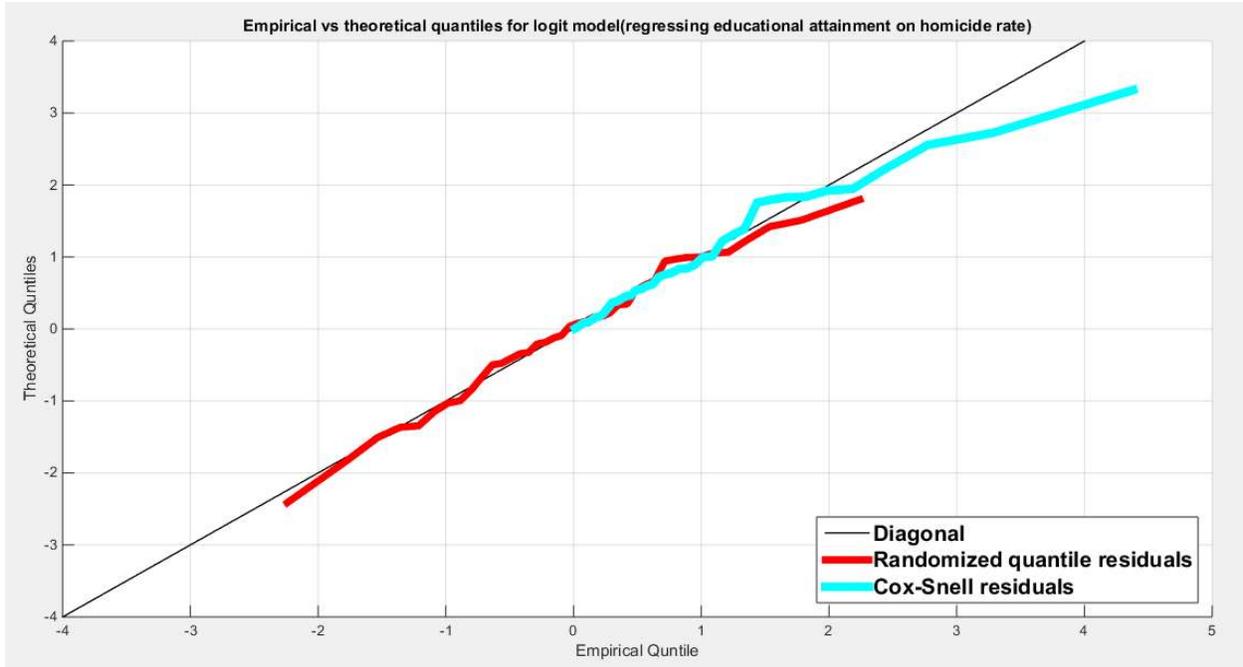

Fig. 30 shows the QQ plot of the empirical quantiles and the theoretical qunatiles for both types of residuals.

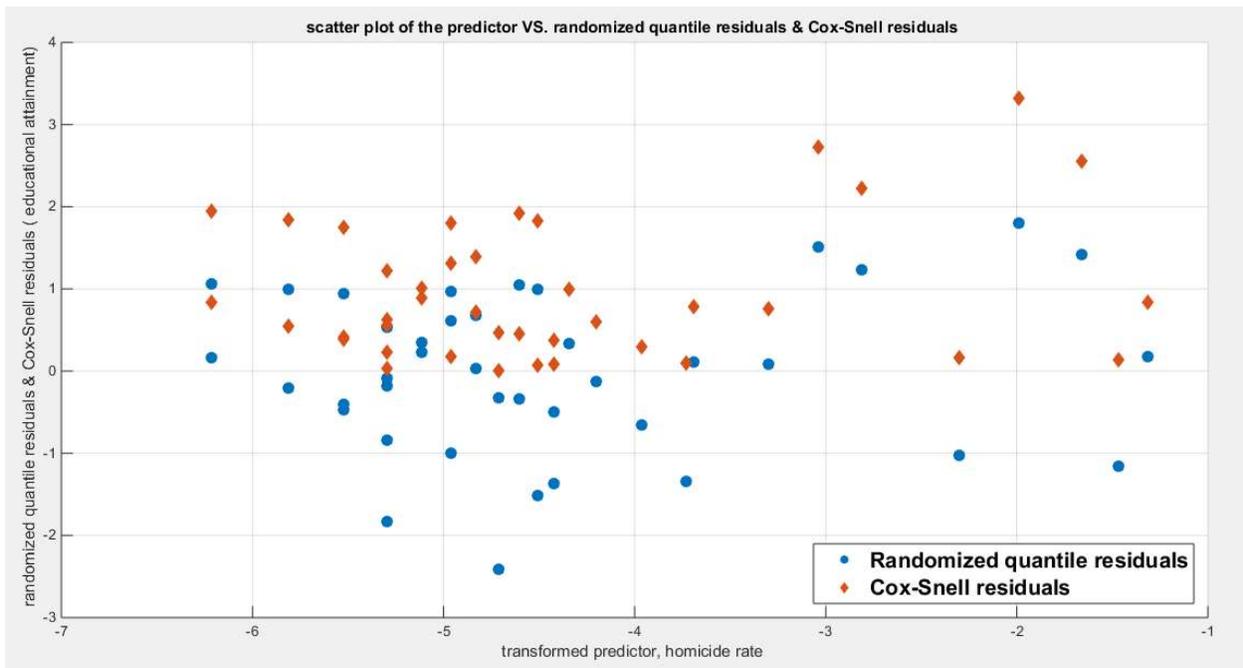

Fig. 31 shows the scatter plot of residuals of both types against transformed predictors.

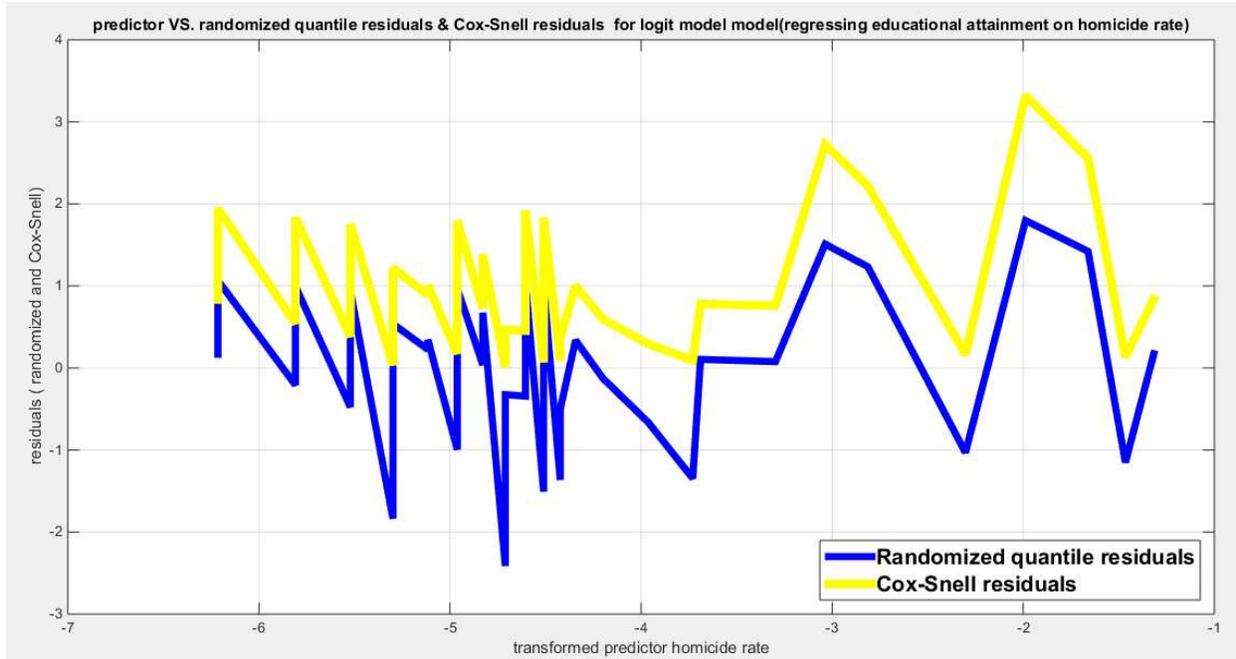

Fig. 32 shows the plot of residuals of both types against transformed predictors.

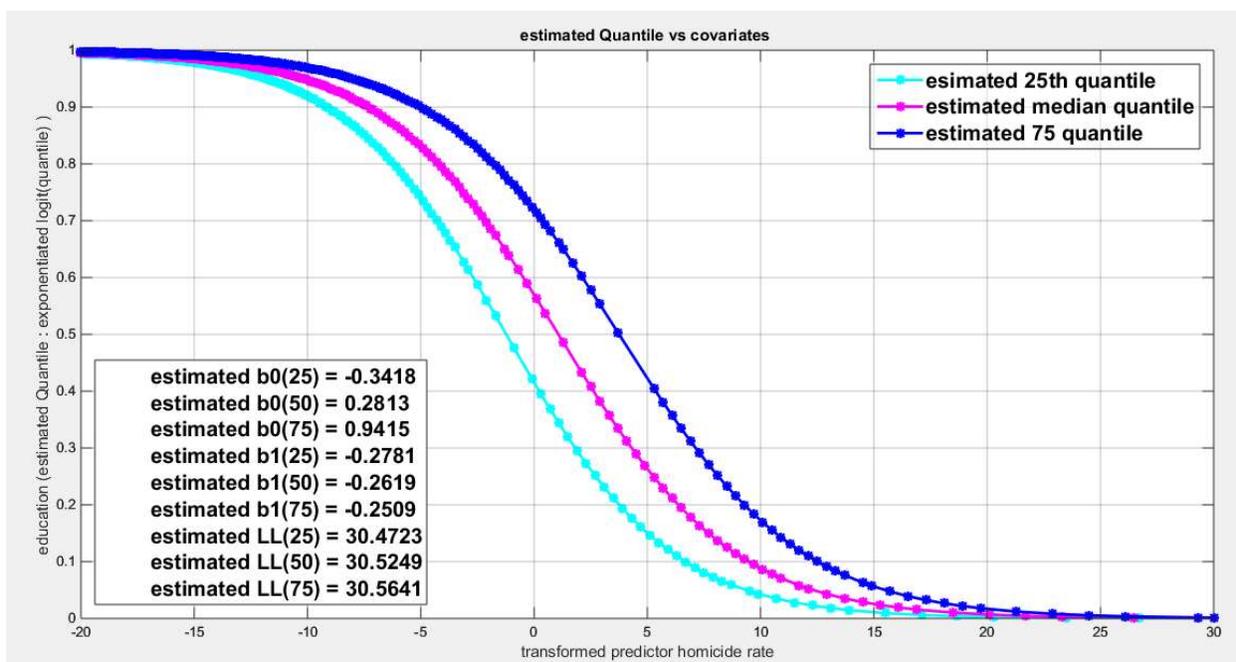

Fig. 33 shows parallel quantile curves across $25^{th}$, $50^{th}$ (median), $75^{th}$ percentiles, suggesting that the predictor exerts a uniform influence on the response consistent with homoscedasticity. ( for the logit link)

The marginal correlations between the variables (the response and the predictors) are shown in Table 7. The educational attainment shows positive and statistical significant correlation with the employment rate. The employment rate exhibits positive and statistical significant correlation with the life satisfaction. However; it depicts a negative and a statistical significant correlation with both the air pollution and the homicide rate. The life satisfaction has a negative and a statistically significant correlation with the homicide rate.

**Table 7: The marginal correlation matrix Kendall tau coefficient and associated p-value:**

|  | Educational Attainment(Y) | Employment Rate(X1) | Air Pollution(X2) | Life Satisfaction(X3) | Homicide Rate(X4) |
|---|---|---|---|---|---|
| **Educational Attainment(Y)** | 1 | 0.3184 P=0.005 | -0.1306 P=0.2477 | 0.1485 P=0.1939 | -0.0515 P=0.6567 |
| **Employment Rate (X1)** | 0.3184 P=0.005 | 1 | -0.4372 P=0.0001 | 0.5293 P=0.0000033 | -0.3137 P=0.0058 |
| **Air Pollution (X2)** | -0.1306 P=0.2477 | -0.4372 P=0.0001 | 1 | -0.4863 P=0.000016 | 0.1374 P=0.2244 |
| **Life Satisfaction(X3)** | 0.1485 P=0.1939 | 0.5293 P=0.0000033 | 0.5293 P=0.0000033 | 1 | -0.3655 P=0.0013 |
| **Homicide Rate (X4)** | -0.0515 P=0.6567 | -0.3137 P=0.0058 | 0.1374 P=0.2244 | -0.3655 P=0.0013 | 1 |

The condition indices obtained from standardized transformed X'X are 3.3505, 2.9709, 1.9377 and1. The VIF for the employment rate is 2.7441, for air pollution is 1.9916, for life satisfaction is 2.8607 and for homicide rate is 1.5120. So as the largest condition index is 3.3505 less than 10 and the VIF values are less than 5 so there are no evidence of significant multi-collinearity between the predictors.

The signs of the coefficients of the marginal correlations match those signs of the conditional correlations coefficients when regressing the educational attainment response variable on one predictor at a time. The employment rate is positively dependent with the education and the homicide rate is negatively dependent with education attainment. As will be shown later, in multiple regression analysis, this consistency is not applied to the air pollution, the life satisfaction, and the homicide rate as the sign of the coefficient flips. The air pollution may be a proxy for the industrial urbanization which increases the educational attainment; hence, the sign of the air pollution coefficient in the median regression equation may be positive rather than negative and this is called suppression effect of the proxy variable. Life satisfaction may be a proxy for the less motivated drive for education in some rich and high economic population, so this may decrease its effect on educational attainment and reverse the sign of the coefficient of the life satisfaction predictor. Homicide rate may be a proxy for urbanization where there are high rates of the gun ownership, high rates of the drug trafficking, gang violence, and organized crime. The kendall correlation of both air pollution and homicide rate are similar and this may indicate urbanization. So apparently increased homicide rate leads to increased educational attainment.

The author added these four predictors in one equation and used the different link functions, then removed each one at a time and calculated the LRT to assess the significance of this particular predictor while controlling for other predictors. The results are summarized in Table 8. Each row represents a model. The first row is the full model. The second model represents the model with removed predictor (Rx1 stands for the removed first predictor from the full model) and this is applicable to the remaining rows. The estimated standard error is recorded below each estimated coefficient for each variable. The colored area represents the removed predictor from the model. LRT is recorded with its associated p-value below it. The Log-likelihood (LL) is also recorded. There is a column documenting the sign preservation if all the predictors keep their sign consistent as in the simple regression models. The author used the logit link function for the full model and each of the nested and reduced models. All the models presented in Table 8 show the model adequacy diagnostic tests for residual types, the RQ and the CS residuals. Table 9 shows the AIC, CAIC, BIC, and HQIC for different models.

Table 8: the coefficients of parametric median regression analysis with removal of different predictors and associated standard error below each estimated coefficient value.

| | intercept | X1 Coeff. | X2 Coeff. | X3 Coeff. | X4 Coeff. | LL | LRT P-value | Preserve sign |
|---|---|---|---|---|---|---|---|---|
| Full model | 3.0176 5.0230 | 5.674 1.7695 | 0.4823 0.3893 | -0.6381 1.8157 | -0.0180 0.1160 | 38.9377 | 25.818 0.00003 | no |
| Rx1 | 8.6049 5.1119 | | -0.0244 0.3980 | 2.8502 1.6224 | -0.1301 0.1082 | 33.4488 | 10.9778 0.0009 | yes |
| Rx2 | 0.4894 4.480 | 4.8450 1.6233 | | -1.0764 1.746 | -0.0126 0.1158 | 38.1756 | 1.5243 0.217 | No |
| Rx3 | 4.7116 1.4354 | 5.3626 1.5301 | 0.5121 0.3816 | | -0.0135 0.1156 | 38.8769 | 0.1217 0.7272 | no |
| Rx4 | 3.2131 4.8766 | 5.7722 1.6607 | 0.4798 0.3889 | -0.6078 1.8096 | | 38.9257 | 0.0241 0.8765 | no |
| Rx1,x2 | 8.8170 3.7723 | | | 2.9109 1.2874 | -0.1312 0.1068 | 33.4469 | 0.9816 0.0041 | yes |
| Rx1,x3 | -0.3729 0.6120 | | -0.4772 0.3007 | | -0.1853 0.1044 | 31.7927 | 14.2900 0.00079 | Yes |
| Rx1,x4 | 10.8518 4.9746 | | -0.1062 0.3931 | 3.5325 1.5989 | | 32.7076 | 12.4602 0.002 | yes |
| Rx2,x3 | 3.1996 0.9087 | 4.212 1.2570 | | | -0.0042 0.1155 | 37.9890 | 1.8975 0.3872 | yes |
| Rx2,x4 | 0.6294 4.3015 | 4.9139 1.4984 | | -1.0554 1.7369 | | 38.1697 | 1.5361 0.4639 | no |
| Rx3,x4 | 4.7993 1.2272 | 5.4494 1.3473 | 0.5091 0.3806 | | | 38.8700 | 0.1355 0.9345 | no |
| Reduced model | 1.3793 | | | | | 26.0287 | | |

The first column of the table 8 shows the predictors that is removed from the regression equation. So Rx1 means X1 has been removed from the full model thus the other three predictors are the only ones involved in the regression. The model is named by the variable or variables removed. The shaded areas are the predictors removed from each model. The coefficients of the remaining predictors are recorded with the estimated standard error below each vlaue. The log likelihood (LL), the Likelihood Ratio Test (LRT) and its associated p value are also shown. In the next table the AIC, CAIC, BIC, HQIC and the pseudo R squared are illustrated. When X2 or X3 are removed from the full model, only the employment rate (X1) and homicide rate (X4) preserve their signs. When X1 is removed from the full model all the remaining predictors preserve their signs and LL is significantly decreased which denotes the importance of the X1 and this is reflected by high LRT with p value less than 0.05. Also removing X1 with any of the other predictors significantly lowers the LL and and hence increased the LRT significantly. As the employment rate is highly significant so it cannot be removed and hence the models that do not involve X1 are not of good choice to be considered even if the remaining predictors preserve their signs. Also the model with X1 removed with any other predictors like X2, X3 or X4 have LL more or less equal to the model where only X1 is removed; ranging from about 31 to 33 and also this is applicable to AIC, CAIC, and BIC. What about other models like the model where X2 and X4 are removed or X3 and X4 are removed? These models do not preserve the signs of the remaining predictors although their LL is high around 38 like the full model. This is in contrast to the model where the X2 and X3 are removed, the signs of the remaining predictors are consistent and preserved, the LL is high 37.989 nearly approximating LL of the model containing the X1 alone and adding X4 to X1 gives a model with nearly equal LL of 37.989, but the AIC and BIC differs because they penalize for the parameters. The model with only X1 has AIC of -71.9766 while the model that contains both X1 and X4 has AIC equals to -69.9779.  Moreover; the LL of the full model with 5 parameters is 38.9377 which is not far high from 37.989 the value of the LL of the model containing the X1 alone or of the model containing the X1 and X4. So it is a matter of choosing a model with high LL involving the significant predictors and keeping the sign of the coefficient consistent with the signs of the marginal and the conditional correlation coefficients taking into consideration the penalty paid for the number of the parameters. And this can have a logical and meaningful interpretation, the more the employment rate is, the more the educational attainment is and the less the homicide rate is, the more the educational attainment is. This also matches the sign of the coefficient obtained from the conditional correlation when regressing the educational attainment on one predictor. However, taking into account the penalty of the number of the parameters and the insignificance of X4, X4 can be removed from the model.  This discussion is applied when using the logit function. When applying this technique adding all predictors then removing one at a time and then more than one at a time for clog-log link function and the log-log link function, the results revealed that the only predictor to keep in the median regression equation is the employment rate and the other predictors can be removed.

Table 9 : AIC, CAIC, BIC, HQIC for different models

|  | AIC | CAIC | BIC | HQIC | R squared |
|---|---|---|---|---|---|
| Full model | -67.8754 | -66.1107 | -59.4310 | -64.8222 | 0.4756 |
| Rx1 | -58.8977 | -57.7548 | -52.1421 | -56.4551 | 0.2400 |
| Rx2 | -68.3511 | -67.2083 | -61.5956 | -65.9086 | 0.0374 |
| Rx3 | -69.7537 | -68.6109 | -62.9982 | -67.3111 | 0.003 |
| Rx4 | -69.8513 | -68.7085 | -63.0958 | -67.4081 | 0.000603 |
| Rx1,x2 | -60.8939 | -60.2272 | -55.8272 | -59.0619 | 0.2401 |
| Rx1,x3 | -57.5855 | -56.9188 | -52.5188 | -55.7535 | 0.3004 |
| Rx1,x4 | -59.4152 | -58.7486 | -54.3486 | -47.5833 | 0.2677 |
| Rx2,x3 | -69.9779 | -69.3113 | -64.9113 | -68.1460 | 0.0463 |
| Rx2,x4 | -70.3393 | -69.6727 | -65.2727 | -68.5074 | 0.0377 |
| Rx3,x4 | -71.7400 | -71.0733 | -66.6733 | -69.9080 | 0.0034 |
| Reduced model | -50.0574 | -49.9521 | -48.3685 | -49.4468 | |

# The Second Response Variable: The Water Quality

Water quality was regressed on five predictors; one at a time then all in one full model. Figure 34 shows the scatter plot that detects the relationship between the response variable and each predictor (transformed). The author presented figures for the logit model, figures that illustrate the estimated curve, plot between the predictor and the residuals, and the QQ plot for the empirical residuals against the theoretical residuals.

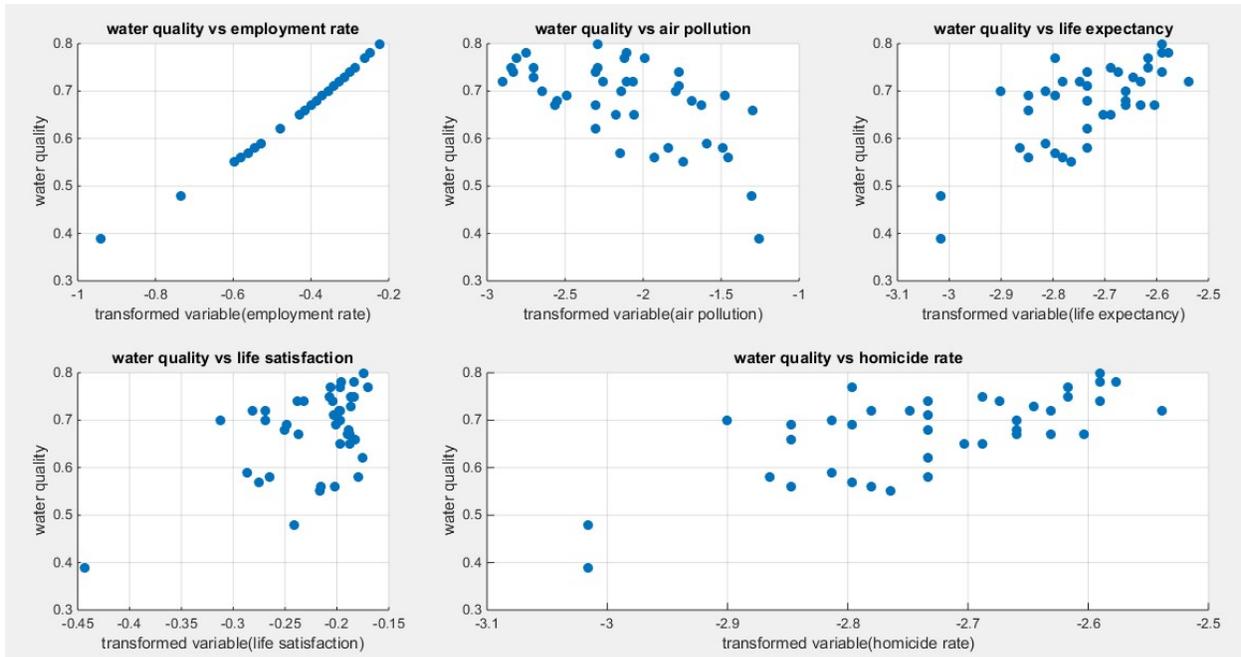

Fig. 34 shows the scatter plot of the response variable and each of the predictor. The relationship is nonlinear relationship except the relationship between the water quality and the employment rate.

Figure 34 shows that the relationship between the water quality and the employment rate is almost linear with positive correlation. On the other hand, the relationship between the water quality and the air pollution is negatively correlated. While the relationship between the water quality and either of the life satisfaction or the life expectancy is positively correlated. The relationship between the water quality and the homicide rate is nonlinear and not that obvious. Tables 10-14 show the results obtained from regressing the water quality on each predictor using different link functions and comparing the statistical indices as regards the estimated coefficients , the Likelihood Ratio Test (LRT ) and its p value, AIC, CAIC, BIC, HQIC and the LL.

Table 10: regressing water quality on employment rate

| | Logit link function | | Log-log complementary | | Log-log median | |
|---|---|---|---|---|---|---|
| B0 | 3.3262 | | 1.4120 | | -3.1982 | |
| B1 | 3.7028 | | 1.8497 | | -3.1702 | |
| LL | 47.9589 | | 48.5677 | | 47.6981 | |
| Wald stat. of b0 | 7.4507 (p < 0.025) | | 6.999 (p < 0.025) | | 8.3472 (p < 0.025) | |
| Wald stat. of b1 | 3.3373 (p < 0.025) | | 3.3947 (p < 0.025) | | 3.4617 (p < 0.025) | |
| AIC | -91.9179 | | -93.1354 | | -91.3962 | |
| CAIC | -91.6021 | | -92.8196 | | -91.0804 | |
| BIC | -88.4907 | | -89.7083 | | -87.9691 | |
| HQIC | -90.6699 | | -91.8874 | | -90.1483 | |
| LRT | 14.9227 (p=1.1201e-4) | | 16.1402(p=5.8822e-5) | | 14.4010 (p=1.4772e-4) | |
| R-squared | 0.3051 | | 0.3254 | | 0.2962 | |
| P-value for randomized quantile residuals | 0.2411 | | 0.2759 | | 0.3517 | |
| p-value for Cox-snell residuals | 0.2411 | | 0.2759 | | 0.3517 | |
| Variance-covariance matrix | 0.1993 | 0.4766 | 0.0407 | 0.1064 | 0.1468 | 0.3355 |
| | 0.4766 | 1.2310 | 0.1064 | 0.2969 | 0.3355 | 0.8387 |
| QR vs. predictor(tau,p) | -0.0174,0.8837 | | -0.0161,0.8926 | | -0.0174,0.8837 | |
| CS vs. predictor(tau,p) | -0.0174,0.8837 | | -0.0161,0.8926 | | -0.0174,0.8837 | |

Table 10 shows that the predictor is significant as likelihood ratio test (LRT) is highly significant; the R squared is also high for this predictor between 0.2962 and 0.3254 across the different link functions. The AIC, CAIC, BIC, HQIC and LL are more or less equal across the different models. The LL is between 47.6981 and 48.5677 across the link functions. The residuals plotted against the predictors show no specific trend and they are randomly scattered. The QQ plot of the randomized quantile residuals shows perfect alignment with the diagonal all through its course in contrast with the Cox Snell residuals that show this perfect alignment at the lower tail and the center. The estimated curve between the estimated median and the transformed predictor is increasing reflecting that the more the employment rate is, the more the percentage expressing increased quality of water supply and cleanliness is. The figure for the clog-log shows the same pattern. The log-log figure has the same pattern. The difference is mainly manifested in the slope of the estimated curve. To assess the assumption of constant variance in the median parametric regression model, residual-based diagnostic tests were conducted using both randomized quantile (RQ) and Cox-Snell (CS) residuals. For each type of residual, an auxiliary regression of the squared residuals on the corresponding predictor was estimated by ordinary least squares, and the null hypothesis of homoscedasticity ($H_0$: constant variance) was tested. The results indicted no significant relationship between the squared residuals and the predictor variable (CS: p=0.538, R-squared=0.0098; RQ: p=0.408, R-squared=0.0176), suggesting that the variance of the residuals remained approximately constant across the range of the predictor. Furthermore, the magnitude of the CS residuals were within a reasonable range (four values between 2.1136 and 3.0478), which supports the absence of heteroscedasticity. These findings provide evidence that the fitted median regression model satisfies the homoscedasticity assumption. These results are from the logit model. Figures 35-41 show the previous results.

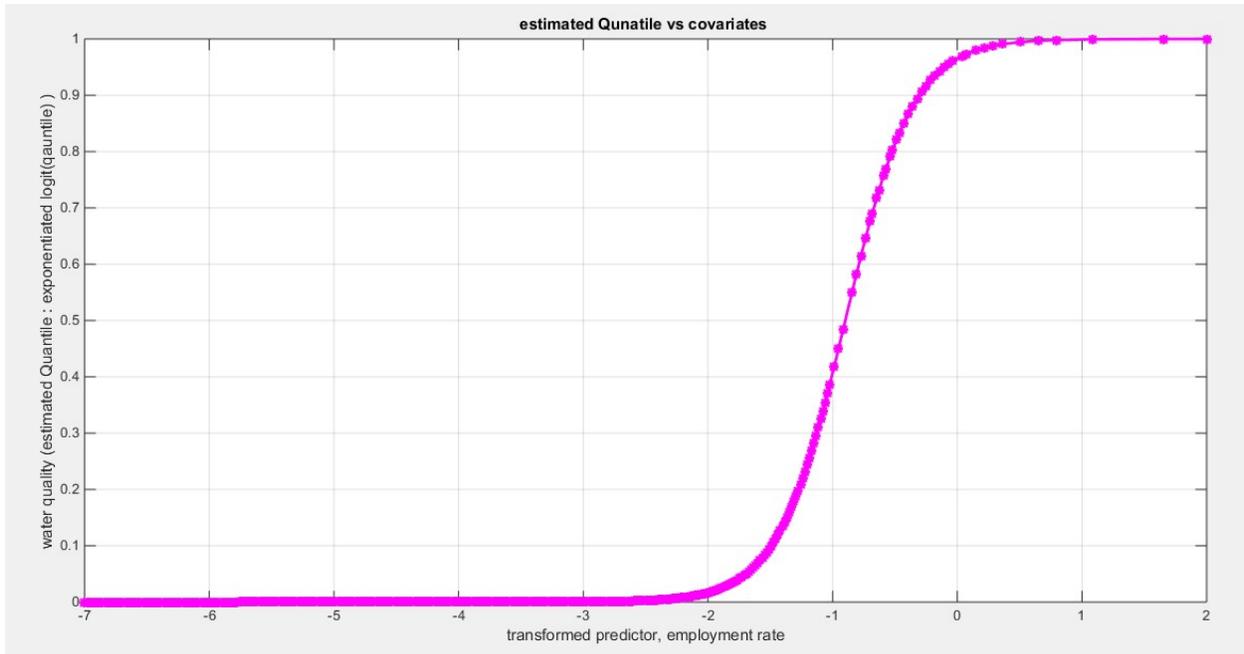

Fig. 35 shows the estimated curve plotting the transformed predictor against the estimated median (for the logit link).

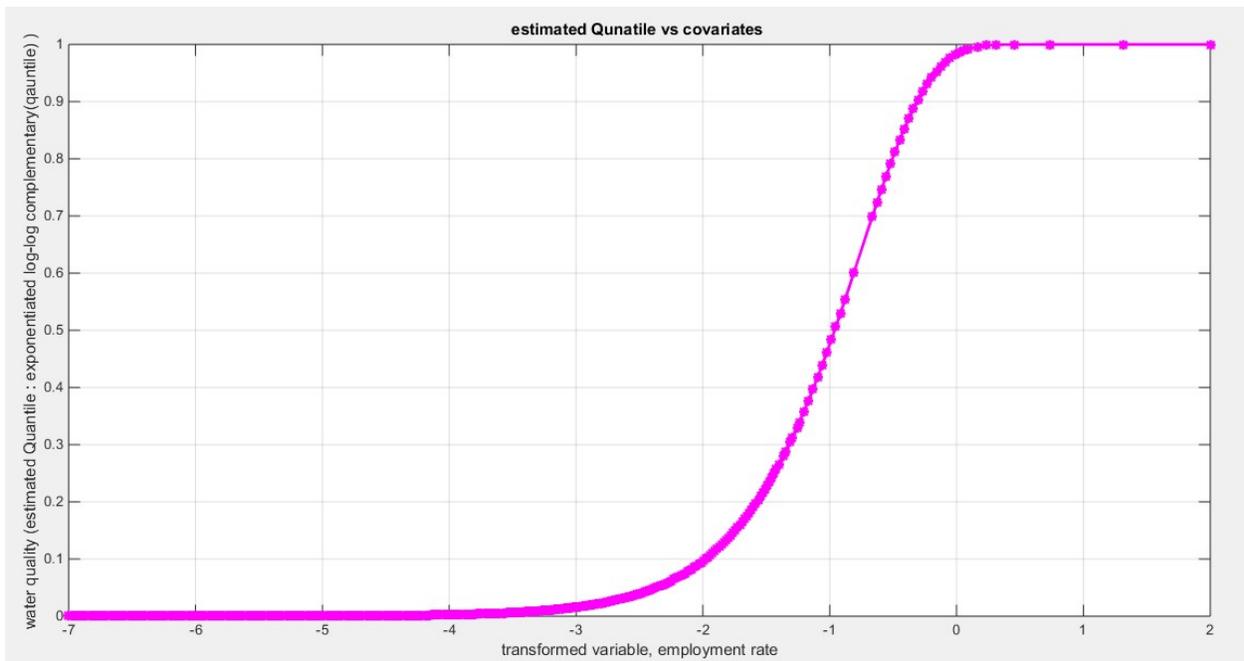

Fig. 36 shows the estimated curve plotting the transformed predictor against the estimated median (for the clog-log link).

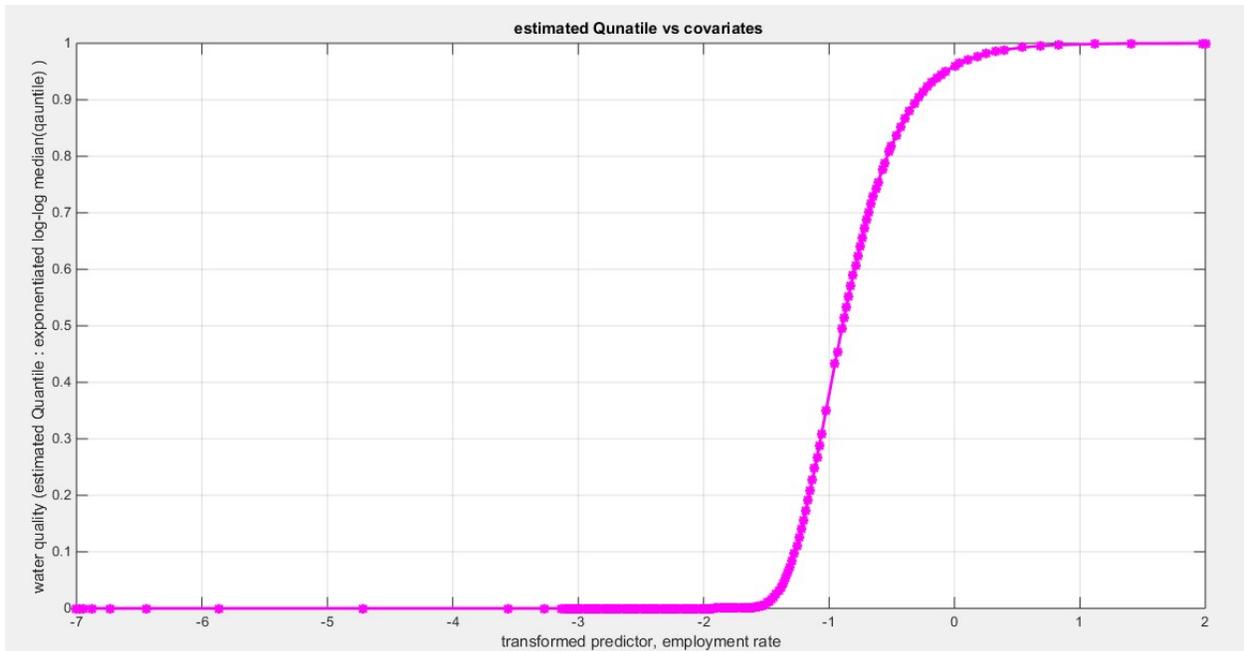

Fig. 37 shows the estimated curve plotting the transformed predictor against the estimated median (for the log-log link).

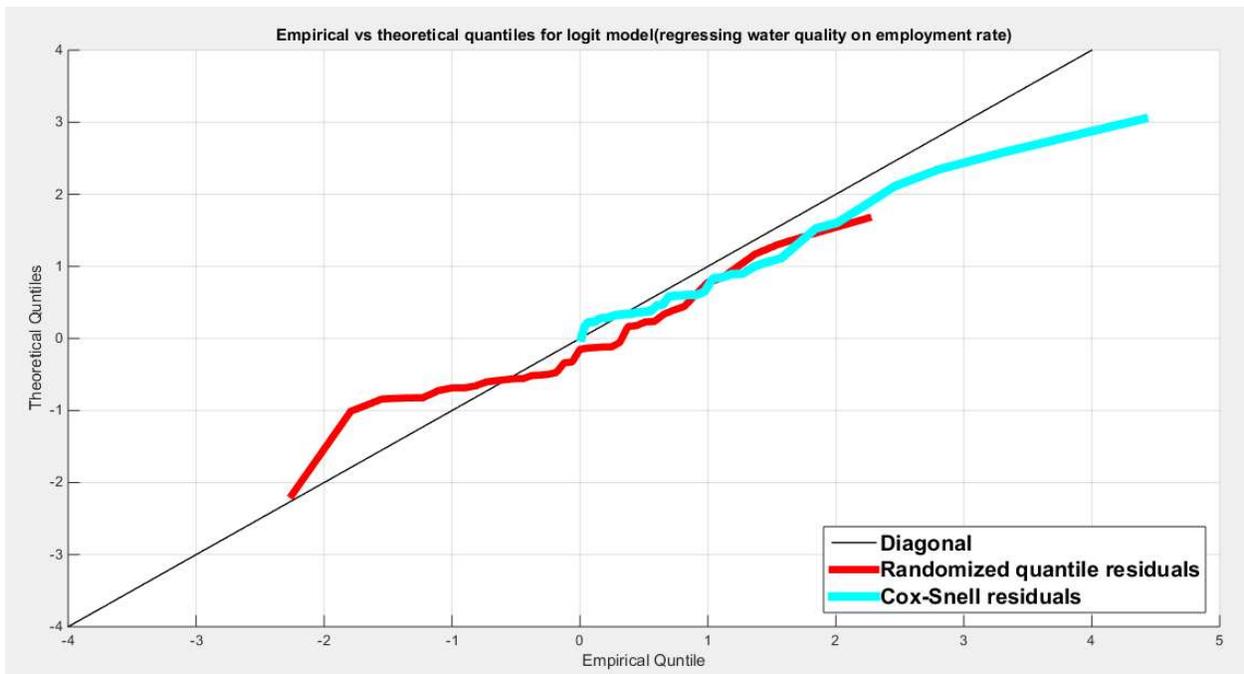

Fig. 38 shows the QQ plot of the empirical quantiles and the theoretical qunatiles for both types of residuals.

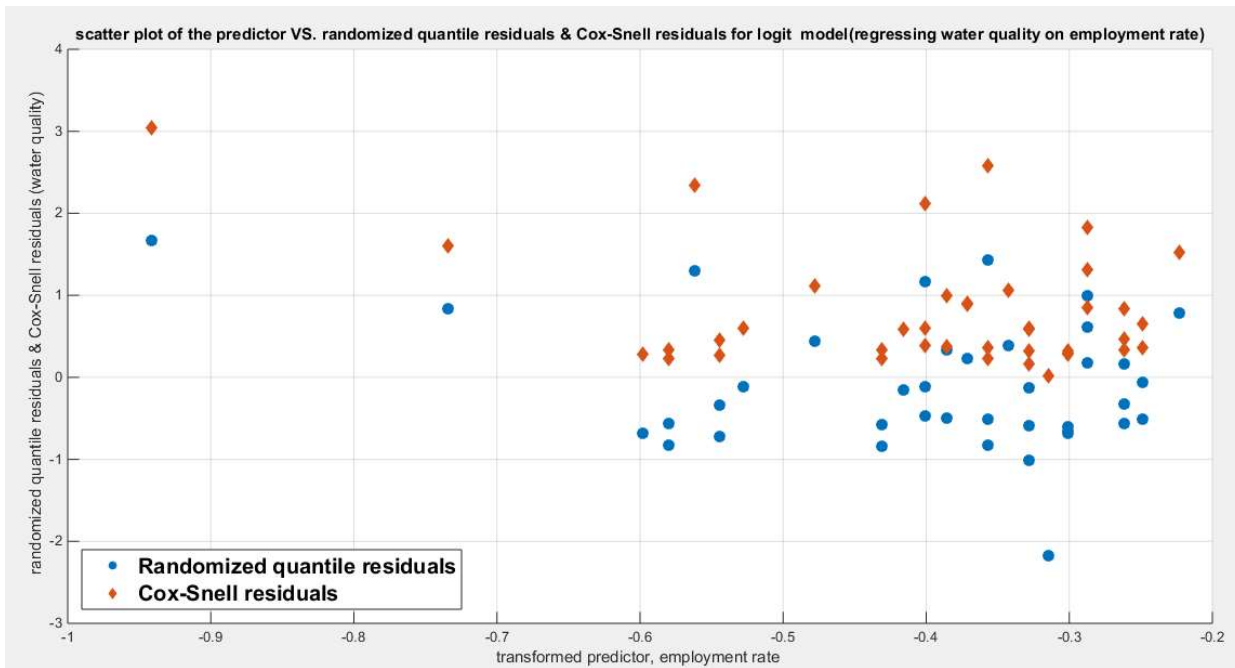

Fig. 39 shows the scatter plot of residuals of both types against transformed predictors.

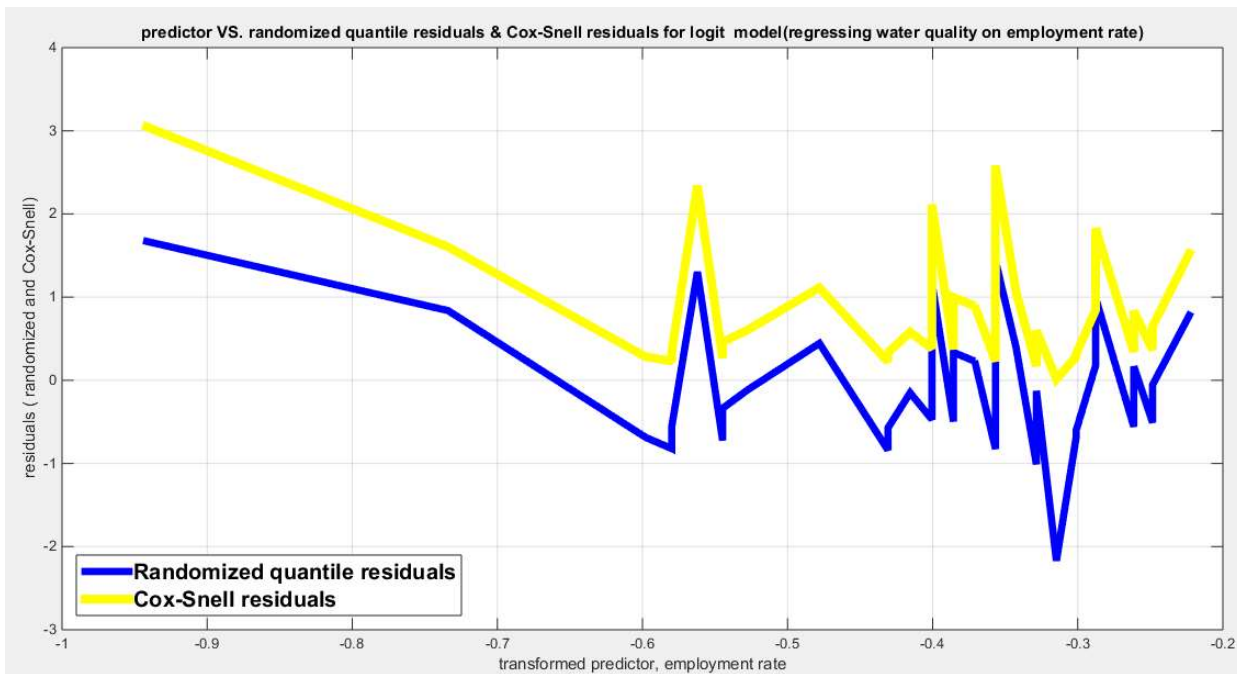

Fig. 40 shows the plot of residuals of both types against transformed predictors.

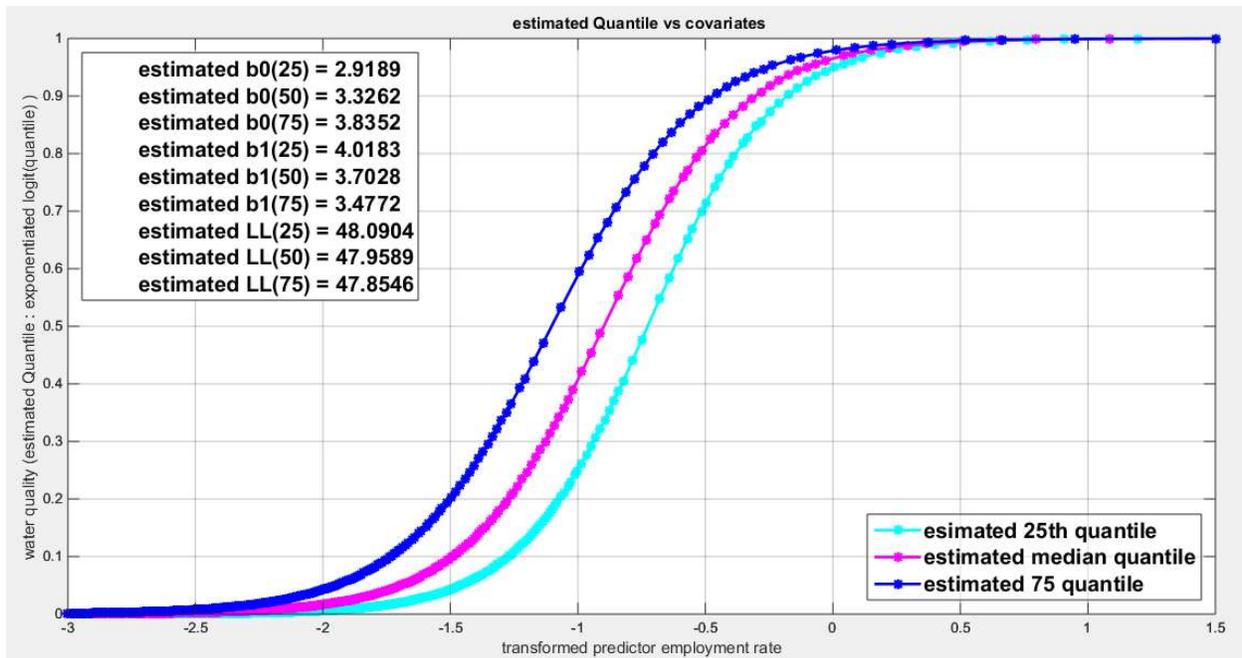

Fig. 41 shows parallel quantile curves across 25th , 50th ( median), 75th percentiles, suggesting that the predictor exerts a uniform influence on the response consistent with homoscedasticity.

Table 11: Regressing water quality on air pollution

| | Logit link function | | Log-log complementary | | Log-log median | |
|---|---|---|---|---|---|---|
| B0 | -0.1521 | | -0.2213 | | -0.0937 | |
| B1 | -0.9411 | | -0.4233 | | 0.8630 | |
| LL | 46.2983 | | 46.5111 | | 46.2508 | |
| Wald stat. of b0 | 0.249(p>0.025) | | 0.7878 (p>0.025) | | 0.1704(p>0.025) | |
| Wald stat. of b1 | 3.369(p<0.025) | | 3.4562 (p<0.025) | | 3.3798(p<0.025) | |
| AIC | -88.5967 | | -89.0223 | | -88.5016 | |
| CAIC | -88.2809 | | -88.7065 | | -88.1858 | |
| BIC | -85.1695 | | -85.5951 | | -85.0745 | |
| HQIC | -87.3487 | | -87.7743 | | -87.2537 | |
| LRT | 11.6015(p-val.=0.0007) | | 12.0271(p-val=0.00052) | | 11.5064(0.0007) | |
| R-squared | 0.2465 | | 0.2442 | | 0.2447 | |
| P-value for randomized quantile residuals | 0.7791 | | 0.8083 | | 0.778 | |
| p-value for Cox-snell residuals | 0.7791 | | 0.8083 | | 0.778 | |
| Variance-covariance matrix | 0.3706 | 0.1666 | 0.0789 | 0.0339 | 0.3024 | 0.1375 |
| | 0.1666 | 0.0780 | 0.0339 | 0.015 | 0.1375 | 0.0652 |
| QR vs. predictor(t,p) | 0.0159,0.8927 | | 0.0159,0.8927 | | 0.0159,0.8927 | |
| CS vs. predictor(t,p) | 0.0159,0.8927 | | 0.0159,0.8927 | | 0.0159,0.8927 | |

Table 11 shows that the predictor is significant as likelihood ratio test (LRT) is highly significant; the R squared is also high for this predictor between 0.24 across the different link functions. The AIC, CAIC, BIC, HQIC and LL are more or less equal between the different models. The LL is around 46 across the link functions. The residuals plotted against the predictors show no specific trend and they are randomly scattered. The QQ plot of the randomized quantile residuals shows perfect alignment with the diagonal all through its course in contrast with the Cox Snell residuals that show this perfect alignment at the lower tail and the center. The estimated curve between the estimated median and the transformed predictor is decreasing reflecting that the more the air pollution is, the less the percentage expressing the increased quality of the water supply and cleanliness is. The figure for the clog-log shows the same pattern. The log-log figure has the same pattern. The difference is mainly manifested in the slope of the estimated curve.an. To assess the assumption of constant variance in the median parametric regression model, residual-based diagnostic tests were conducted using both randomized quantile (RQ) and Cox-Snell (CS) residuals. For each type of residual, an auxiliary regression of the squared residuals on the corresponding predictor was estimated by ordinary least squares, and the null hypothesis of homoscedasticity ($H_0$: constant variance) was tested. The results indicted no significant relationship between the squared residuals and the predictor variable (CS: p=0.0751, R-squared=0.079; RQ: p=0.0378, R-squared=0.106), suggesting that the variance of the residuals remained approximately constant across the range of the predictor. Furthermore, the magnitude of the CS residuals were within a reasonable range (four values between 2.0575 and 2.8284), which supports the absence of heteroscedasticity. These findings provide evidence that the fitted median regression model satisfies the homoscedasticity assumption. These results are from the logit model. Figures 42-48 show the previous results.

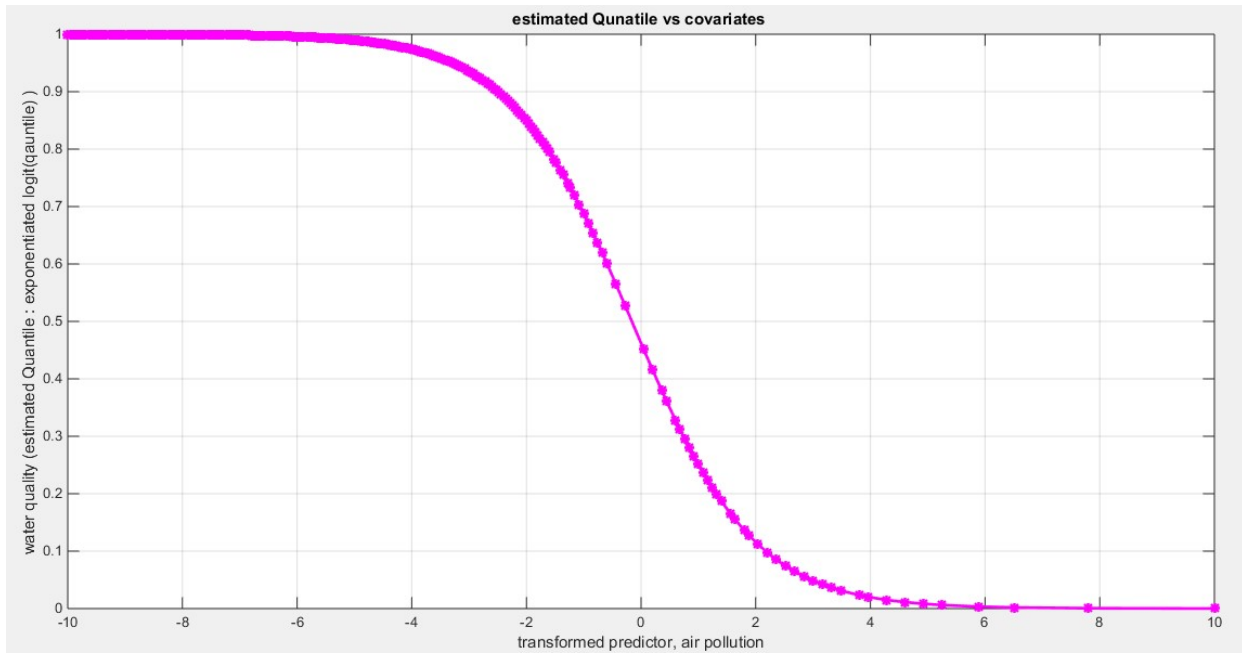

Fig. 42 shows the estimated curve plotting the transformed predictor against the estimated median (for the logit link).

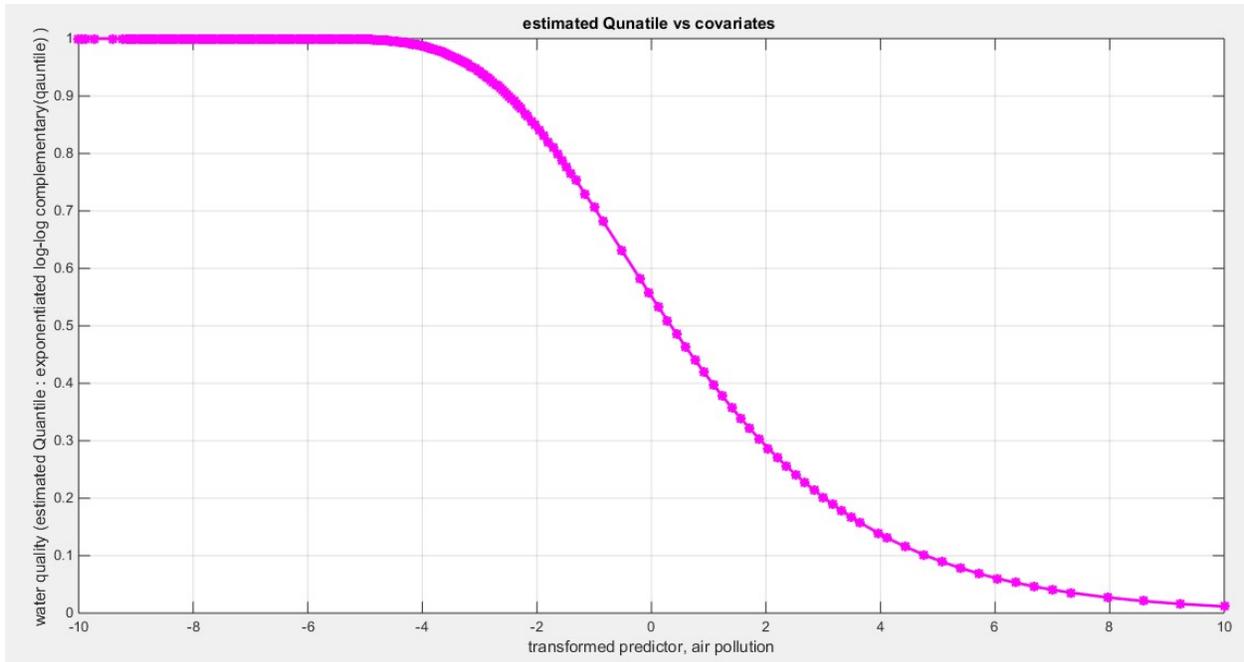

Fig. 43 shows the estimated curve plotting the transformed predictor against the estimated median (for the clog-log link).

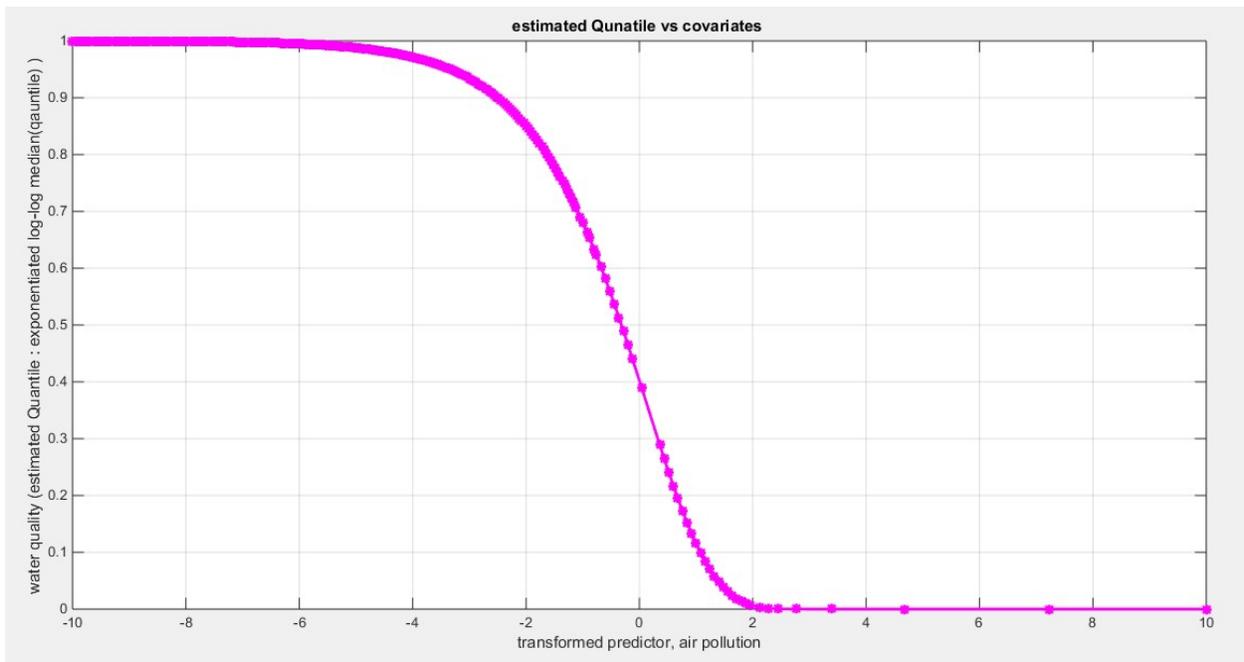

Fig. 44 shows the estimated curve plotting the transformed predictor against the estimated median (for the log-log link).

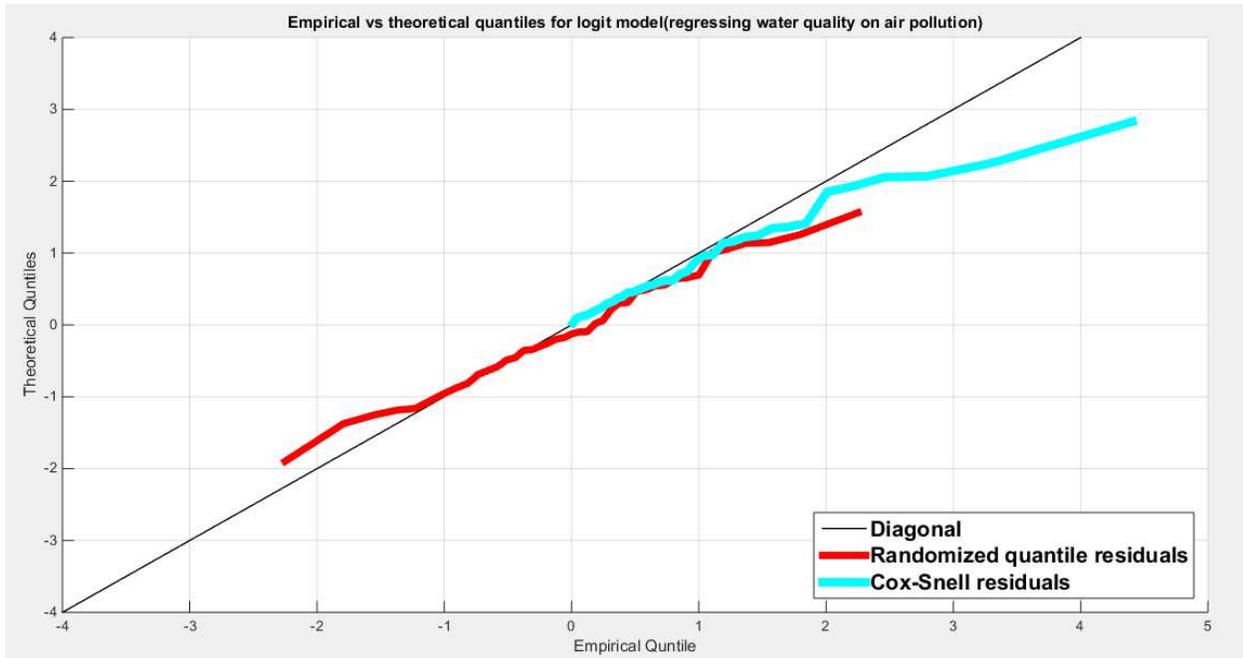

Fig. 45 shows the QQ plot of the empirical quantiles and the theoretical qunatiles for both types of residuals.

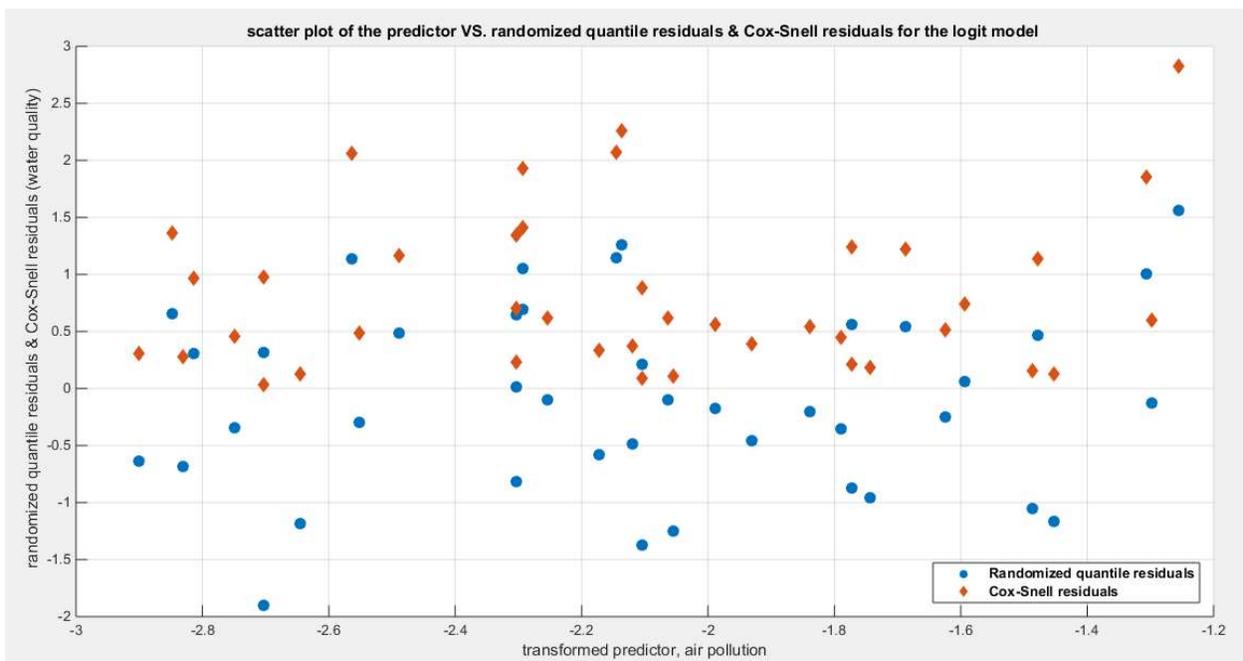

Fig. 46 shows the scatter plot of residuals of both types against transformed predictors.

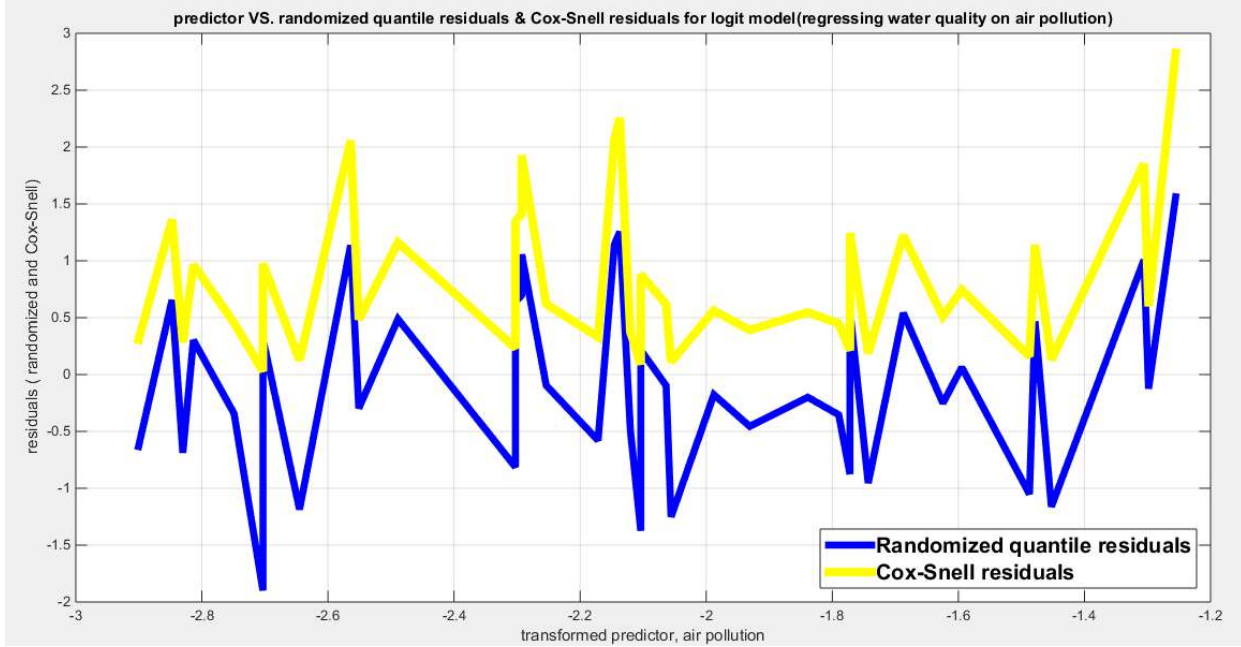

Fig. 47 shows the plot of residuals of both types against transformed predictors.

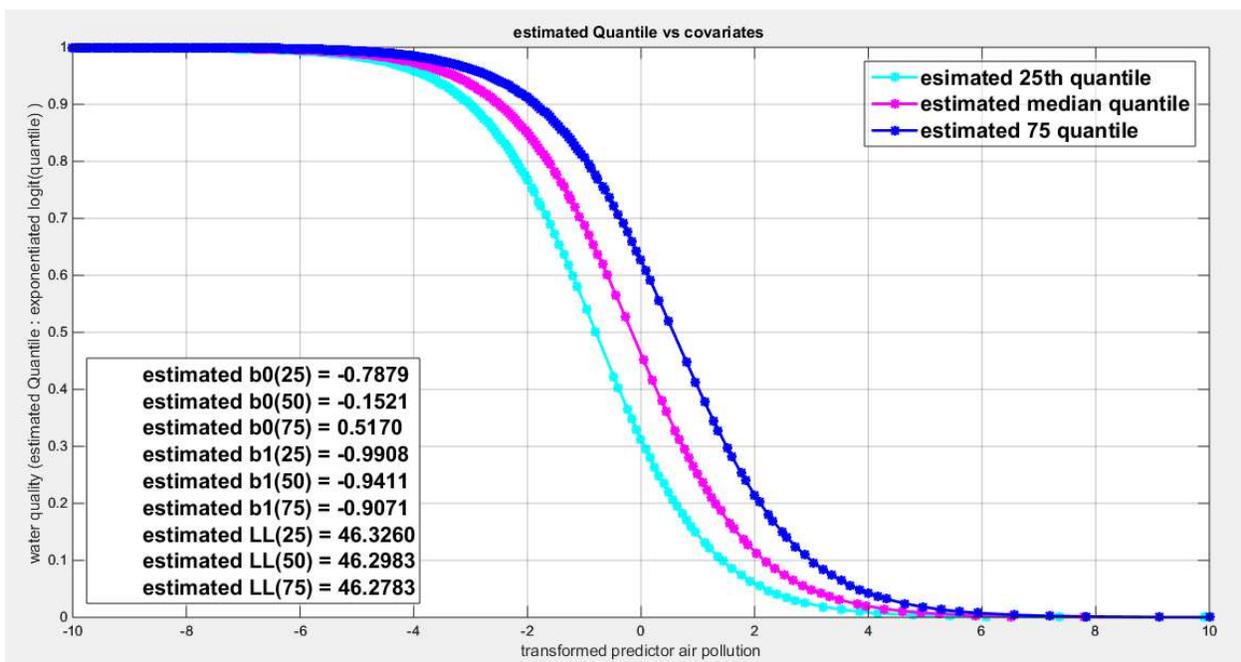

Fig. 48 shows parallel quantile curves across 25th, 50th ( median), 75th percentiles, suggesting that the predictor exerts a uniform influence on the response consistent with homoscedasticity.

Table 12: regressing water quality on the life expectancy

| | Logit link function | | Log-log complementary | | Log-log median | |
|---|---|---|---|---|---|---|
| B0 | 3.0944 | | 1.2917 | | -3.0225 | |
| B1 | 5.8750 | | 2.8480 | | -5.1829 | |
| LL | 43.2027 | | 43.3006 | | 43.1548 | |
| Wald stat. of b0 | 4.8933 ( p < 0.025) | | 4.2819 ( p < 0.025) | | 5.4846 ( p < 0.025) | |
| Wald stat. of b1 | 2.0702 ( p < 0.025) | | 2.0367 ( p < 0.025) | | 2.1251 ( p < 0.025) | |
| AIC | -82.4053 | | -82.6011 | | -82.3097 | |
| CAIC | -82.0896 | | -82.2853 | | -81.9939 | |
| BIC | -78.9782 | | -79.1740 | | -78.8825 | |
| HQIC | -81.1574 | | -81.3532 | | -81.0617 | |
| LRT | 5.410 (p0.02) | | 5.6059 (p=0.0179) | | 5.3145 (p=0.0211) | |
| R-squared | 0.1236 | | 0.1278 | | 0.1216 | |
| P-value for randomized quantile residuals | 0.6821 | | 0.6190 | | 0.7129 | |
| p-value for Cox-Snell residuals | 0.6821 | | 0.6190 | | 0.7129 | |
| Variance-covariance matrix | 0.3999 | 1.7612 | 0.0910 | 0.4154 | 0.3037 | 1.3156 |
| | 1.7612 | 8.0536 | 0.4154 | 1.9553 | 1.3156 | 5.9479 |
| QR vs. predictor(tau,p) | -0.1093, 0.3225 | | -0.1093, 0.3225 | | -0.1093, 0.3225 | |
| CS vs. predictor(tau,p) | -0.1093, 0.3225 | | -0.1093, 0.3225 | | -0.1093, 0.3225 | |

Table 12 shows that the predictor is significant as likelihood ratio test (LRT) is significant; the R squared is also high for this predictor around 0.12 across the different link functions but it is less than the previous two predictors, the employment rate and the air pollution. The AIC, CAIC, BIC, HQIC and LL are more or less equal between the different models. The LL is around 43 across the link functions and it is also less than the previous two predictors. The residuals plotted against the predictors show no specific trend and they are randomly scattered. The QQ plot of the randomized quantile residuals shows perfect alignment with the diagonal all through its course in contrast with the Cox Snell residuals that show this perfect alignment at the lower tail and the center. The estimated curve between the estimated median and the transformed predictor is increasing reflecting that the more the life expectancy, the more the percentage expressing the increased quality of water supply and cleanliness is. The figure for the clog-log shows the same pattern. The log-log figure has the same pattern. The difference is mainly manifested in the slope of the estimated curve. To assess the assumption of constant variance in the median parametric regression model, residual-based diagnostic tests were conducted using both randomized quantile (RQ) and Cox-Snell (CS) residuals. For each type of residual, an auxiliary regression of the squared residuals on the corresponding predictor was estimated by ordinary least squares, and the null hypothesis of homoscedasticity (H$_0$: constant variance) was tested. The results indicted no significant relationship between the squared residuals and the predictor variable (CS: p=0.157, R-squared=0.0508; RQ: p=0.0796, R-squared=0.0767), suggesting that the variance of the residuals remained approximately constant across the range of the predictor. Furthermore, the magnitude of the CS residuals were within a reasonable range (five values between 2.1570 and 3.4919), which supports the absence of heteroscedasticity. These findings provide evidence that the fitted median regression model satisfies the homoscedasticity assumption. These results are from the logit model. Figures 49-55 show the previous results.

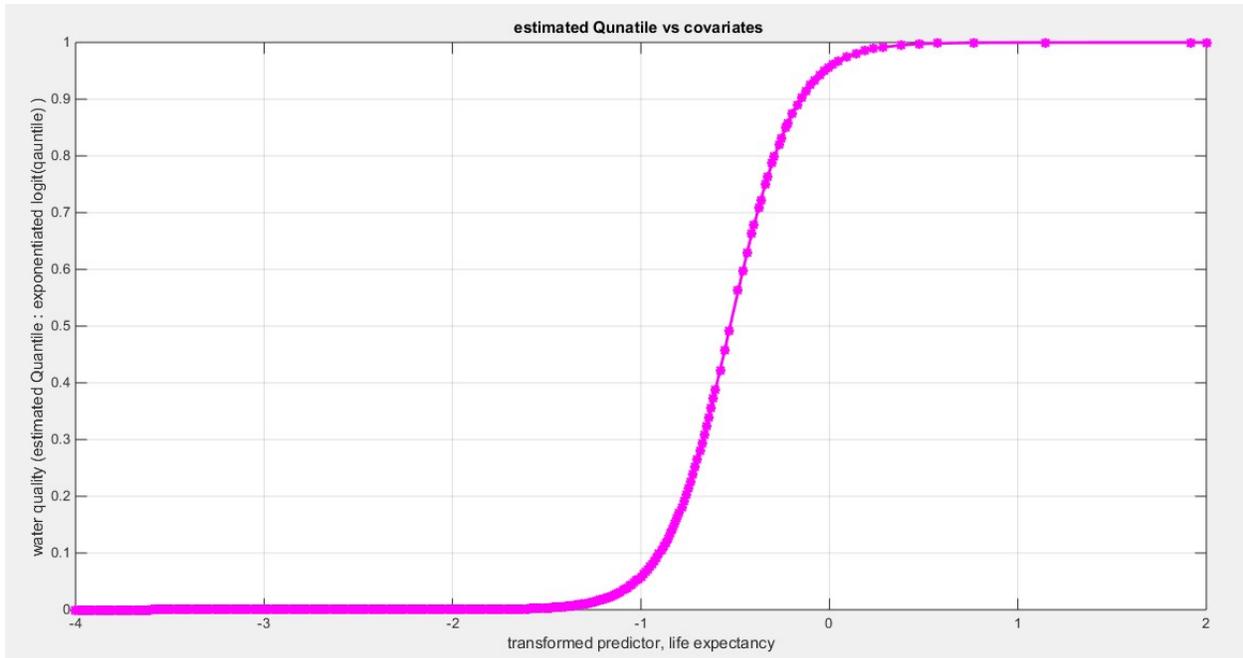

Fig. 49 shows the estimated curve plotting the transformed predictor against the estimated median (for the logit link).

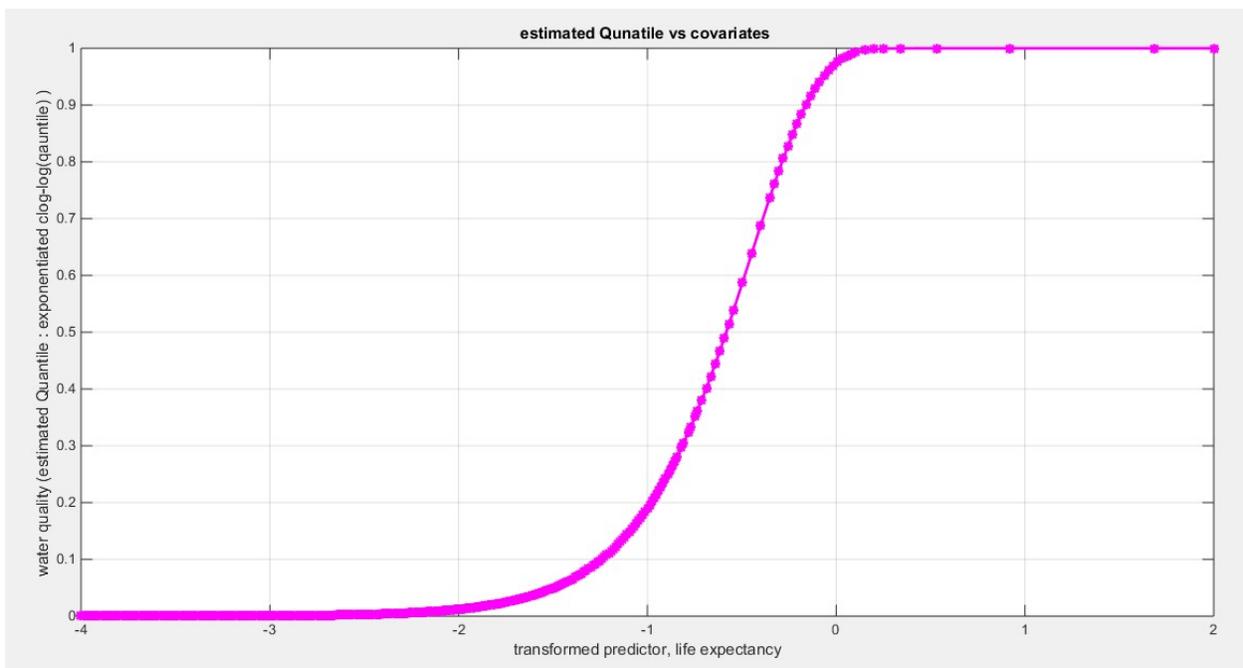

Fig. 50 shows the estimated curve plotting the transformed predictor against the estimated median (for the clog-log link).

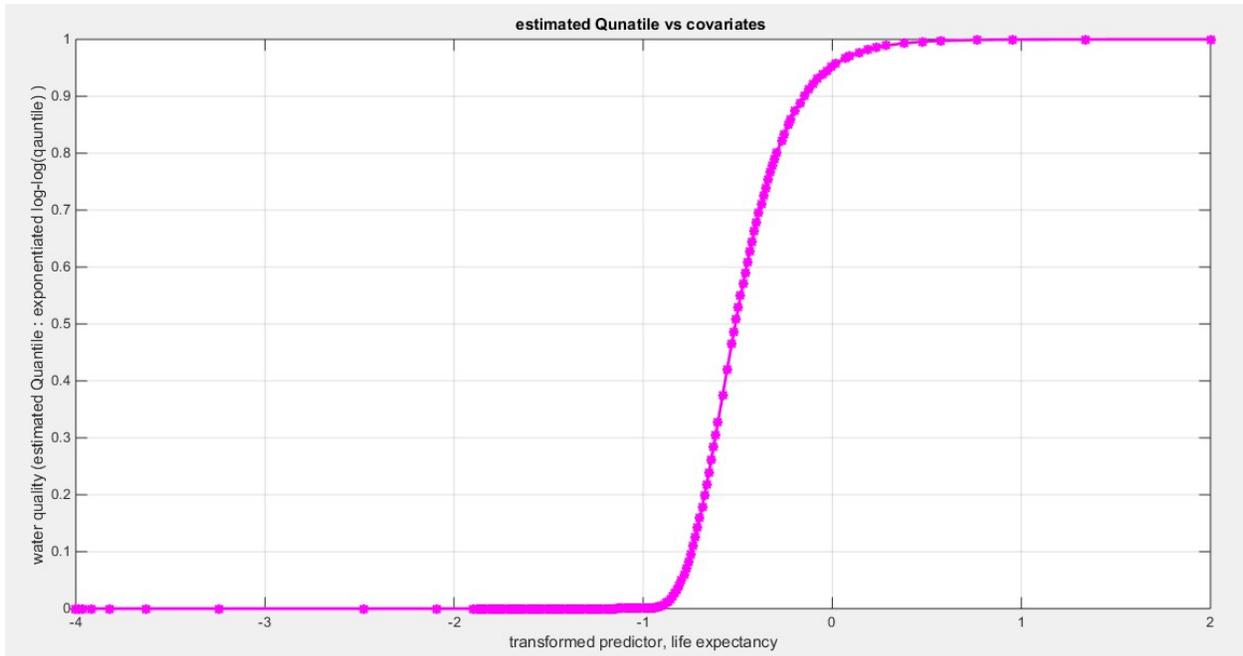

Fig. 51 shows the estimated curve plotting the transformed predictor against the estimated median (for the log-log link).

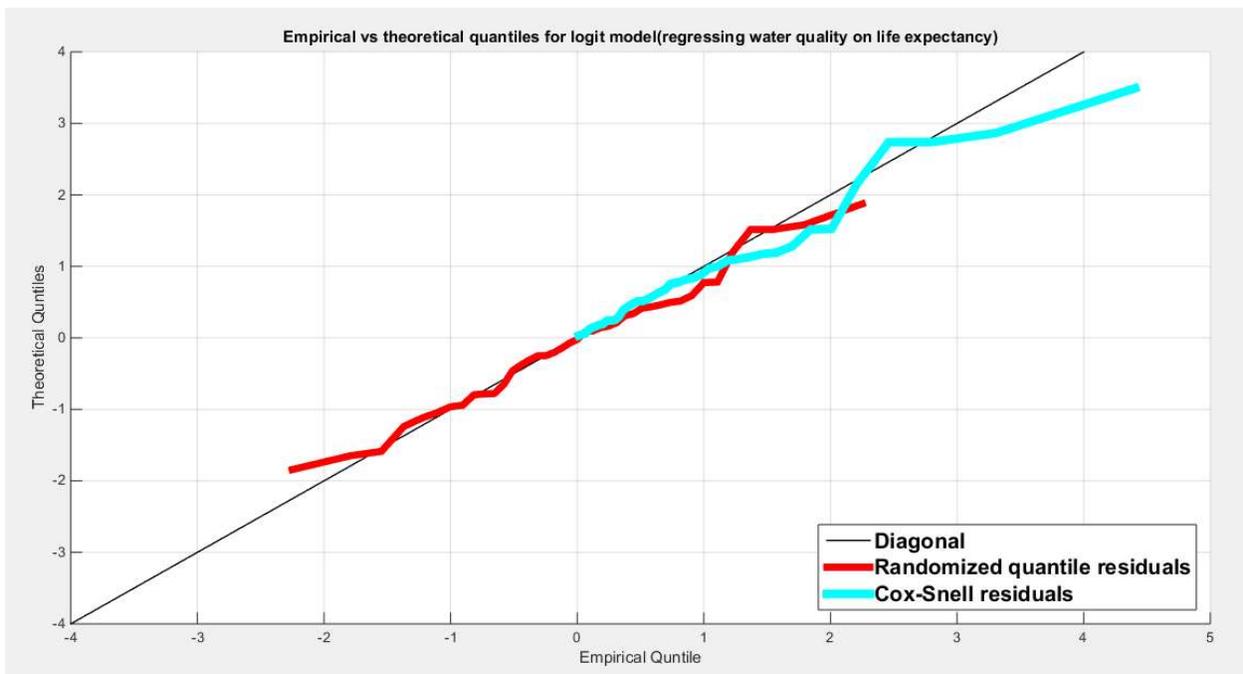

Fig. 52 shows the QQ plot of the empirical quantiles and the theoretical qunatiles for both types of residuals.

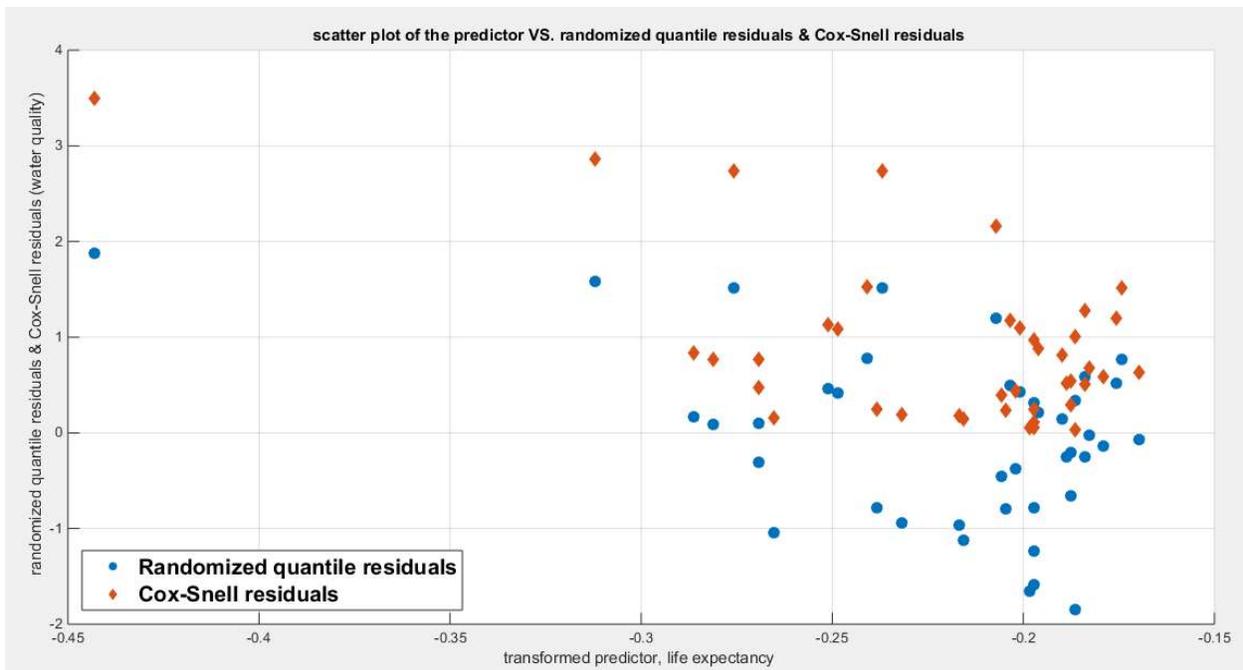

Fig. 53 shows the scatter plot of residuals of both types against transformed predictors.

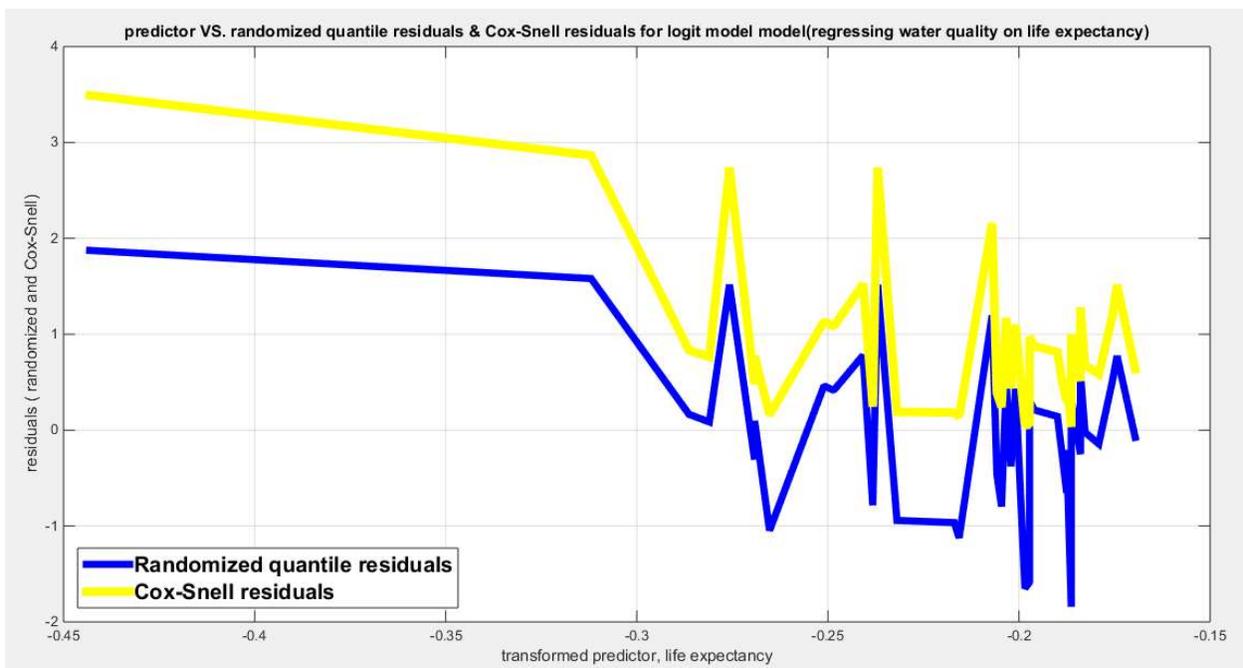

Fig. 54 shows the plot of residuals of both types against transformed predictors.

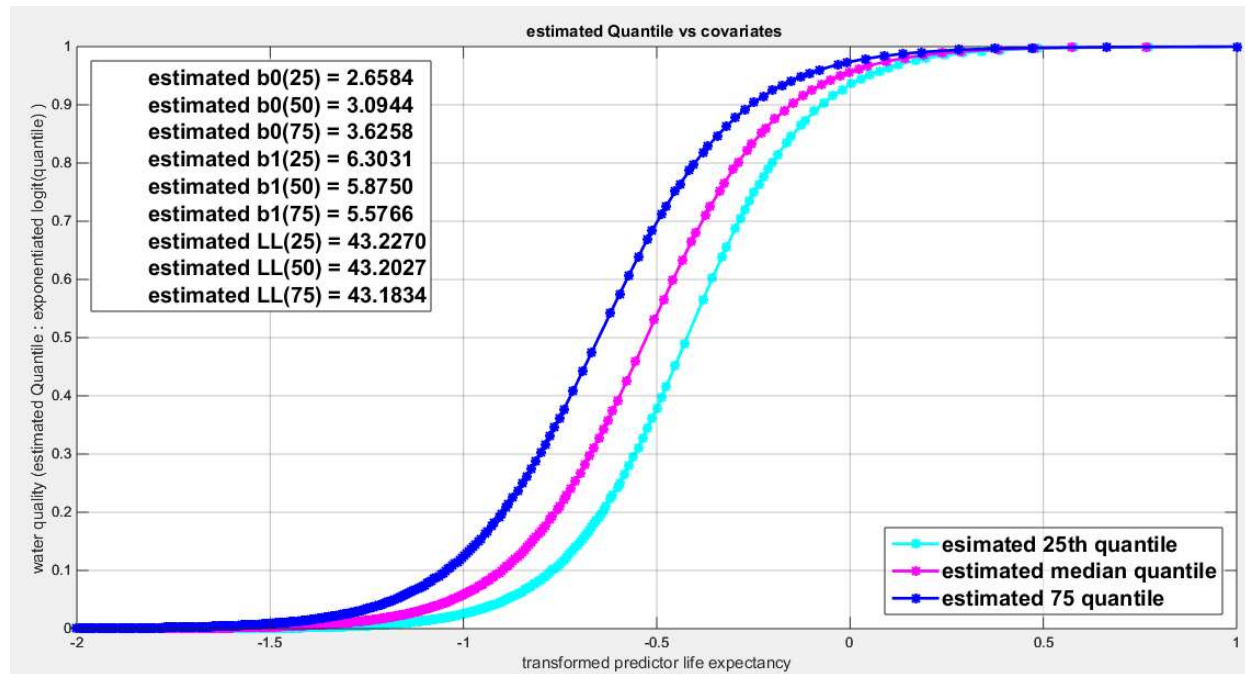

Fig. 55 shows parallel quantile curves across $25^{th}$, $50^{th}$ ( median), $75^{th}$ percentiles, suggesting that the predictor exerts a uniform influence on the response consistent with homoscedasticity.

## Table 13: Regressing water quality on life satisfaction

|  | Logit link function | | Log-log complementary | | Log-log median | |
|---|---|---|---|---|---|---|
| B0 | 15.2157 | | 7.0178 | | -13.8744 | |
| B1 | 4.9010 | | 2.3272 | | -4.3780 | |
| LL | 49.2593 | | 49.8381 | | 49.0980 | |
| Wald stat. of b0 | 4.4078(p<0.025) | | 3.797(p<0.025) | | 4.5692(p<0.025) | |
| Wald stat. of b1 | 3.8571(p<0.025) | | 3.3834(p<0.025) | | 3.9309(p<0.025) | |
| AIC | -94.5187 | | -95.6761 | | -94.1959 | |
| CAIC | -94.2029 | | -95.3603 | | -93.8801 | |
| BIC | -91.0915 | | -92.2490 | | -90.7688 | |
| HQIC | -93.2707 | | -94.4282 | | -92.9479 | |
| LRT | 17.5235(p=2.8378e-5) | | 18.6809(p=1.5452e-5) | | 17.2007 (p=3.3631e-5) | |
| R-squared | 0.3478 | | 0.366 | | 0.3426 | |
| P-value for randomized quantile residuals | 0.3405 | | 0.3309 | | 0.2691 | |
| p-value for Cox-snell residuals | 0.3405 | | 0.3309 | | 0.2691 | |
| Variance-covariance matrix | 11.9160 | 4.3834 | 3.4159 | 1.2708 | 9.2202 | 3.3796 |
|  | 4.3834 | 1.6145 | 1.2708 | 0.4731 | 3.3796 | 1.2404 |
| QR vs. predictor(tau,p) | (-0.1481,0.1838) | | (-0.1481,0.1838) | | (-0.1481,0.1838) | |
| CS vs. predictor(tau,p) | (-0.1481,0.1838) | | (-0.1481,0.1838) | | (-0.1481,0.1838) | |

Table 13 shows that the predictor is significant as likelihood ratio test (LRT) is highly significant; the R squared is also high for this predictor. It is between 0. 3426 and 0.366  across the different link functions. So it is more than the previous three predictors, the employment rate, the air pollution, and life expectancy. The AIC, CAIC, BIC, HQIC and LL are more or less equal between the different models. The LL is around 49 across the different link functions hence, it is more than the previous three predictors. The residuals plotted against the predictors show no specific trend and they are randomly scattered. The QQ plot of the randomized quantile residuals shows perfect alignment with the diagonal all through its course in contrast with the Cox Snell residuals that show this perfect alignment at the lower tail and the center. The estimated curve between the estimated median and the transformed predictor is increasing reflecting that the more the life satisfaction is, the more the percentage expressing the increased quality of water supply and cleanliness is. The figure for the clog-log shows the same pattern. The log-log figure has the same pattern. The difference is mainly manifested in the slope of the estimated curve. To assess the assumption of constant variance in the median parametric regression model, residual-based diagnostic tests were conducted using both randomized quantile (RQ) and Cox-Snell (CS) residuals. For each type of residual, an auxiliary regression of the squared residuals on the corresponding predictor was estimated by ordinary least squares, and the null hypothesis of homoscedasticity ($H_0$: constant variance) was tested. The results indicted no significant relationship between the squared residuals and the predictor variable (CS: p=0.19, R-squared=0.0435; RQ: p=0.0789, R-squared=0.077), suggesting that the variance of the residuals remained approximately constant across the range of the predictor. Furthermore, the magnitude of the CS residuals were within a reasonable range (two values 2.1412 and 2.8338), which supports the absence of heteroscedasticity. These findings provide evidence that the fitted median regression model satisfies the homoscedasticity assumption. These results are from the logit model. Figures 56-62 show the previous results.

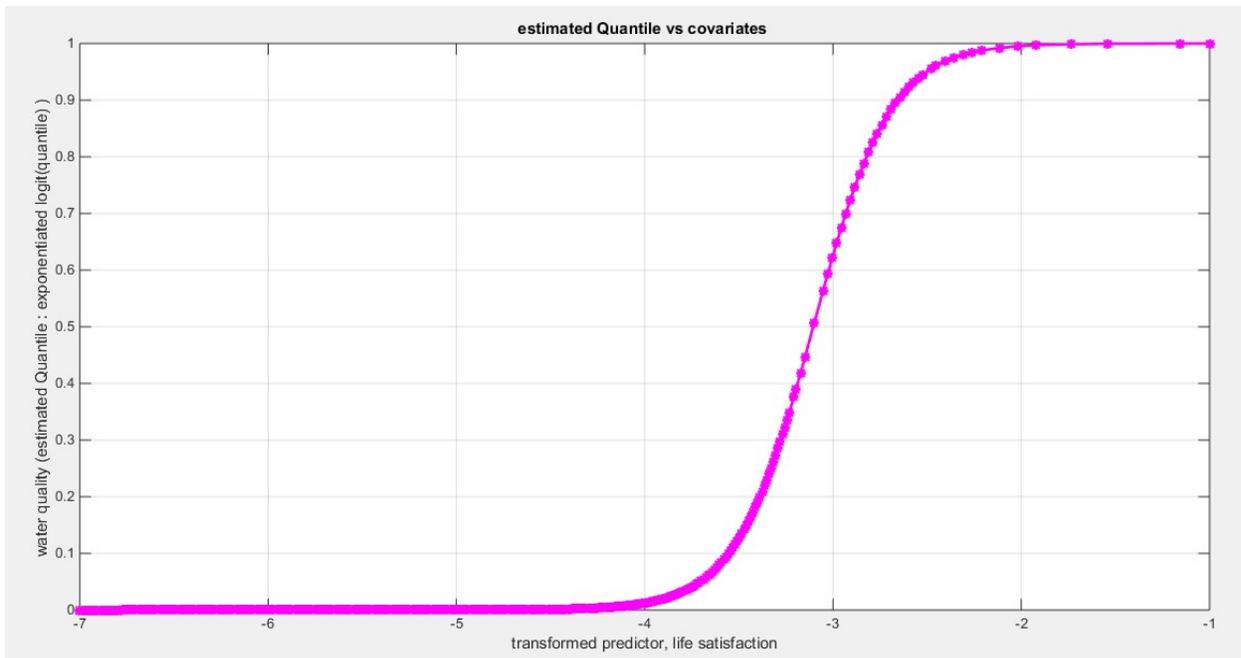

Fig. 56 shows the estimated curve plotting the transformed predictor against the estimated median (for the logit link).

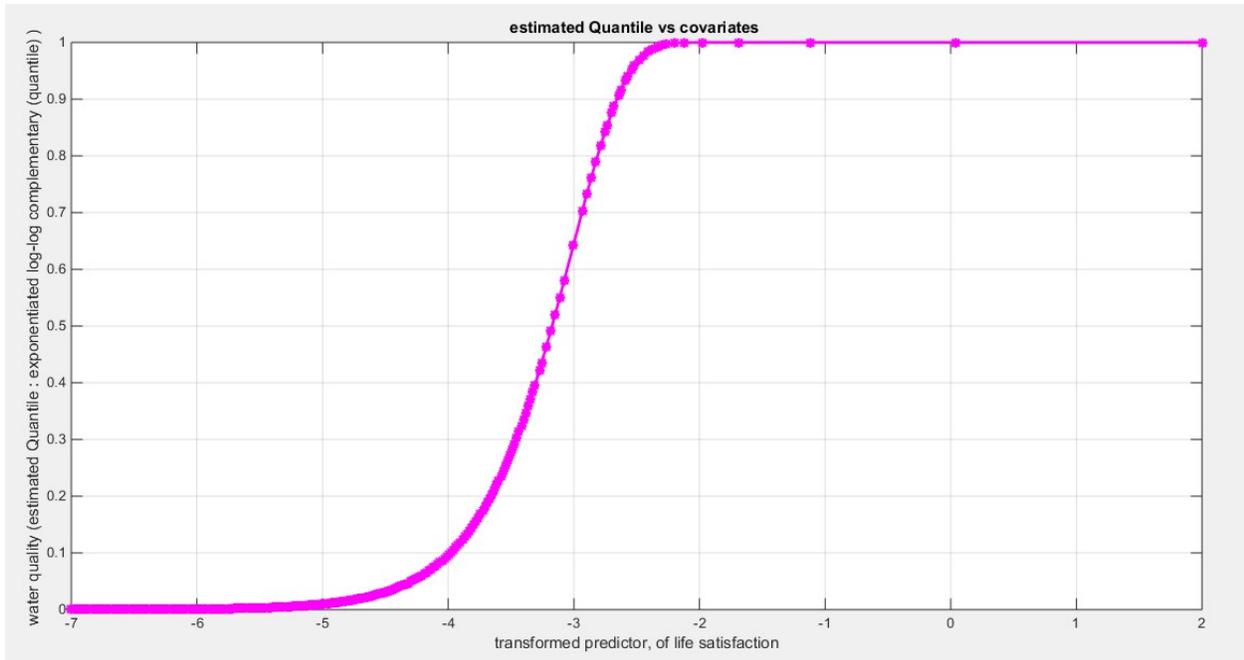

Fig. 57 shows the estimated curve plotting the transformed predictor against the estimated median (for the clog-log link).

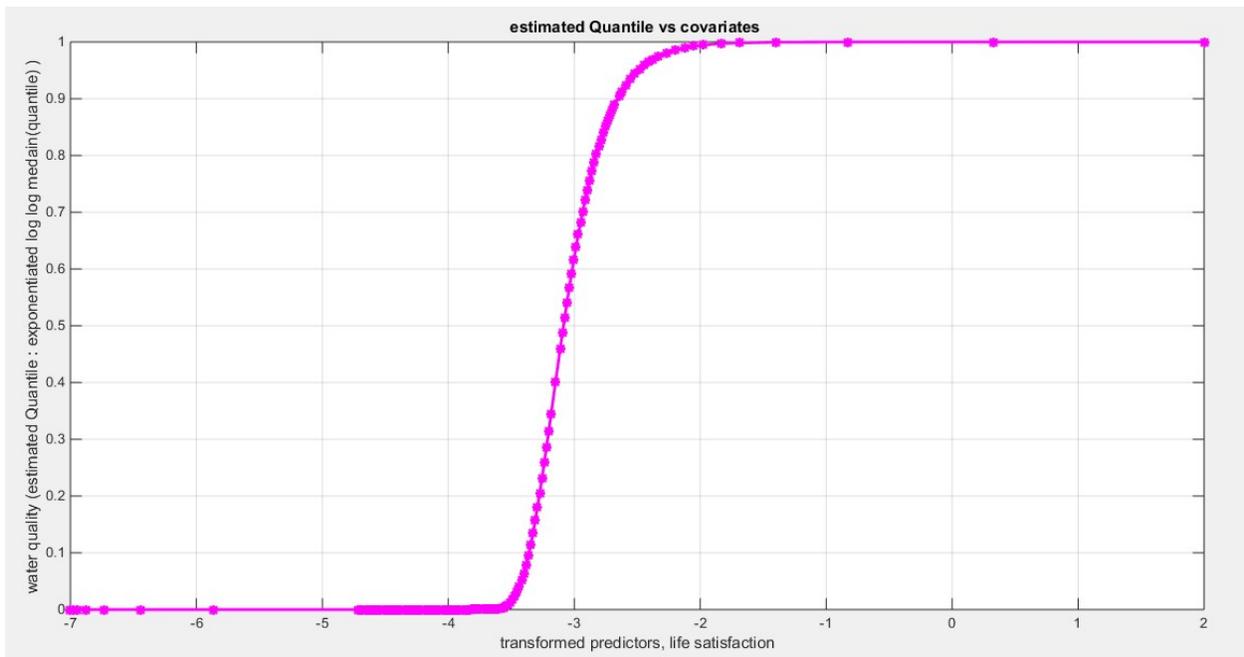

Fig. 58 shows the estimated curve plotting the transformed predictor against the estimated median (for the log-log link).

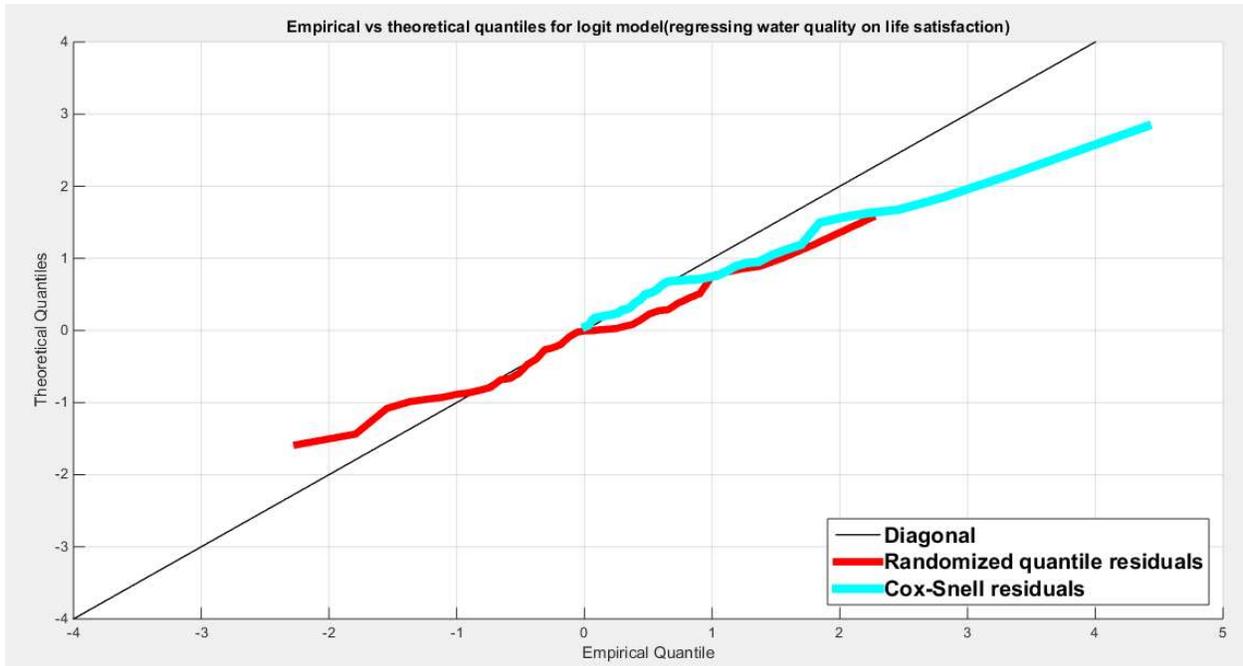

Fig. 59 shows the QQ plot of the empirical quantiles and the theoretical qunatiles for both types of residuals.

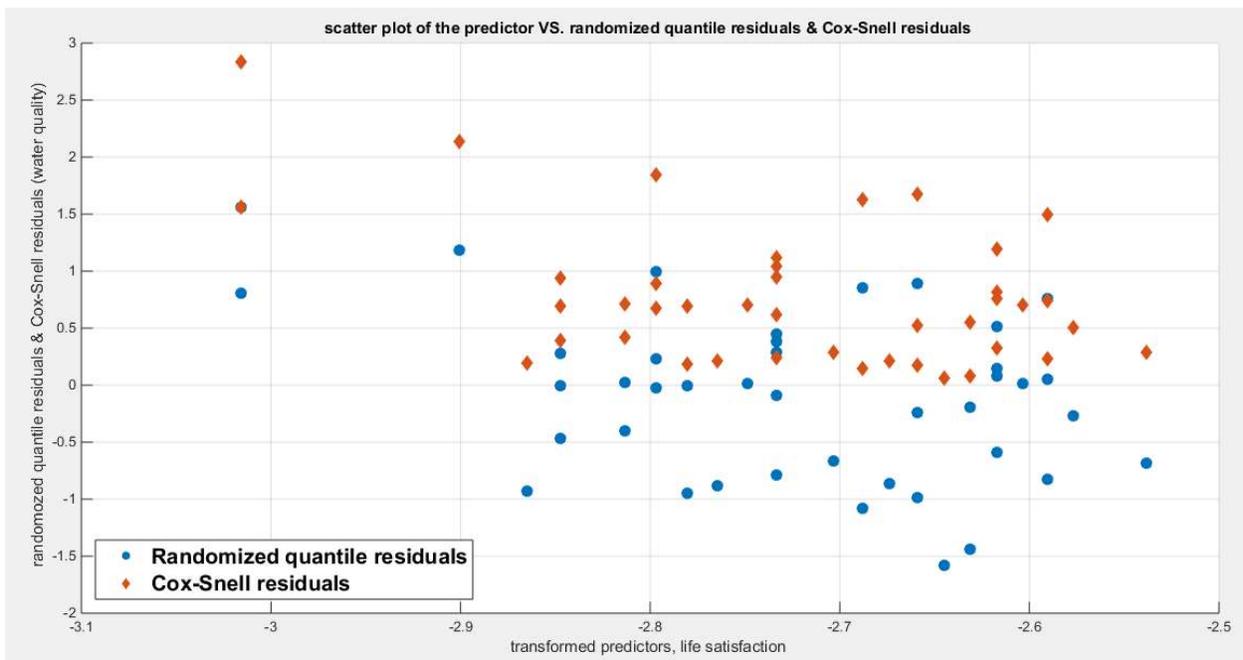

Fig. 60 shows the scatter plot of residuals of both types against transformed predictors.

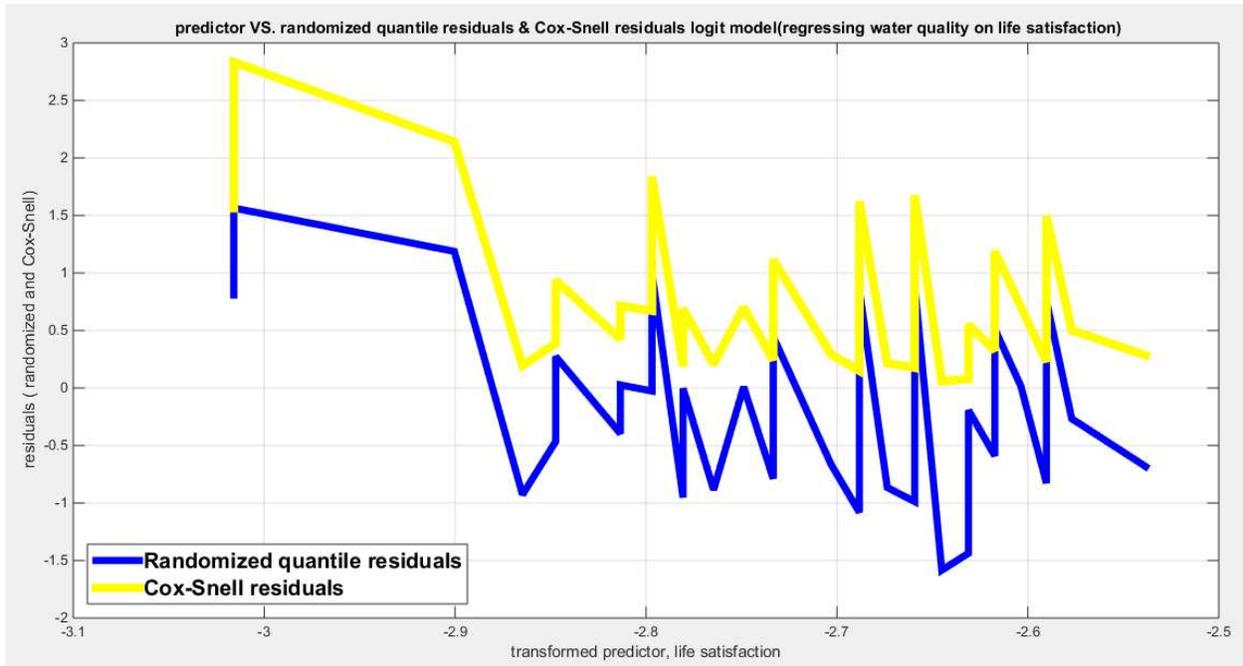

Fig. 61 shows the plot of residuals of both types against transformed predictors.

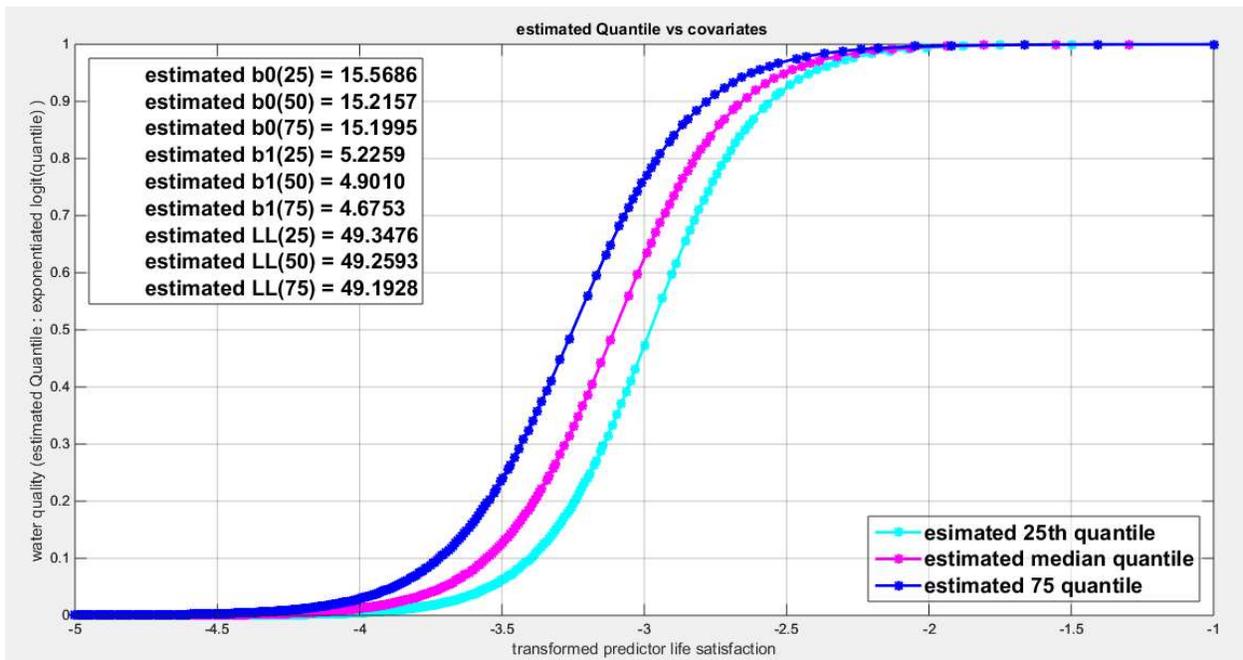

Fig. 62 shows parallel quantile curves across 25th, 50th (median), 75th percentiles, suggesting that the predictor exerts a uniform influence on the response consistent with homoscedasticity.

Table 14: regressing water quality on homicide rate

| | Logit link function | | Log-log complementary | | Log-log median | |
|---|---|---|---|---|---|---|
| B0 | 0.8551 | | 0.2276 | | -1.0197 | |
| B1 | -0.2107 | | -0.0975 | | 0.1918 | |
| LL | 42.8787 | | 42.9256 | | 42.8631 | |
| Wald stat. of b0 | 1.8153(p>0.025) | | 0.9905(p>0.025) | | 2.4231(p<0.025) | |
| Wald stat. of b1 | 2.0862(p<0.025) | | 2.0330(p<0.025) | | 2.1053(p<0.025) | |
| AIC | -81.7574 | | -81.8512 | | -81.7263 | |
| CAIC | -81.4416 | | -81.5355 | | -81.4105 | |
| BIC | -78.3303 | | -78.4241 | | -78.2991 | |
| HQIC | -80.5094 | | -80.6033 | | -80.4783 | |
| LRT | 4.7622 (p=0.0291) | | 4.8561(p=0.0275) | | 4.7311 (p=0.0296) | |
| R-squared | 0.1097 | | 0.1117 | | 0.1090 | |
| P-value for randomized quantile residuals | 0.5577 | | 0.5471 | | 0.5629 | |
| p-value for Cox-snell residuals | 0.5577 | | 0.5471 | | 0.5629 | |
| Variance-covariance matrix | 0.2219 | 0.0460 | 0.0528 | 0.0107 | 0.1771 | 0.0370 |
| | 0.0460 | 0.0102 | 0.0107 | 0.0023 | 0.0370 | 0.0083 |
| QR vs. predictor(tau,p) | (0.0087, 0.9461) | | (0.0087, 0.9461) | | (0.0087, 0.9461) | |
| CS vs. predictor(tau,p) | (0.0087, 0.9461) | | (0.0087, 0.9461) | | (0.0087, 0.9461) | |

Table 14 shows that the predictor is significant as likelihood ratio test (LRT) is significant; the R squared is also high for this predictor. It is between 0.1090 and 0.1117 across the different link functions. It is less than the previous four predictors, the employment rate, the air pollution, life expectancy, and life satisfaction. The AIC, CAIC, BIC, HQIC and LL are more or less equal across the different models. The LL is around 42 across the different link functions hence, it is the least value among the LL values of the previous four predictors. The residuals plotted against the predictors show no specific trend and they are randomly scattered. The QQ plot of the randomized quantile residuals shows perfect alignment with the diagonal all through its course in contrast with the Cox Snell residuals that show this perfect alignment at the lower tail and the center. The estimated curve between the estimated median and the transformed predictor is decreasing reflecting that the more the homicide rate is, the less the percentage expressing the increased quality of water supply and cleanliness is. The figure for the clog-log shows the same pattern. The log-log figure has the same pattern. The difference is mainly manifested in the slope of the estimated curve. To assess the assumption of constant variance in the median parametric regression model, residual-based diagnostic tests were conducted using both randomized quantile (RQ) and Cox-Snell (CS) residuals. For each type of residual, an auxiliary regression of the squared residuals on the corresponding predictor was estimated by ordinary least squares, and the null hypothesis of homoscedasticity ($H_0$: constant variance) was tested. The results indicted no significant relationship between the squared residuals and the predictor variable (CS: p=0.403, R-squared=0.018; RQ: p=0.309, R-squared=0.0265), suggesting that the variance of the residuals remained approximately constant across the range of the predictor. Furthermore, the magnitude of the CS residuals were within a reasonable range (five values between 2.1434 and 3.6064), which supports the absence of heteroscedasticity. These findings provide evidence that the fitted median regression model satisfies the homoscedasticity assumption. These results are from the logit model. Figures 56-62 show the previous results.

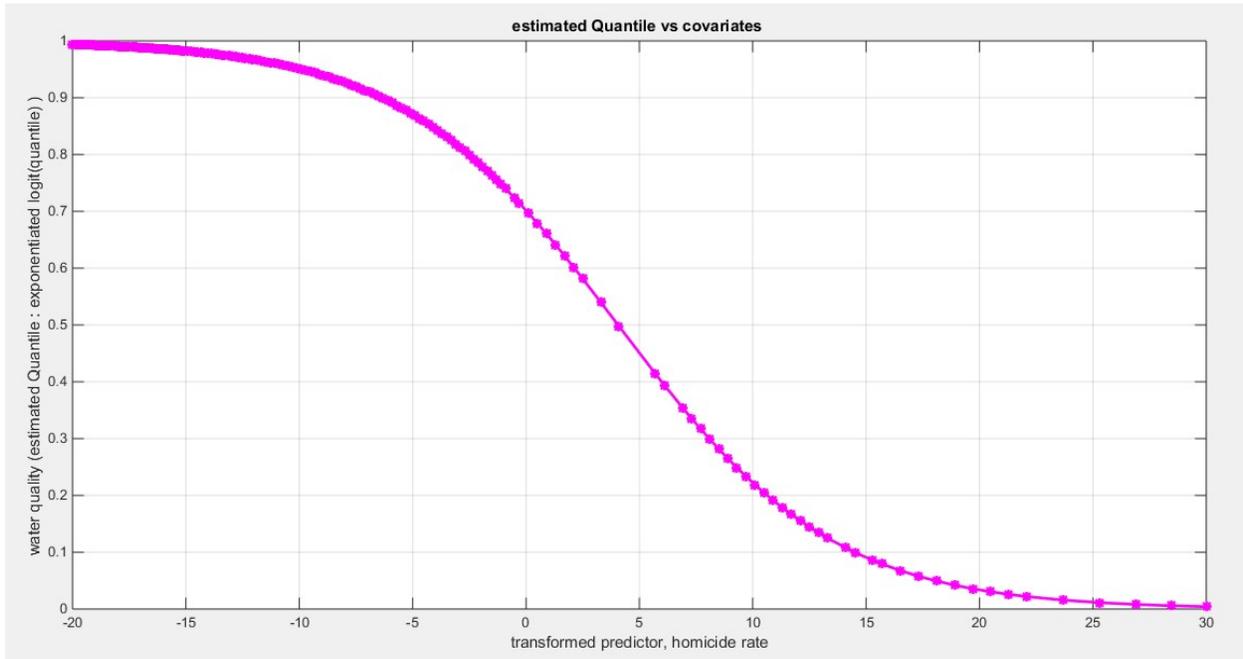

Fig. 63 shows the estimated curve plotting the transformed predictor against the estimated median (for the logit link).

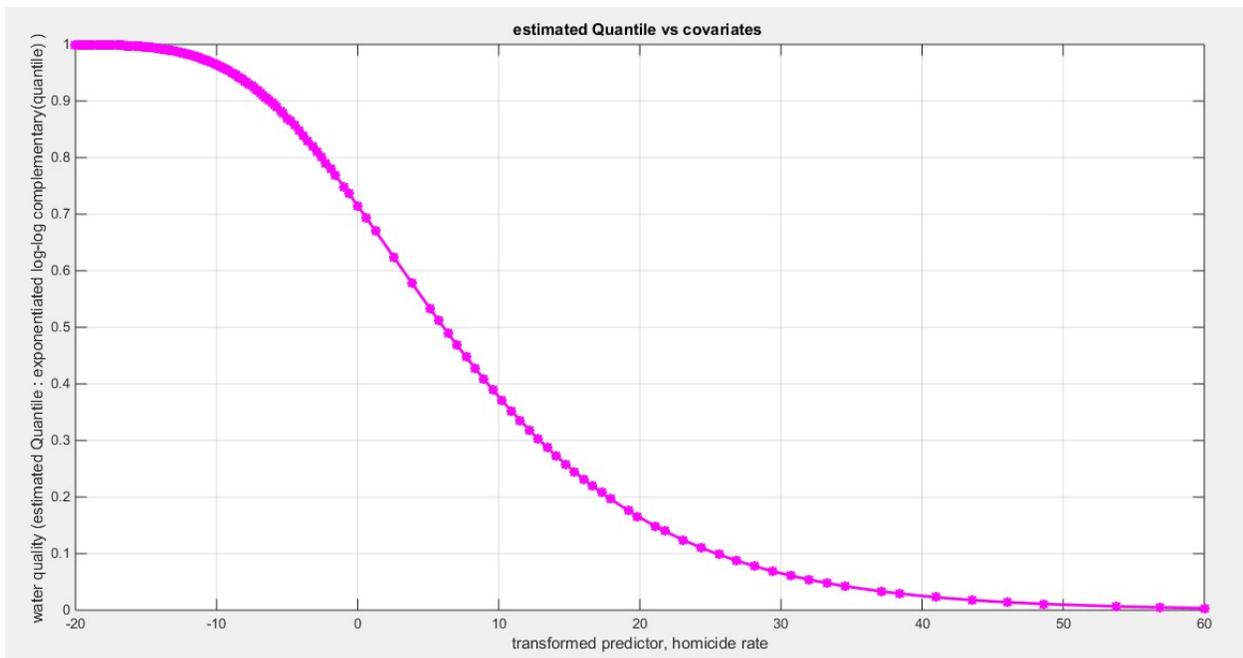

Fig. 64 shows the estimated curve plotting the transformed predictor against the estimated median (for the clog-log link).

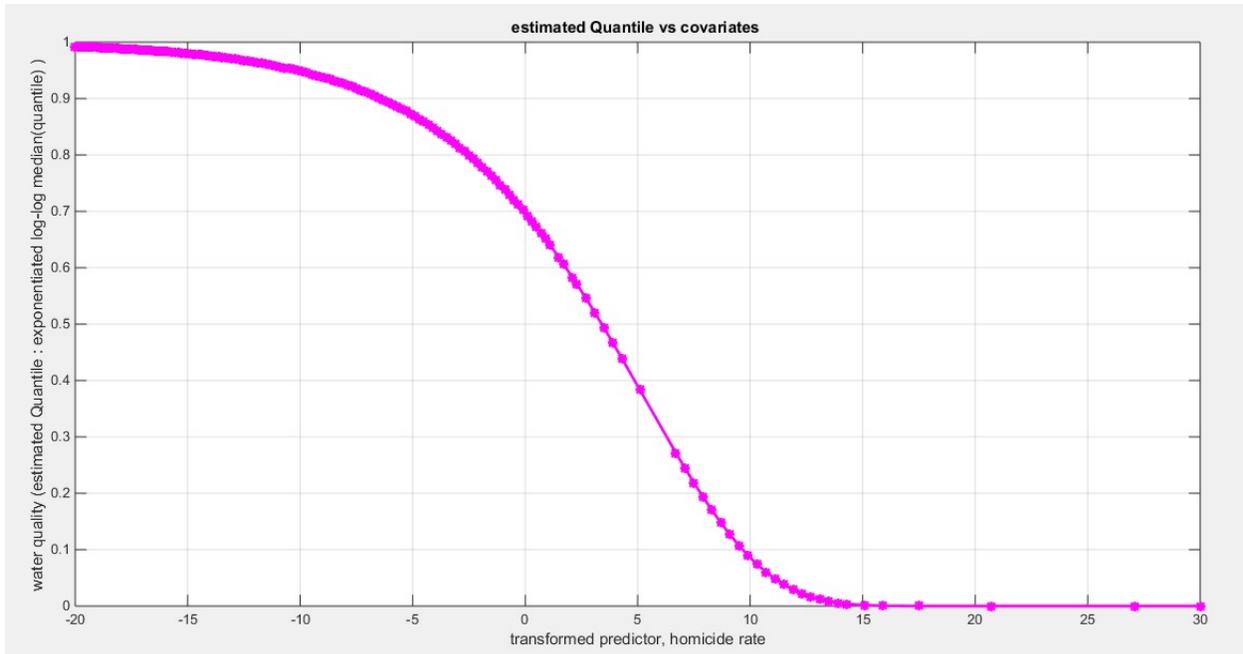

Fig. 65 shows the estimated curve plotting the transformed predictor against the estimated median (for the log-log link).

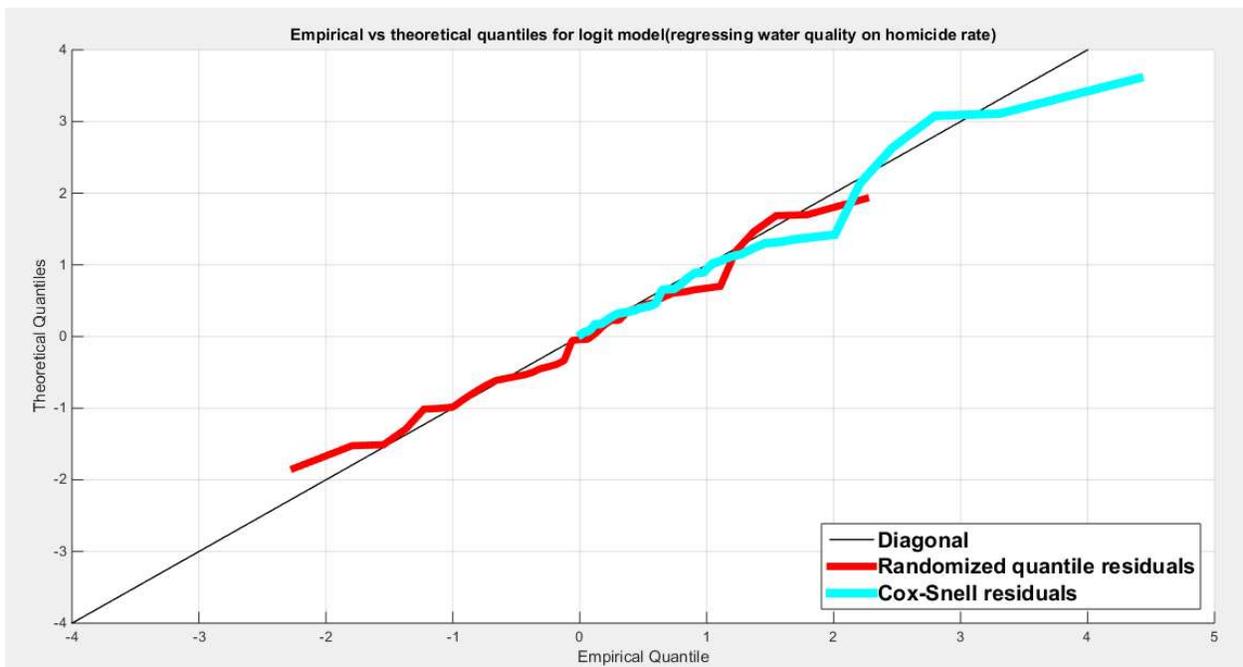

Fig. 66 shows the QQ plot of the empirical quantiles and the theoretical quantiles for both types of residuals.

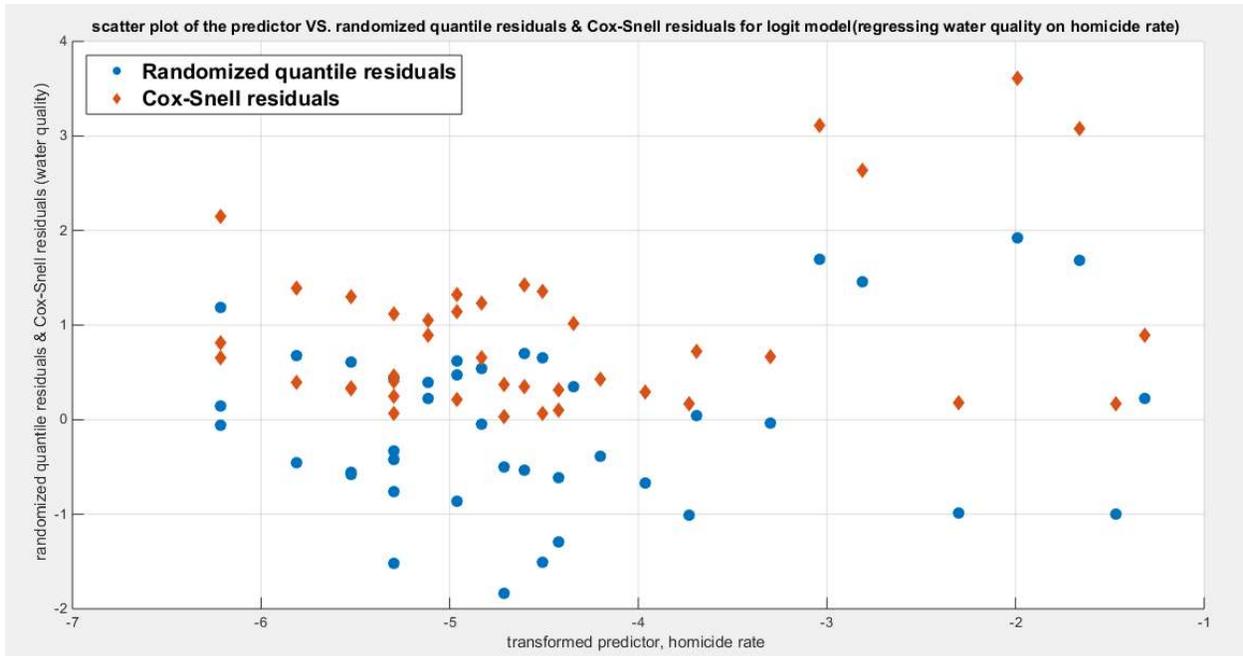

Fig. 67 shows the scatter plot of residuals of both types against transformed predictors.

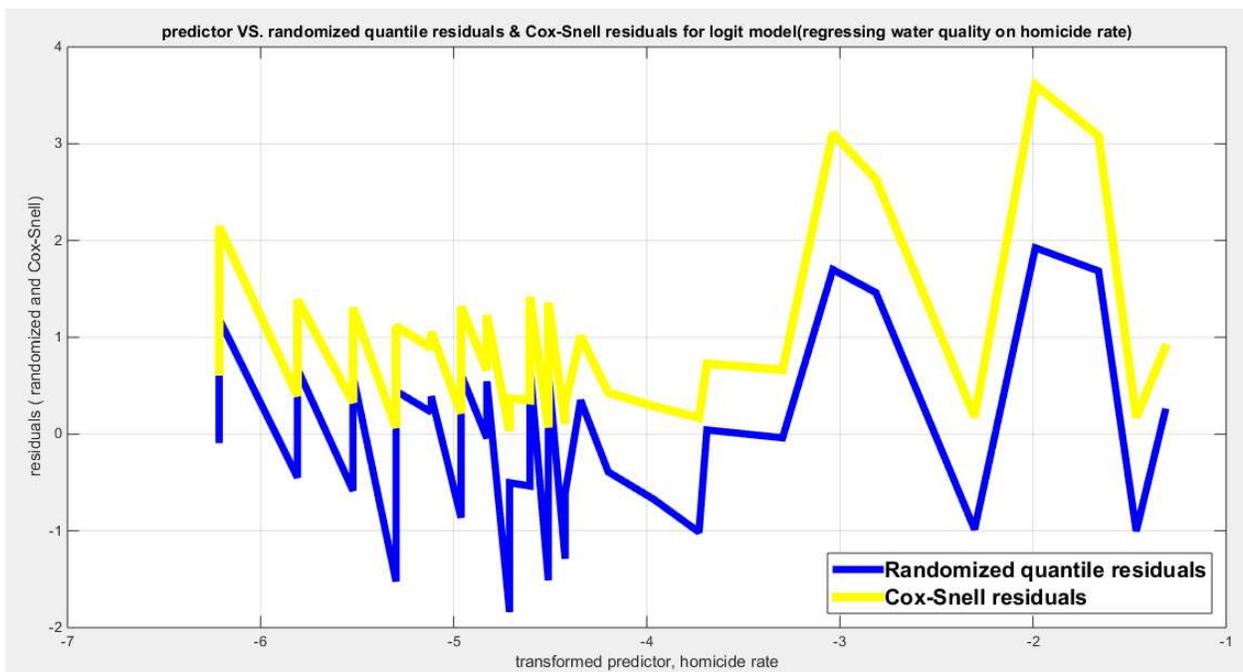

Fig. 68 shows the plot of residuals of both types against transformed predictors

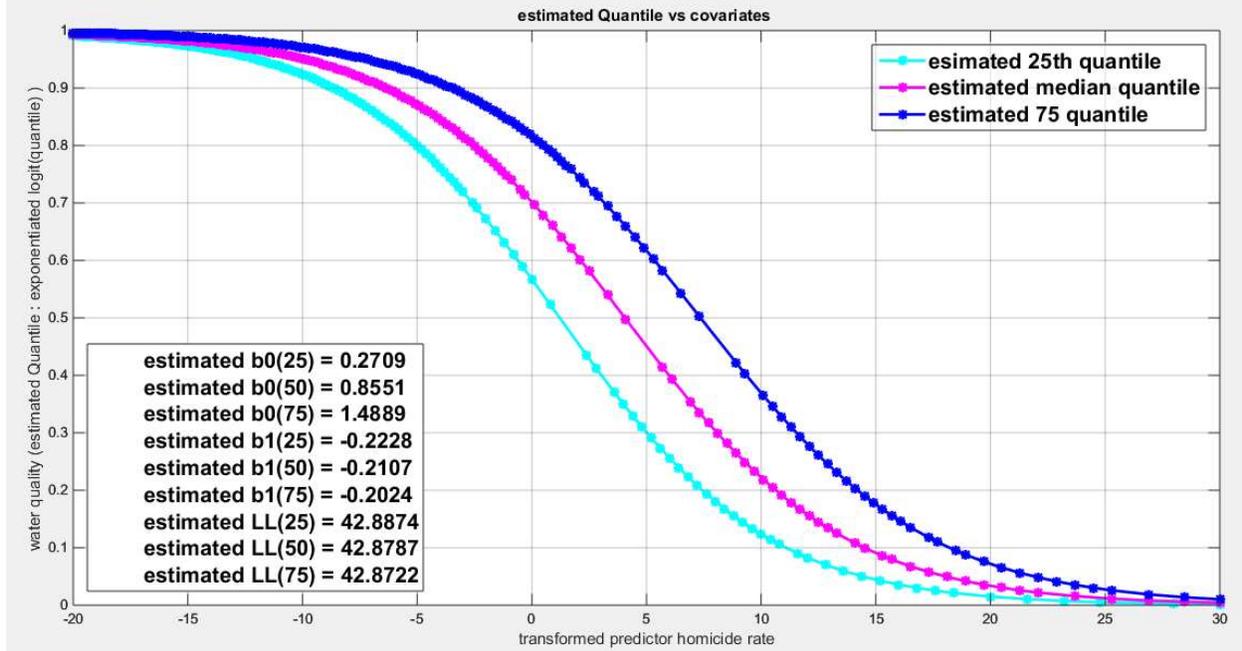

Fig. 69 shows parallel quantile curves across 25th, 50th ( median), 75th percentiles, suggesting that the predictor exerts a uniform influence on the response consistent with homoscedasticity.

The marginal correlations between the variables (the response and the predictors) are shown in Table 15. While the water quality is significantly and positively correlated with employment rate, life expectancy, and life satisfaction; it is significantly and negatively correlated with the air pollution and the homicide rate. The employment rate is significantly and negatively correlated with air pollution and homicide rate while it is significantly and positively correlated with life satisfaction. Moreover, the air pollution is significantly and negatively correlated with the life expectancy and the life satisfaction. Both the life expectancy and the life satisfaction are significantly and negatively correlated with homicide rate.

Table 15: The marginal correlation matrix Kendall tau coefficient and associated p-value:

| | water quality(Y) | Employment Rate(X1) | Air Pollution(X2) | Life expectancy(X3) | Life Satisfaction(X3) | Homicide Rate(X4) |
|---|---|---|---|---|---|---|
| **water quality(Y)** | 1 | 0.5942 P=0.0000 | -0.4144 P=0.0002 | 0.2859 P=0.0099 | 0.5347 P=0.0000 | -0.2762 P=0.0135 |
| **Employment Rate (X1)** | 0.5942 P=0.0000 | 1 | -0.4135 P=0.0002 | 0.2208 P=0.0474 | 0.4886 P=0.0000 | -0.3365 P=0.0027 |
| **Air Pollution(X2)** | -0.4144 P=0.0002 | -0.4135 P=0.0002 | 1 | -0.2711 P=0.0138 | -0.4876 P=0.0000 | 0.1207 P=0.2796 |
| **Life expectancy(X3)** | 0.2859 P=0.0099 | 0.2208 P=0.0474 | -0.2711 P=0.0138 | 1 | 0.4020 P=0.0003 | -0.4327 P=0.0001 |
| **Life Satisfaction(X4)** | 0.5347 P=0.0000 | 0.4886 P=0.0000 | -0.4876 P=0.0000 | 0.4020 P=0.0003 | 1 | -0.3251 P=0.0039 |
| **Homicide Rate (X5)** | -0.2762 P=0.0135 | -0.3365 P=0.0027 | 0.1207 P=0.2796 | -0.4327 P=0.0001 | -0.3251 P=0.0039 | 1 |

The condition indices obtained from the standardized transformed X'X are 3.9624, 3.2842, 2.9637 and 1.9594 and 1. The VIF for the employment rate is 2.6446, for air pollution is 1.9961, for life expectancy is 2.4788, for life satisfaction is 3.0426 and for homicide rate is 2.1171. So as the largest condition index is 3.9624 less than 10 and the VIF for each predictor is less than 5 so there is no evidence of significant multi-collinearity between the predictors.

The signs of the coefficients of the marginal correlations match those signs of the conditional correlations coefficients when regressing the water quality response variable on each predictor, one at a time. As will be shown later, in multiple regression  analysis the employment rate and life satisfaction are positively dependent with the water quality in full model or models with whatever X is removed and the air pollution is negatively dependent with the water quality. These three variables have consistent signs of their regression coefficients regardless the X removed while life expectancy and homicide rate show different patterns of flipping their regression coefficients signs according to the X removed.  Life expectancy may be a proxy for the loneliness that drives elderly to minimize the quality of the water supply and cleanliness so the sign of the regression coefficient flips.

The author added these five predictors in one equation and used the different link functions, then removed each one at a time and calculated the LRT to assess the significance of this particular predictor while controlling for other predictors. The results are summarized in Tables 16-17. The description of the rows and columns of these tables is the same as the description of the rows and columns of the tables of the first response variable.

Table 16 : the coefficients of parametric median regression analysis with removal of different predictors  with the associated standard error below each estimated coefficient value.

| | intercept | X1 Coeff. | X2 Coeff. | X3 Coeff. | X4 Coeff. | X5 Coeff. | Preserve sign |
|---|---|---|---|---|---|---|---|
| Full model | 2.3456 5.1937 | 1.8747 1.6166 | -0.6055 0.3925 | -1.3509 5.2189 | 0.6654 1.9920 | -0.1079 0.1694 | no |
| Rx1 | 11.5211 5.3377 | | -0.3957 0.3498 | -2.6892 4.3807 | 4.2041 1.8840 | -0.0849 0.1377 | no |
| Rx2 | 12.1562 4.6809 | 2.1342 1.6621 | | -0.0876 5.735 | 3.4491 2.1164 | 0.0121 0.1636 | no |
| Rx3 | 10.4164 5.1585 | 1.7830 1.5112 | -0.235 0.3681 | | 3.0561 1.7246 | -3.0644e-4 0.1199 | yes |
| Rx4 | 3.7434 2.6968 | 3.1112 1.6677 | -0.3404 0.3669 | 4.9364 5.1481 | | 0.0616 0.1791 | no |
| Rx5 | 10.6081 5.2286 | 1.7581 1.4859 | -0.2341 0.3616 | -0.3006 4.0885 | 3.1531 2.089 | | no |
| Rx3,x5 | 10.4226 4.5705 | 1.7841 1.4483 | -0.2348 0.3614 | | 3.0575 1.6295 | | yes |
| Reduced model | 1.7644 | | | | | | |

Table 17 : AIC, CAIC, BIC, and HQIC for different models

| | LRT & P value | LL | AIC | CAIC | BIC | HQIC | R squared |
|---|---|---|---|---|---|---|---|
| Full model | 17.6445 0.0034 | 49.3198 | -86.6396 | -84.1691 | -76.3582 | -82.8957 | 0.3497 |
| Rx1 | 1.6786 1 | 50.1591 | -90.3183 | -88.6040 | -81.7504 | -87.1983 | 0.0418 |
| Rx2 | 2.3382 1 | 50.4889 | -90.9778 | -89.2635 | -82.4099 | -87.8579 | 0.0587 |
| Rx3 | 2.7486 1 | 50.6941 | -91.3882 | -89.6739 | -82.8204 | -88.2683 | 0.0693 |
| Rx4 | 0.4608 1 | 49.5502 | -89.1004 | -87.3861 | -80.5325 | -85.9805 | 0.0113 |
| Rx5 | 2.7540 1 | 50.6968 | -91.3936 | -89.6793 | -82.8257 | -88.2737 | 0.0695 |
| Rx3,x5 | 2.7486 1 | 50.6941 | -93.3882 | -92.2771 | -86.5339 | -90.8923 | 0.0693 |
| Reduced model | | 40.4976 | -78.9952 | -78.8926 | -77.2816 | -78.3712 | |

Table 17 shows that the full model is statistically significant than the intercept model alone. But the removal of any of the predictors leads to insignificant LRT indicating that each predictor is significant when involved alone in the regression equation. But the model of the 4 variables (any four combinations) is insignificant from the full model of 5 variables. So the removed variable is insignificant when controlling for others. Moreover, removal of the variable leads to reversal of the sign of some coefficients except removal of the third variable (the life expectancy) where the signs of the remaining variables are consistent with the signs of the marginal Kendall. For this model, the model with the removed life expectancy; the coefficient of the fifth variable which is the homicide rate is almost negligible (3.0644e-4). So the fifth variable can also be removed. Furthermore, the Log Likelihood of the model with the removed X3 is 50.6941 and it is similar to LL of the model with both X3 and X5 removed. In other words, the model incorporating the employment rate, the air pollution, and the life satisfaction can be considered a good model, it has LL=50.6941 which is higher than the LL obtained by simple regression of any one of them alone. However, it is not statistically significant from the model containing only the life satisfaction which has LL equals to 49.2593. Considering the AIC, it is -94.5187 for the simple regression containing the life satisfaction alone, while for the 3 variables model, it is -93.3882 and this is the penalty paid for the addition of the two parameters reflecting the effect of the employment rate and the air pollution. Also to be noticed is that the signs of both the employment rate and the air pollution are consistent in all the models as shown from the tables. Also removing the employment rate from the 3 variable model yields an AIC equals to -93.8338 and removing the air pollution from the 3 variable model gives an AIC equals to -94.9638 while removing the life satisfaction predictor from the 3 variable model gives an AIC equals to -91.650. This denotes the importance of the life satisfaction over both the employment rate and the air pollution. Although, the life satisfaction is important but the presence of the employment rate in the regression equation yields an AIC equals to -94.9638; which is slightly more negative than the presence of life satisfaction alone in the regression equation whose AIC is -94.5187. Therefore, the life satisfaction is the most important variable to be engaged in the model but also adding the employment rate slightly enhances the AIC but not the BIC. Because in communities where the population has a high welfare and can freely express their opinion in the services offered to them they can definitely influence the quality of water. Hence; the higher the life satisfaction is, the higher the quality of water supply and cleanliness are. This is also applicable to the employment rate. Table 18 shows the statistics of the three model variable and the effect of removing each variable on the LL, AIC, and BIC.

Table 18 : estimated coefficients for the 3 variable model

|  | intercept | X1 Coeff. | X2 Coeff. | X4 Coeff. | Preserve sign | LL | AIC | BIC |
|---|---|---|---|---|---|---|---|---|
| Full model | 10.4226 4.5705 | 1.7841 1.4483 | -0.2348 0.3614 | 3.0575 1.6295 | yes | 50.6941 | -93.3882 | -86.5339 |
| Rx1 | 11.8060 4.5433 | | -0.0349 0.3442 | 3.9526 1.5088 | yes | 49.9194 | -93.8338 | -88.6981 |
| Rx2 | 11.9091 3.9652 | 2.1051 1.3650 | | 3.3754 1.5583 | Yes | 50.4819 | -94.9638 | -89.8231 |
| Rx4 | 2.0178 1.1068 | 2.7988 1.3038 | -0.4542 0.3469 | | yes | 48.8251 | -91.650 | -86.5094 |

# Third Response Variable: Quality of Support Network

The quality of support network was regressed on three predictors; one at a time then all in one full model. Figure 70 shows the scatter plot that detects the relationship between the response variable and each predictor (transformed). The author presented figures for the logit model, figures that illustrate the estimated curve, plot between the predictor and the residuals, and the QQ plot for the empirical residuals against the theoretical residuals. The relationship between the response variable and each of the predictor is nonlinear. Tables 19-21 show the results obtained from regressing the quality of support network on each predictor using different link functions and comparing the statistical indices as regards the estimated coefficients , the Likelihood Ratio Test (LRT ) and its p value, AIC, CAIC, BIC, HQIC and the LL.

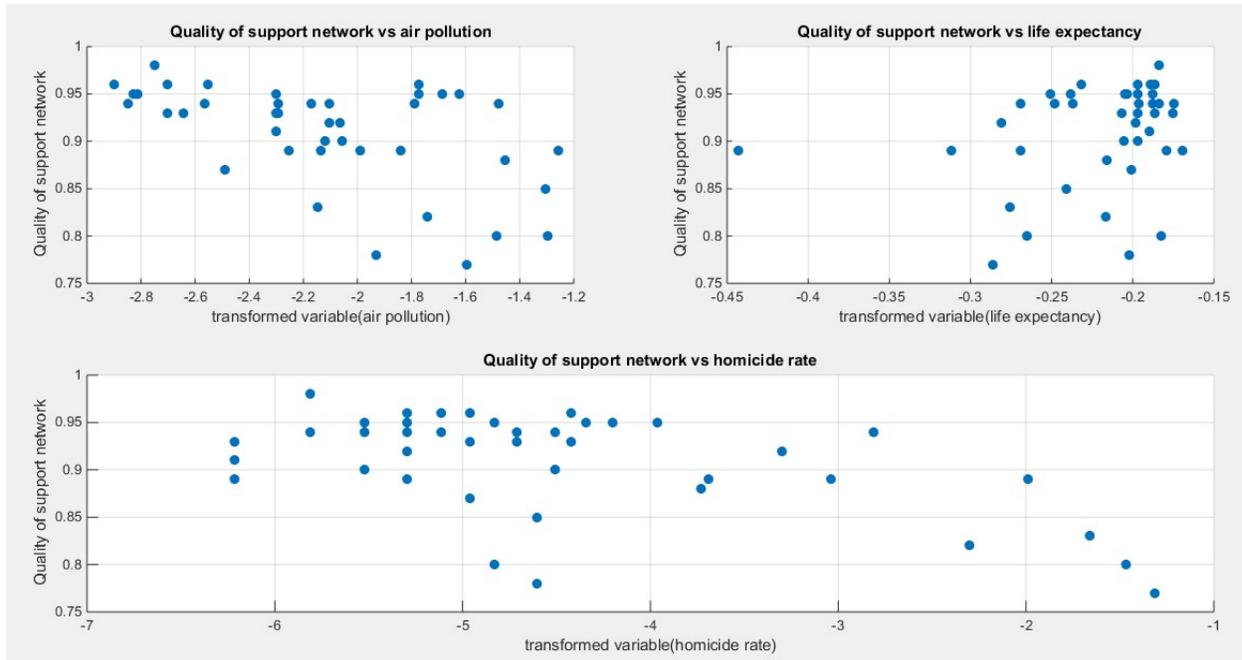

Fig. 70 shows the scatter plot of the response variable and each of the predictor. The relationship is nonlinear.

Table 19: Regressing quality of support networks on air pollution

|  | Logit link function | | Log-log complementary | | Log-log median | |
|---|---|---|---|---|---|---|
| B0 | 0.8736 | | 0.3501 | | -0.9847 | |
| B1 | -0.7754 | | -0.2817 | | 0.7426 | |
| LL | 71.2863 | | 71.3626 | | 71.2751 | |
| Wald stat. of b0 | 1.5465 (p> 0.025) | | 1.6559 (p > 0.025) | | 1.8329 (p > 0.025) | |
| Wald stat. of b1 | 2.9779 (p <0.025) | | 3.0029 (p < 0.025) | | 2.9824 (p < 0.025) | |
| AIC | -138.5726 | | -138.7252 | | -138.5502 | |
| CAIC | -138.2568 | | -138.4094 | | -138.2345 | |
| BIC | -135.1454 | | -135.2981 | | -135.1231 | |
| HQIC | -137.3246 | | -137.4772 | | -137.3023 | |
| LRT | 8.9221 (p-val.=0.0028) | | 9.0747 (p-val.=0.0026) | | 8.8998(p-val.=0.0029) | |
| R-squared | 0.1956 | | 0.1986 | | 0.1951 | |
| P-value for randomized quantile residuals | 0.1421 | | 0.1356 | | 0.1439 | |
| p-value for Cox-snell residuals | 0.1421 | | 0.1356 | | 0.1439 | |
| Variance-covariance matrix | 0.3191 | 0.1439 | 0.0447 | 0.0194 | 0.2886 | 0.1308 |
| | 0.1439 | 0.0678 | 0.0194 | 0.0088 | 0.1308 | 0.0620 |
| QR vs. predictor(tau,p) | (0.0159, 0.8927) | | (0.0159, 0.8927) | | (0.0159, 0.8927) | |
| CS vs. predictor(tau,p) | (0.0159, 0.8927) | | (0.0159, 0.8927) | | (0.0159, 0.8927) | |

Table 19 shows that the predictor is significant as likelihood ratio test (LRT) is highly significant; the R squared is also high for this predictor around 0.19 across the different link functions. The AIC, CAIC, BIC, HQIC and LL are more or less equal across the different models. The LL is around 71 across the link functions. The residuals plotted against the predictors show no specific trend and they are randomly scattered. The QQ plot of the randomized quantile residuals shows perfect alignment with the diagonal all through its course in contrast with the Cox Snell residuals that show this perfect alignment at the lower tail and the center. The estimated curve between the estimated median and the transformed predictor is decreasing reflecting that the more the air pollution is, the less the percentage expressing increased quality of support is. The figure for the clog-log shows the same pattern. The log-log figure has the same pattern. The difference is mainly manifested in the slope of the estimated curve. To assess the assumption of constant variance in the median parametric regression model, residual-based diagnostic tests were conducted using both randomized quantile (RQ) and Cox-Snell (CS) residuals. For each type of residual, an auxiliary regression of the squared residuals on the corresponding predictor was estimated by ordinary least squares, and the null hypothesis of homoscedasticity ($H_0$: constant variance) was tested. The results indicted no significant relationship between the squared residuals and the predictor variable (CS: p=0.294, R-squared=0.0282; RQ: p=0.203, R-squared=0.0412), suggesting that the variance of the residuals remained approximately constant across the range of the predictor. Furthermore, the magnitude of the CS residuals was within a reasonable range, which supports the absence of heteroscedasticity. These findings provide evidence that the fitted median regression model satisfies the homoscedasticity assumption. These results are from the logit model. Figures 71-77 show the previous results.

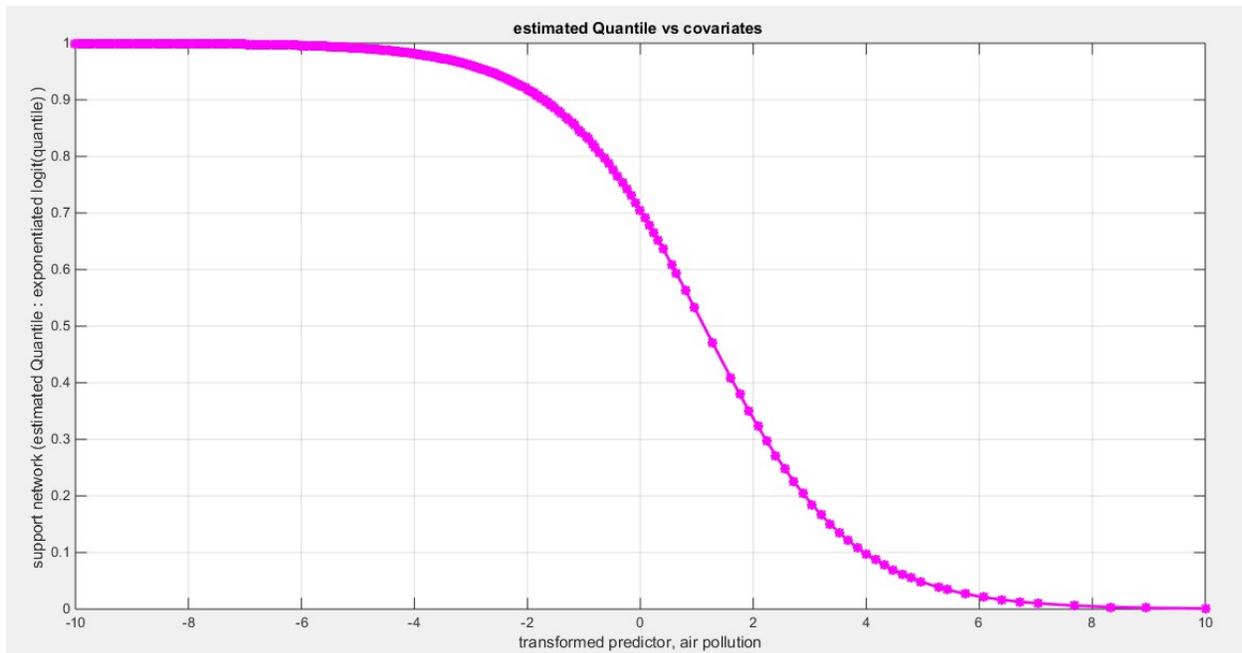

Fig. 71 shows the estimated curve plotting the transformed predictor against the estimated median (for the logit link).

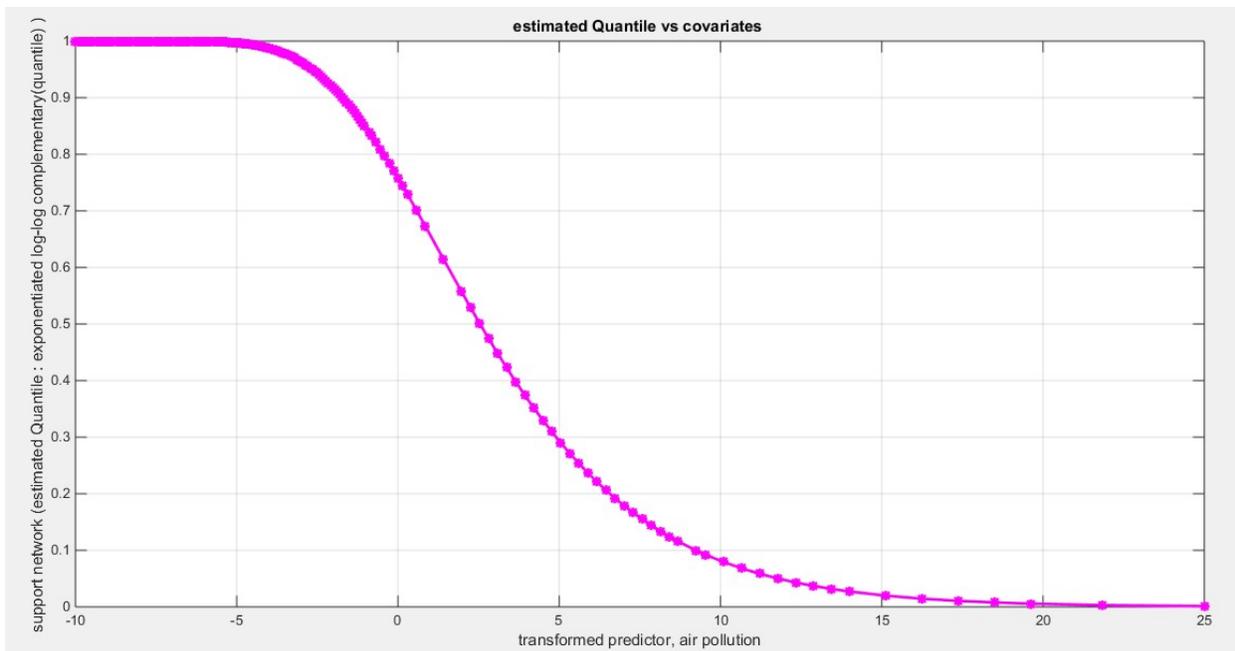

Fig. 72 shows the estimated curve plotting the transformed predictor against the estimated median (for the clog-log link).

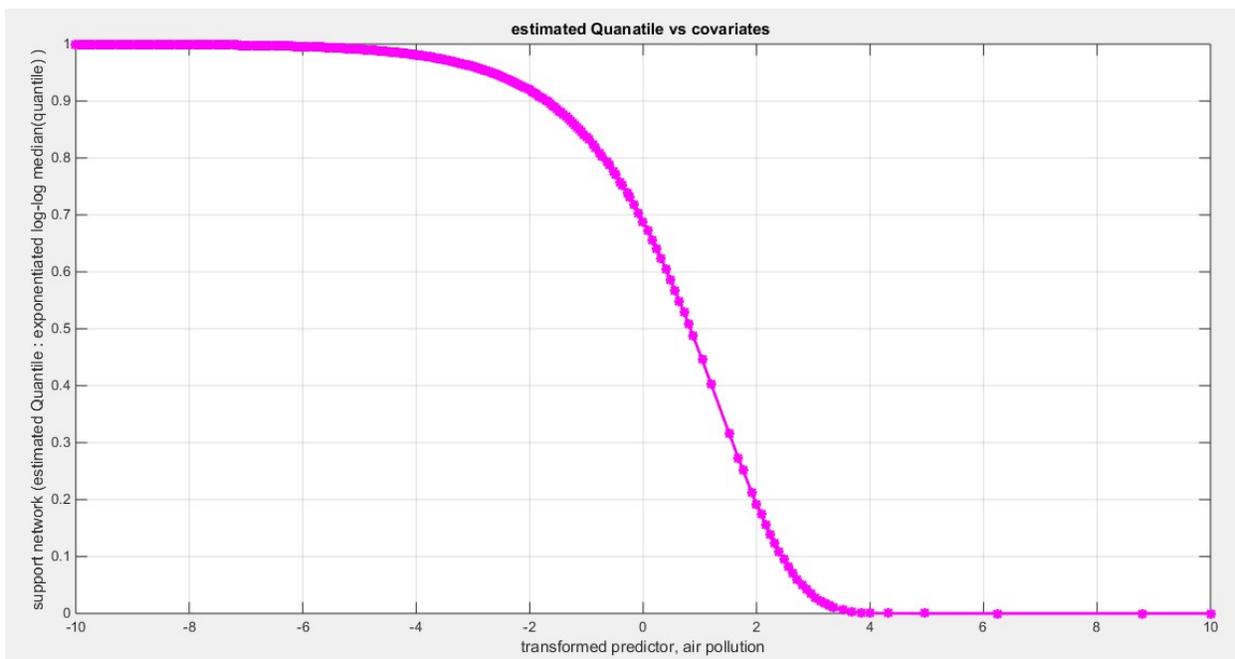

Fig. 73 shows the estimated curve plotting the transformed predictor against the estimated median (for the log-log link).

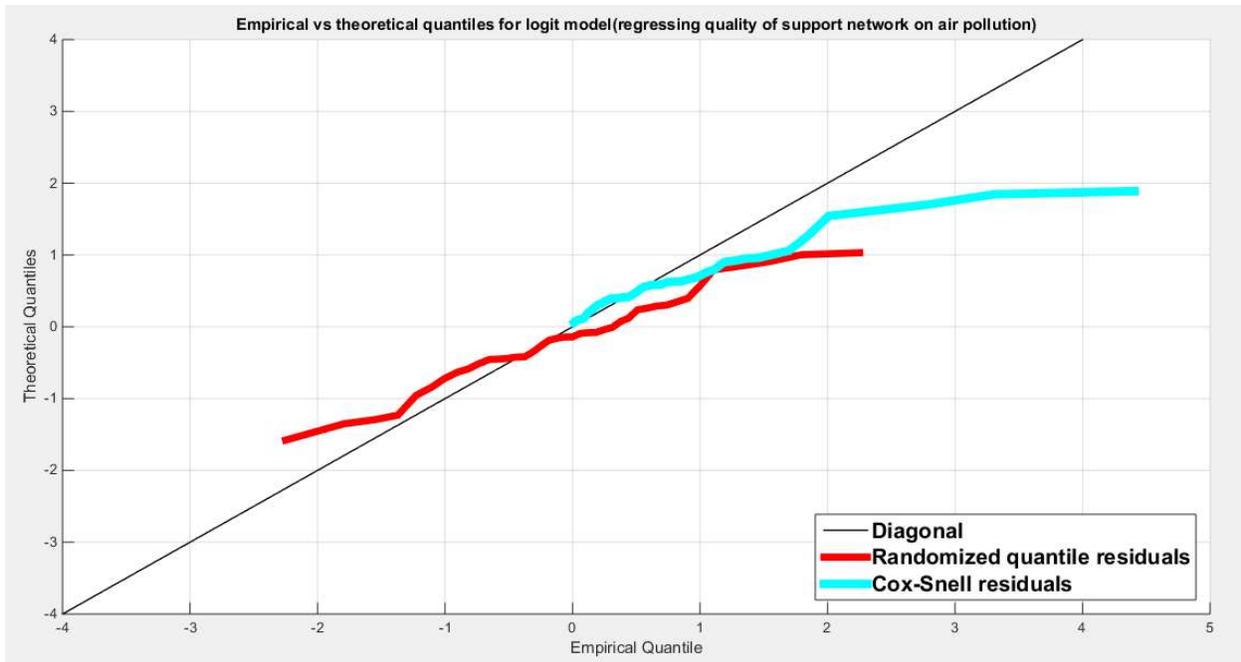

Fig. 74 shows the QQ plot of the empirical quantiles and the theoretical quantiles for both types of residuals.

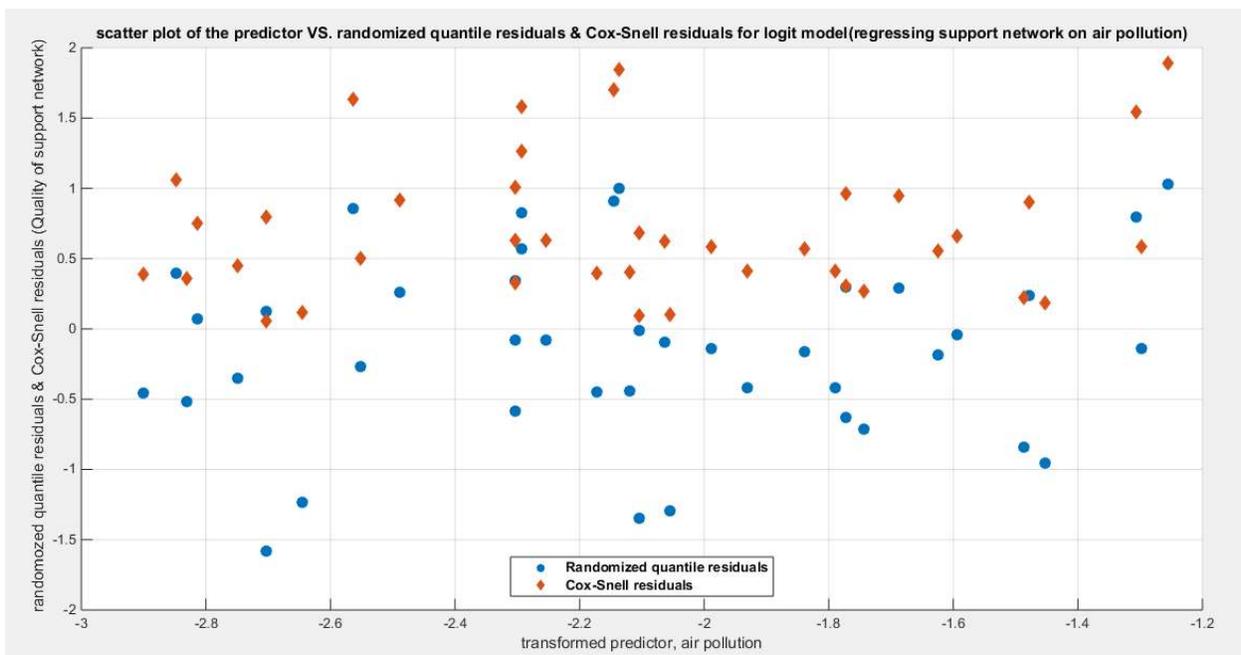

Fig. 75 shows the scatter plot of residuals of both types against transformed predictors.

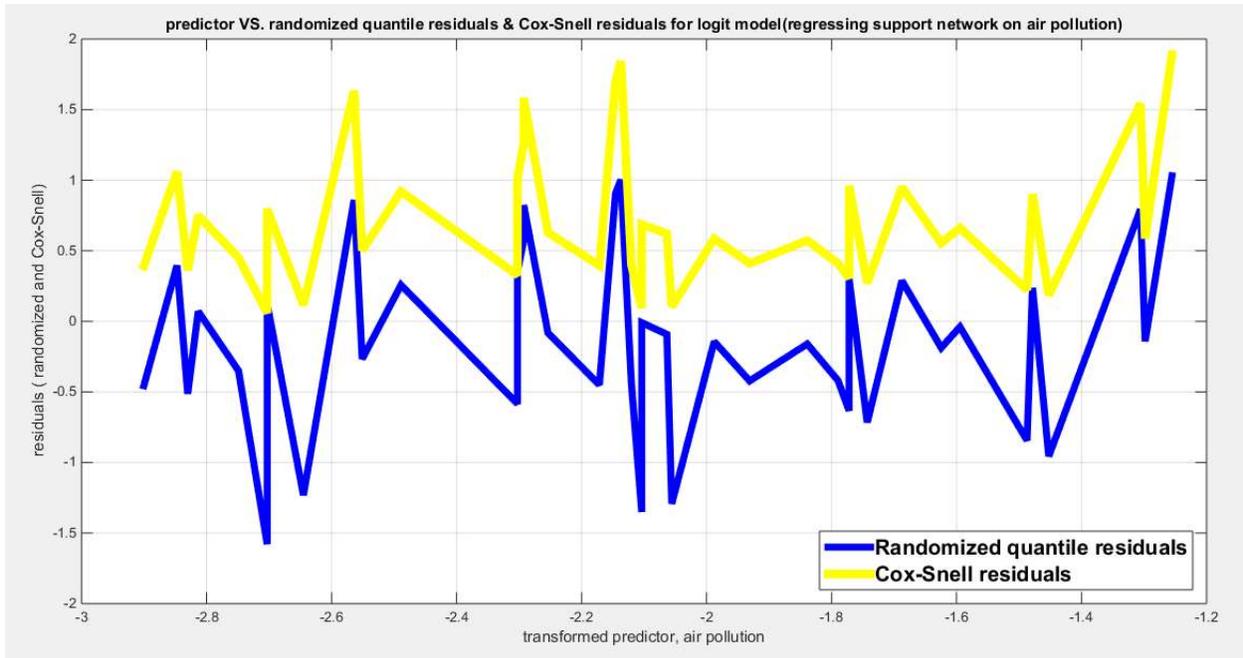

Fig. 76 shows the plot of residuals of both types against transformed predictors

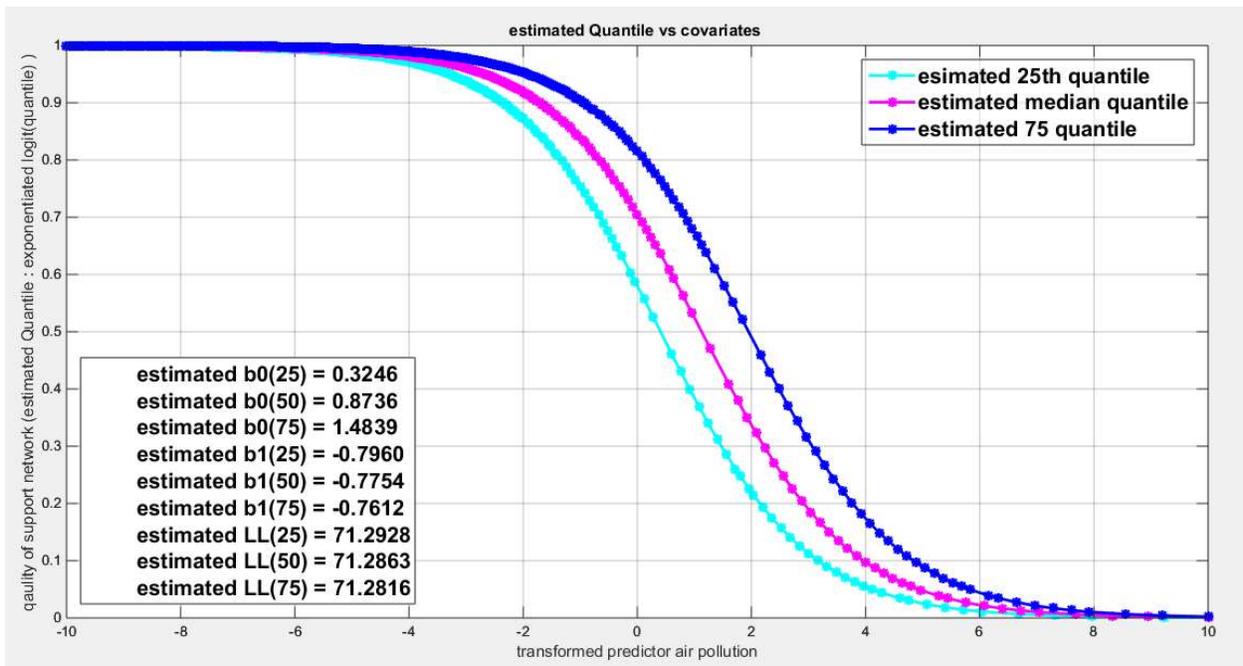

Fig. 77 shows parallel quantile curves across 25th , 50th ( median), 75th percentiles, suggesting that the predictor exerts a uniform influence on the response consistent with homoscedasticity.

Table 20: regressing quality of support network on the life expectancy

| | Logit link function | | Log-log complementary | | Log-log median | |
|---|---|---|---|---|---|---|
| B0 | 3.2216 | | 1.2317 | | -3.2201 | |
| B1 | 3.4077 | | 1.3446 | | -3.2133 | |
| LL | 67.7959 | | 67.8400 | | 67.7874 | |
| Wald stat. of b0 | 5.5193 (p < 0.025) | | 5.4066 (p < 0.025) | | 5.8605 (p < 0.025) | |
| Wald stat. of b1 | 1.3129 (p > 0.025) | | 1.2979 (p > 0.025) | | 1.3209 (p > 0.025) | |
| AIC | -131.5917 | | -131.6799 | | -131.5747 | |
| CAIC | -131.2759 | | -131.3641 | | -131.2589 | |
| BIC | -128.1646 | | -128.2528 | | -128.1476 | |
| HQIC | -130.3437 | | -130.4319 | | -130.3268 | |
| LRT | 1.9412 (p=0.1635) | | 2.0294 (p=0.1543) | | 1.9243 (p=0.1654) | |
| R-squared | 0.0462 | | 0.0483 | | 0.0458 | |
| P-value for randomized quantile residuals | 0.1079 | | 0.1033 | | 0.1089 | |
| p-value for Cox-Snell residuals | 0.1079 | | 0.1033 | | 0.1089 | |
| Variance-covariance matrix | 0.3407 | 1.4842 | 0.0519 | 0.2320 | 0.3019 | 1.3083 |
| | 1.4842 | 6.7365 | 0.2320 | 1.0732 | 1.3083 | 5.9173 |
| QR vs. predictor(tau,p) | -0.1104 , 0.3171 | | -0.1104 , 0.3171 | | -0.1104 , 0.3171 | |
| CS vs. predictor(tau,p) | -0.1104 , 0.3171 | | -0.1104 , 0.3171 | | -0.1104 , 0.3171 | |

Table 20 shows that the predictor is insignificant as likelihood ratio test (LRT) is low around 2. The R squared is also low for this predictor around 0.045 across the different link functions. The AIC, CAIC, BIC, HQIC and LL are more or less equal across the different models. The LL is around 67 across the link functions. The residuals plotted against the predictors show no specific trend and they are randomly scattered. The QQ plot shows that the randomized quantile residuals are not perfectly aligned with the diagonal all through its course in similarity with the Cox Snell residuals that do not show this perfect alignment at the lower tail and the center. The estimated curve between the estimated median and the transformed predictor is increasing reflecting that the more the life expectancy is, the more the percentage expressing increased quality of support is. The figure for the clog-log shows the same pattern. The log-log figure has the same pattern. The difference is mainly manifested in the slope of the estimated curve. To assess the assumption of constant variance in the median parametric regression model, residual-based diagnostic tests were conducted using both randomized quantile (RQ) and Cox-Snell (CS) residuals. For each type of residual, an auxiliary regression of the squared residuals on the corresponding predictor was estimated by ordinary least squares, and the null hypothesis of homoscedasticity ($H_0$: constant variance) was tested. The results indicted no significant relationship between the squared residuals and the predictor variable (CS: p=0.746, R-squared=0.00272; RQ: p=0.879, R-squared=0.000601), suggesting that the variance of the residuals remained approximately constant across the range of the predictor. Furthermore, the magnitude of the CS residuals were within a reasonable range (one value 2. 4929), which supports the absence of heteroscedasticity. These findings provide evidence that the fitted median regression model satisfies the homoscedasticity assumption. These results are from the logit model. Figures 78-84 show the previous results.

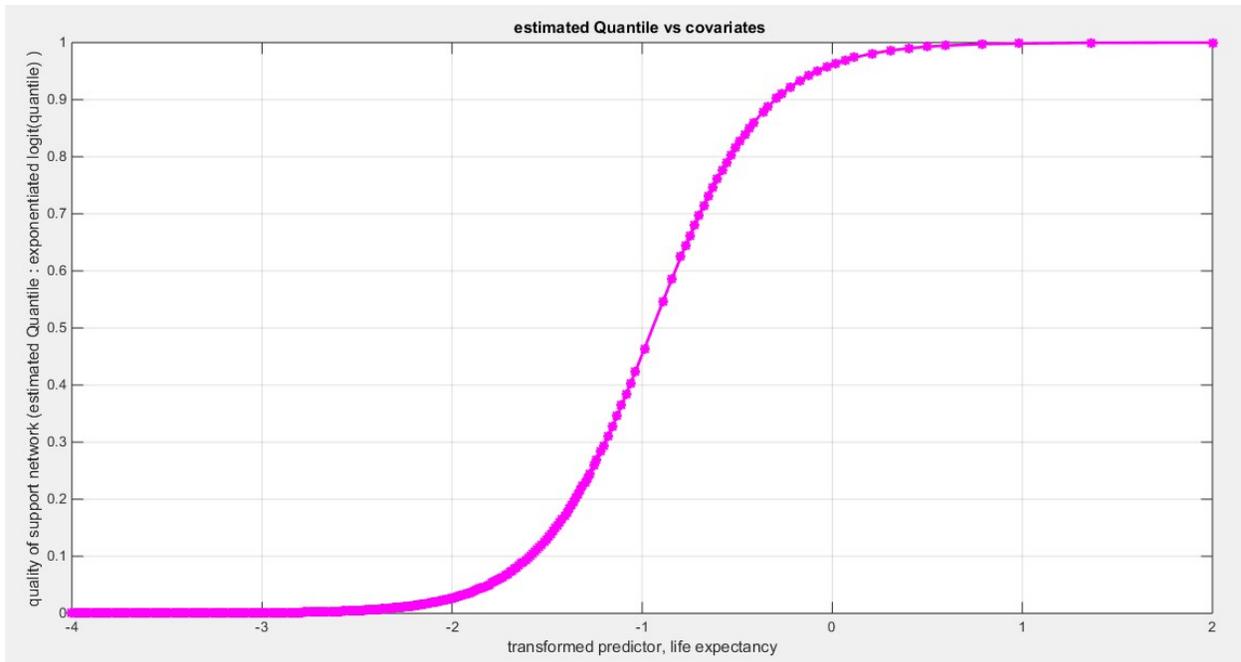

Fig. 78 shows the estimated curve plotting the transformed predictor against the estimated median (for the logit link).

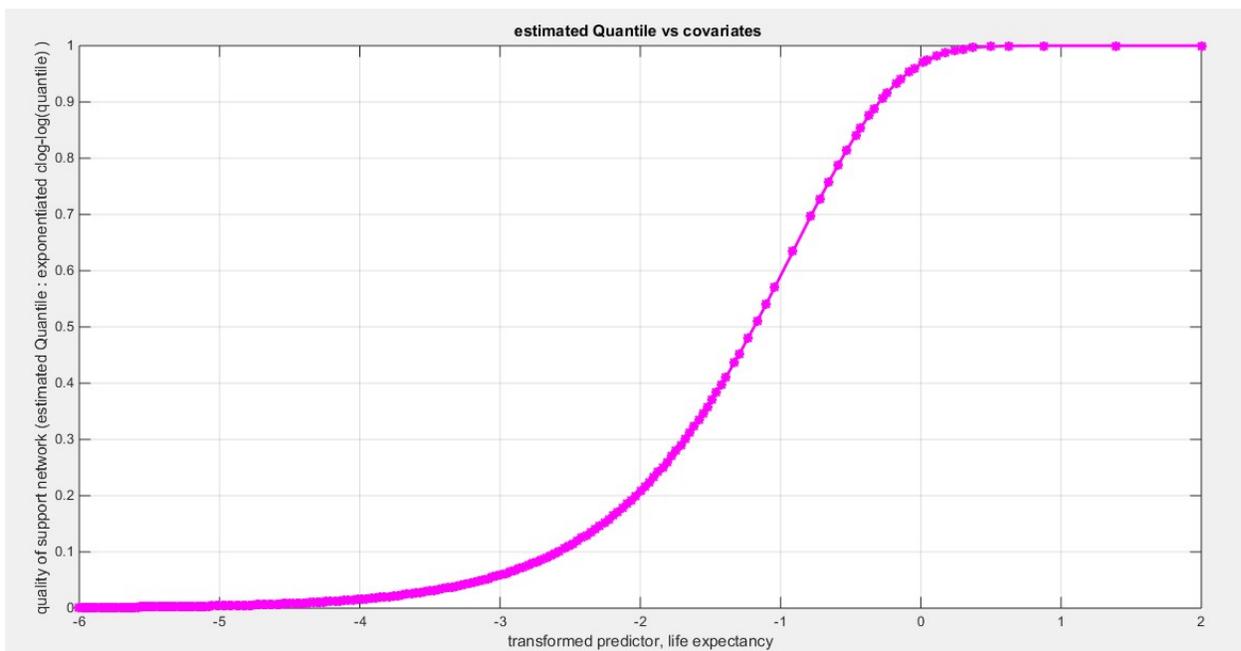

Fig. 79 shows the estimated curve plotting the transformed predictor against the estimated median (for the clog-log link).

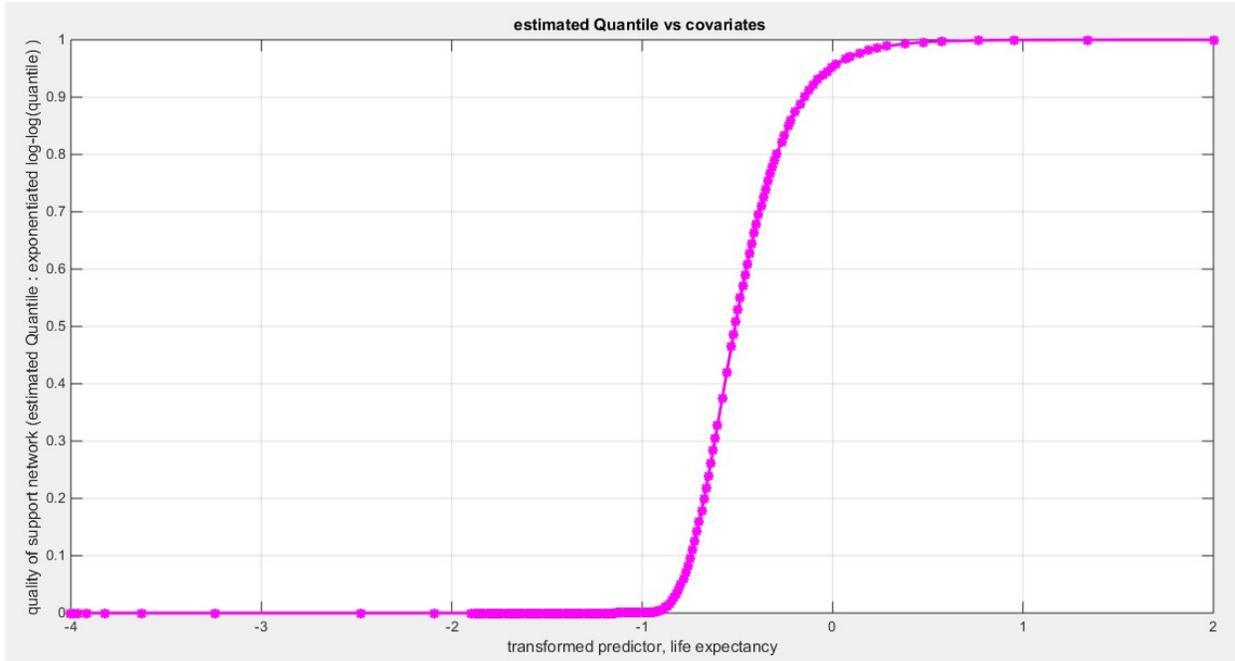

Fig. 80 shows the estimated curve plotting the transformed predictor against the estimated median (for the log-log link).

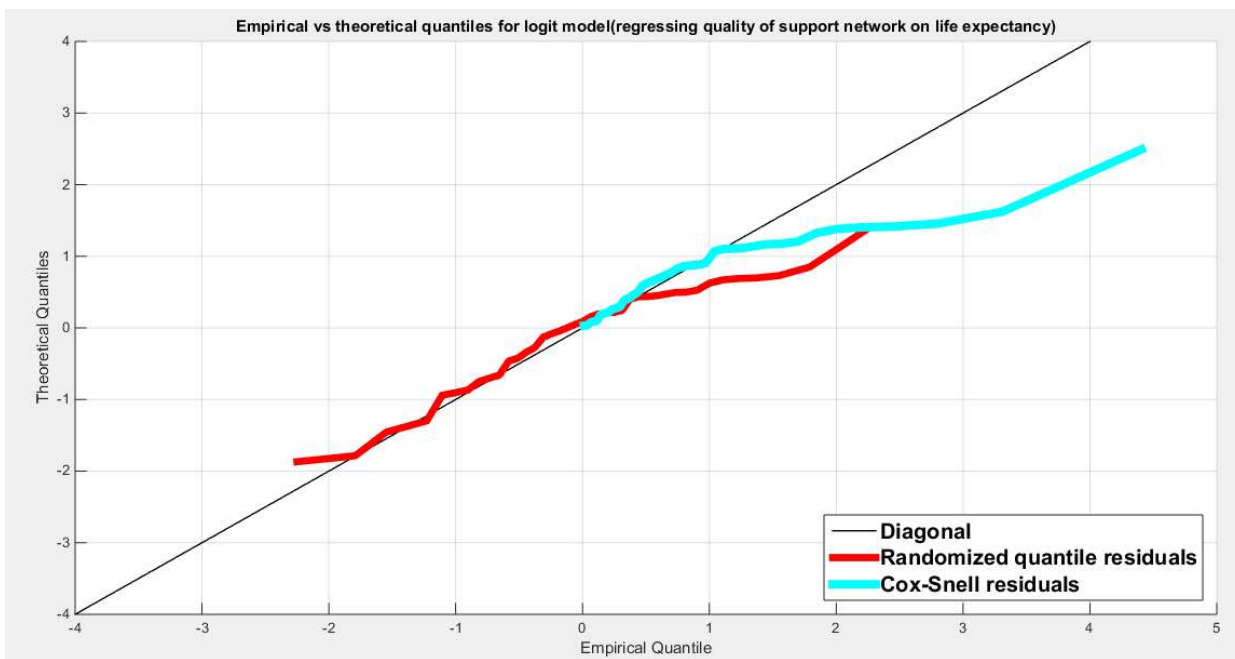

Fig. 81 shows the QQ plot of the empirical quantiles and the theoretical quantiles for both types of residuals.

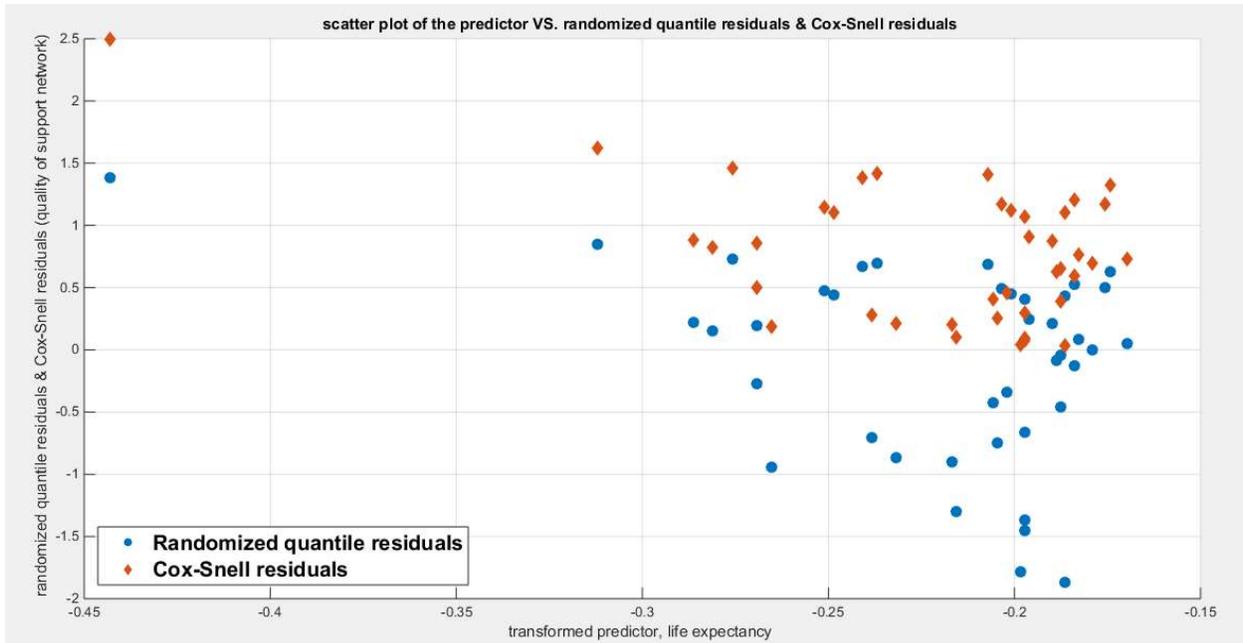

Fig. 82 shows the scatter plot of residuals of both types against transformed predictors.

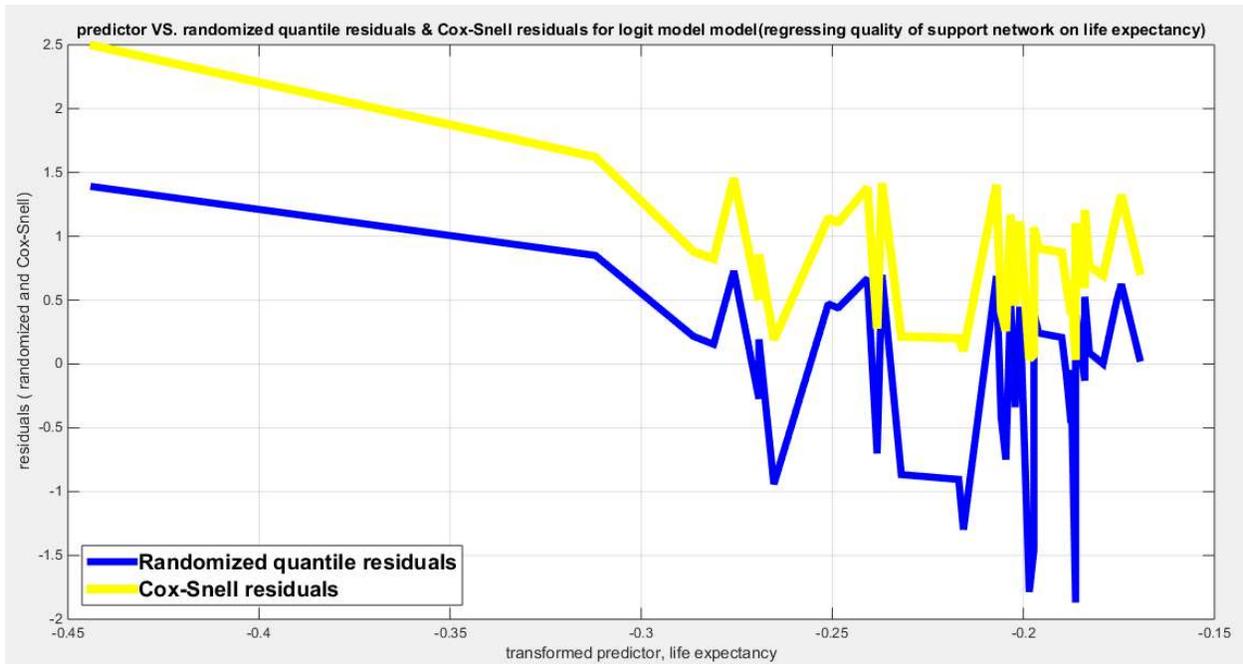

Fig. 83 shows the plot of residuals of both types against transformed predictors

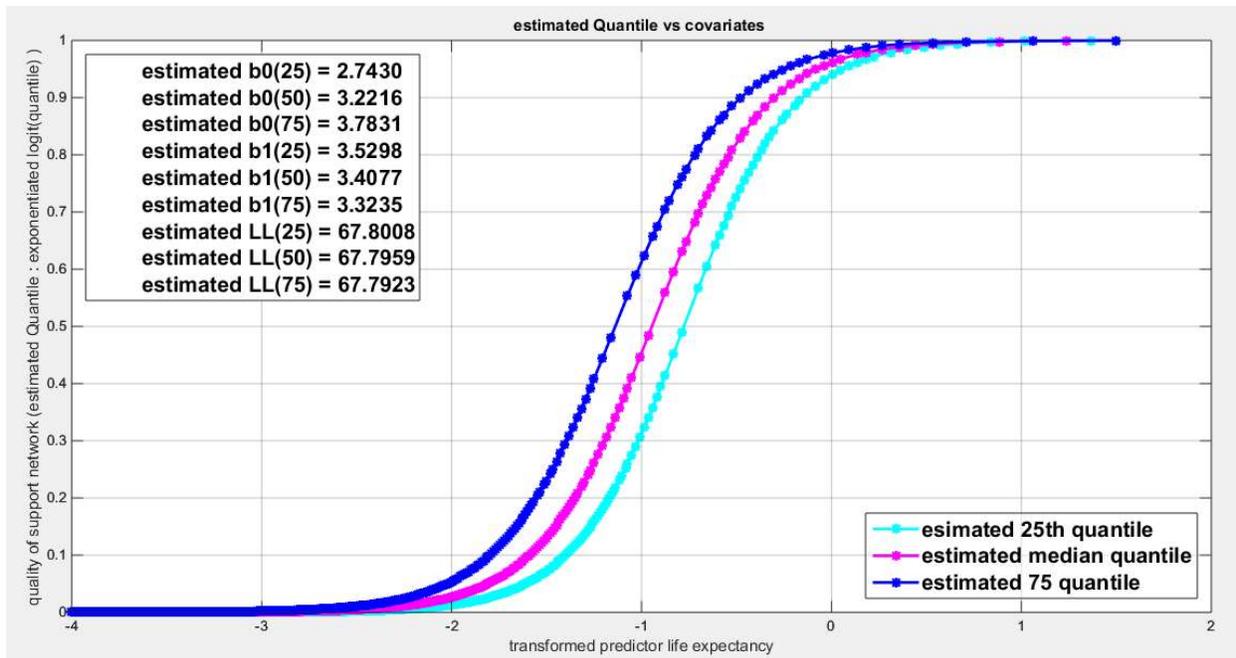

Fig. 84 shows parallel quantile curves across 25th , 50th ( median), 75th percentiles, suggesting that the predictor exerts a uniform influence on the response consistent with homoscedasticity. ( for the logit link).

Table 21: regressing quality of support network on homicide rate

| | Logit link function | | Log-log complementary | | Log-log median | |
|---|---|---|---|---|---|---|
| B0 | 1.5185 | | 0.5899 | | -1.5995 | |
| B1 | -0.2202 | | -0.0792 | | 0.2113 | |
| LL | -70.2822 | | -70.1800 | | -70.2972 | |
| Wald stat. of b0 | 3.7004(p<0.025) | | 3.6655(p<0.025) | | 4.1230(p<0.025) | |
| Wald stat. of b1 | 2.5094(p<0.025) | | 2.3879(p<0.025) | | 2.5256(p<0.025) | |
| AIC | -136.5645 | | -136.3600 | | -136.5944 | |
| CAIC | -136.2487 | | -136.0442 | | -136.2786 | |
| BIC | -133.1373 | | -132.9329 | | -133.1672 | |
| HQIC | -135.3165 | | -135.1121 | | -135.3464 | |
| LRT | 6.9140 (p=0.0086) | | 6.7096 (p=0.0096) | | 6.9439 (p=0.0084) | |
| R-squared | 0.1552 | | 0.1510 | | 0.1558 | |
| P-value for randomized quantile residuals | 0.0844 | | 0.0796 | | 0.0847 | |
| p-value for Cox-snell residuals | 0.0844 | | 0.0796 | | 0.0847 | |
| Variance-covariance matrix | 0.1684 | 0.0346 | 0.0259 | 0.0052 | 0.1505 | 0.0310 |
| | 0.0346 | 0.0077 | 0.0052 | 0.0011 | 0.0310 | 0.0070 |
| QR vs. predictor(tau,p) | (0.0074, 0.9551) | | (0.0074, 0.9551) | | (0.0074, 0.9551) | |
| CS vs. predictor(tau,p) | (0.0074, 0.9551) | | (0.0074, 0.9551) | | (0.0074, 0.9551) | |

Table 21 shows that the predictor is significant as likelihood ratio test (LRT) is high around 6.9. The R squared is also high for this predictor around 0.15 across the different link functions. The AIC, CAIC, BIC, HQIC and LL are more or less equal across the different models. The LL is around 70 across the link functions. The residuals plotted against the predictors show no specific trend and they are randomly scattered. The QQ plot shows that the randomized quantile residuals are not perfectly aligned with the diagonal all through its course in similarity with the Cox Snell residuals that do not show this perfect alignment at the lower tail and the center. The estimated curve between the estimated median and the transformed predictor is decreasing reflecting that the more the homicide rate is, the less the percentage expressing increased quality of support is. The figure for the clog-log shows the same pattern. The log-log figure has the same pattern. The difference is mainly manifested in the slope of the estimated curve. To assess the assumption of constant variance in the median parametric regression model, residual-based diagnostic tests were conducted using both randomized quantile (RQ) and Cox-Snell (CS) residuals. For each type of residual, an auxiliary regression of the squared residuals on the corresponding predictor was estimated by ordinary least squares, and the null hypothesis of homoscedasticity ($H_0$: constant variance) was tested. The results indicted no significant relationship between the squared residuals and the predictor variable (CS: p=0.425, R-squared=0.0164; RQ: p=0.46, R-squared=0.0141), suggesting that the variance of the residuals remained approximately constant across the range of the predictor. Furthermore, the magnitude of the CS residuals were within a reasonable range (one value 2.1671), which supports the absence of heteroscedasticity. These findings provide evidence that the fitted median regression model satisfies the homoscedasticity assumption. These results are from the logit model. Figures 85-91 show the previous results.

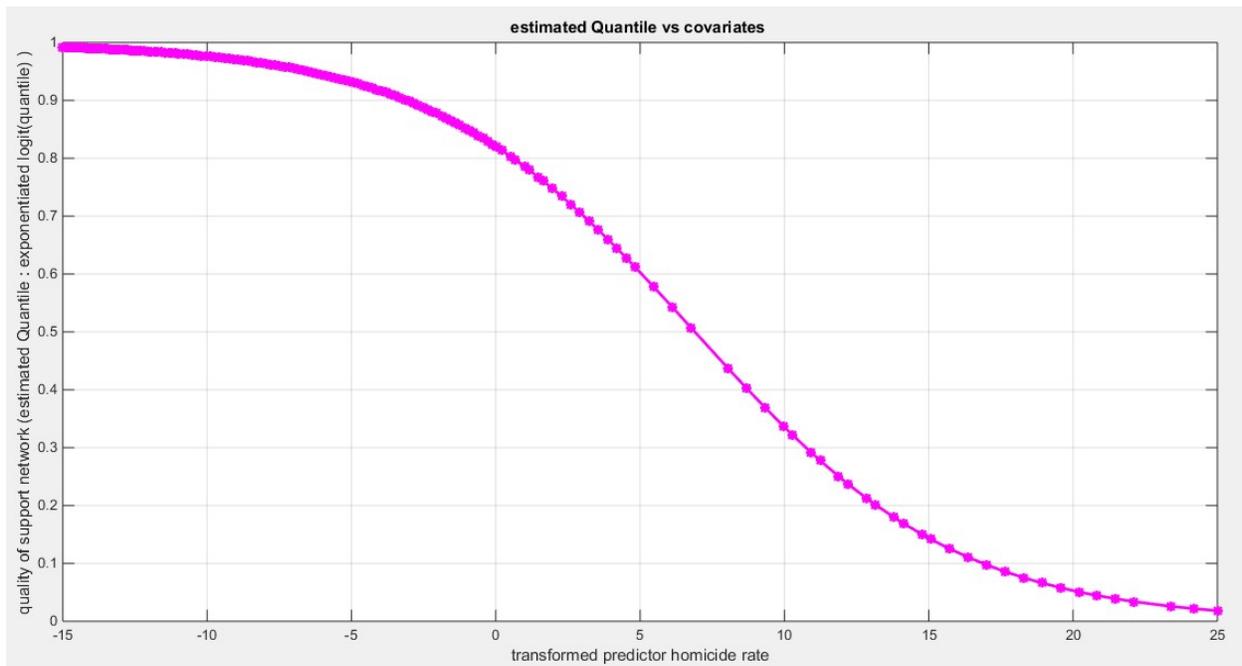

Fig. 85 shows the estimated curve plotting the transformed predictor against the estimated median (for the logit link).

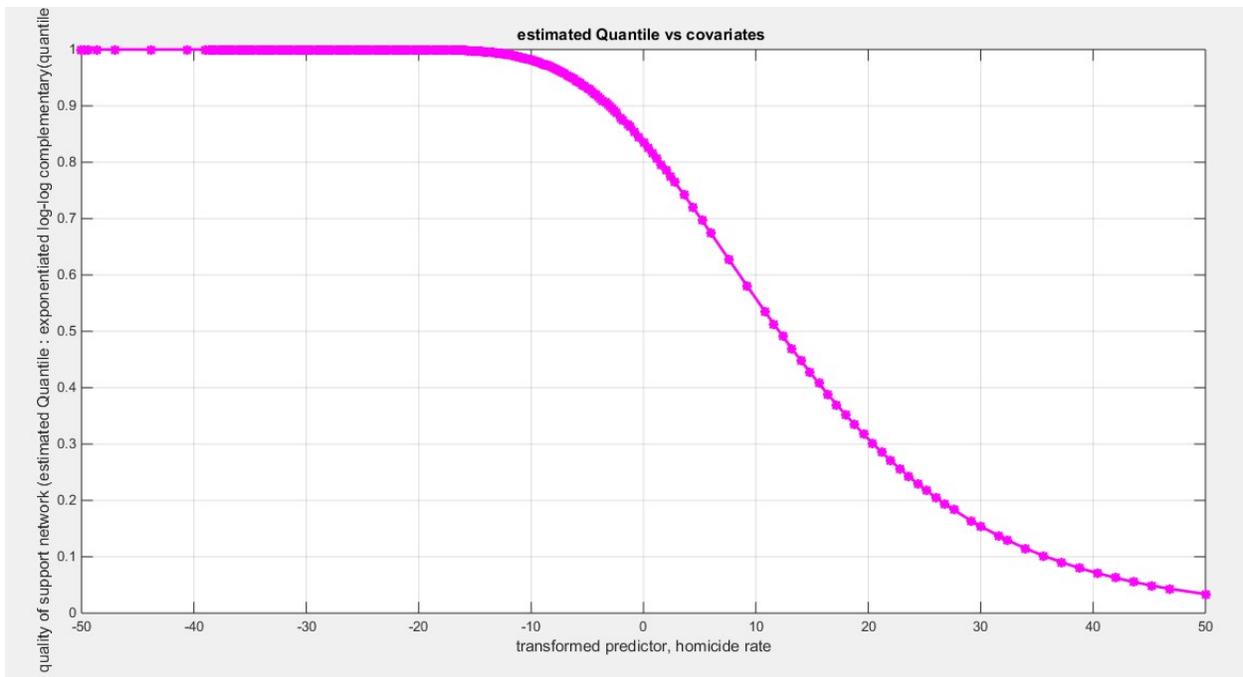

Fig. 86 shows the estimated curve plotting the transformed predictor against the estimated median (for the clog-log link).

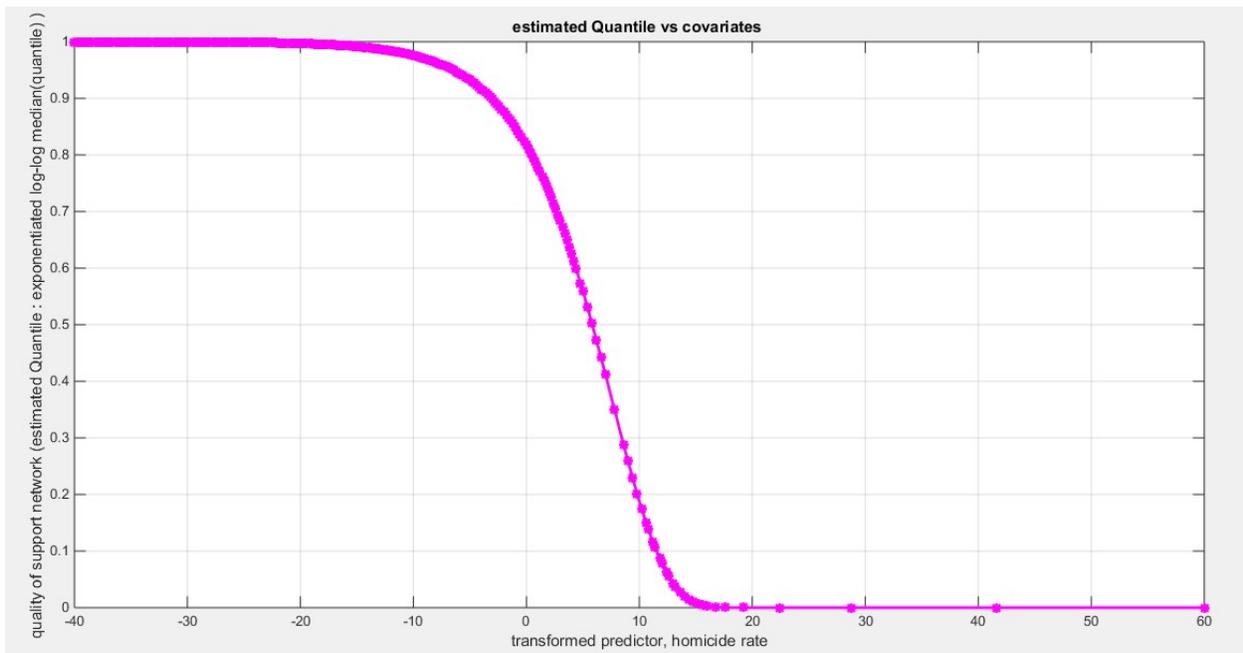

Fig. 87 shows the estimated curve plotting the transformed predictor against the estimated median (for the log-log link).

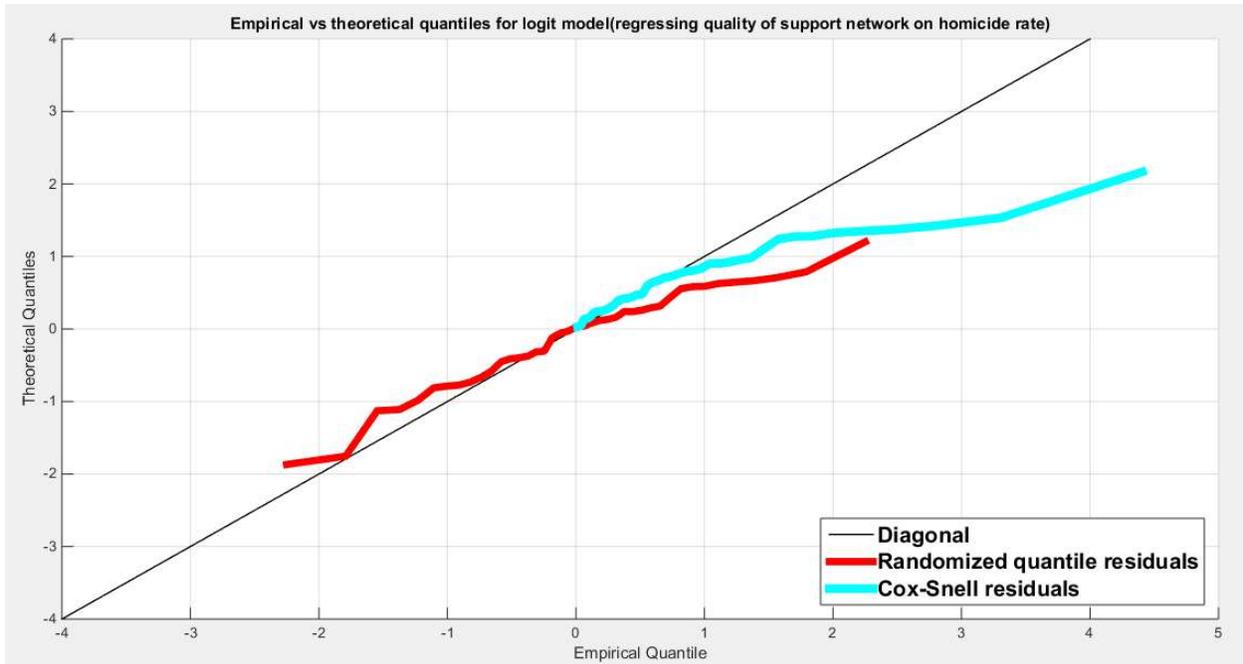

Fig. 88 shows the QQ plot of the empirical quantiles and the theoretical quantiles for both types of residuals.

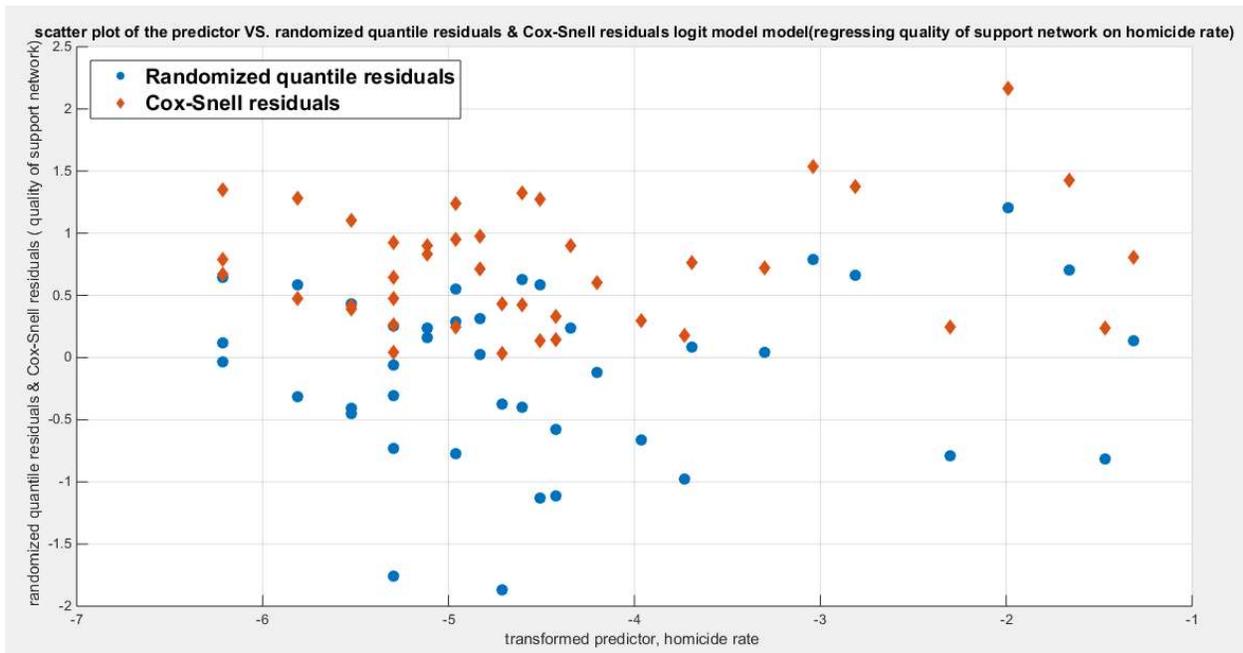

Fig. 89 shows the scatter plot of residuals of both types against transformed predictors.

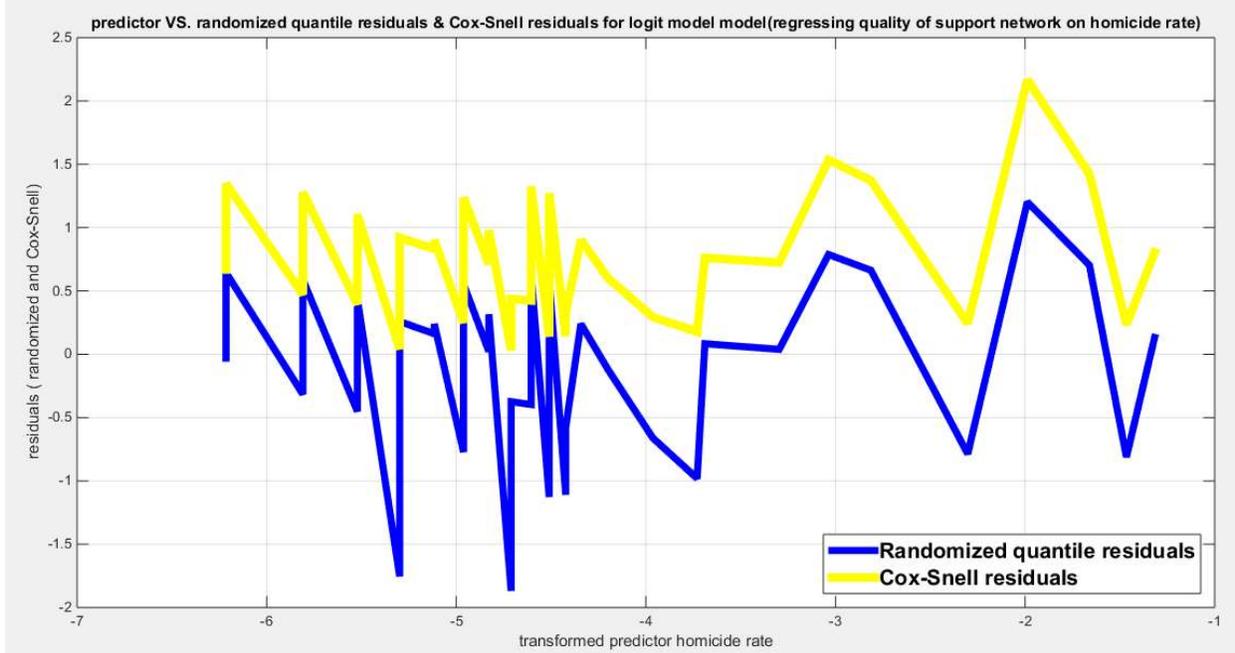

Fig. 90 shows the plot of residuals of both types against transformed predictors

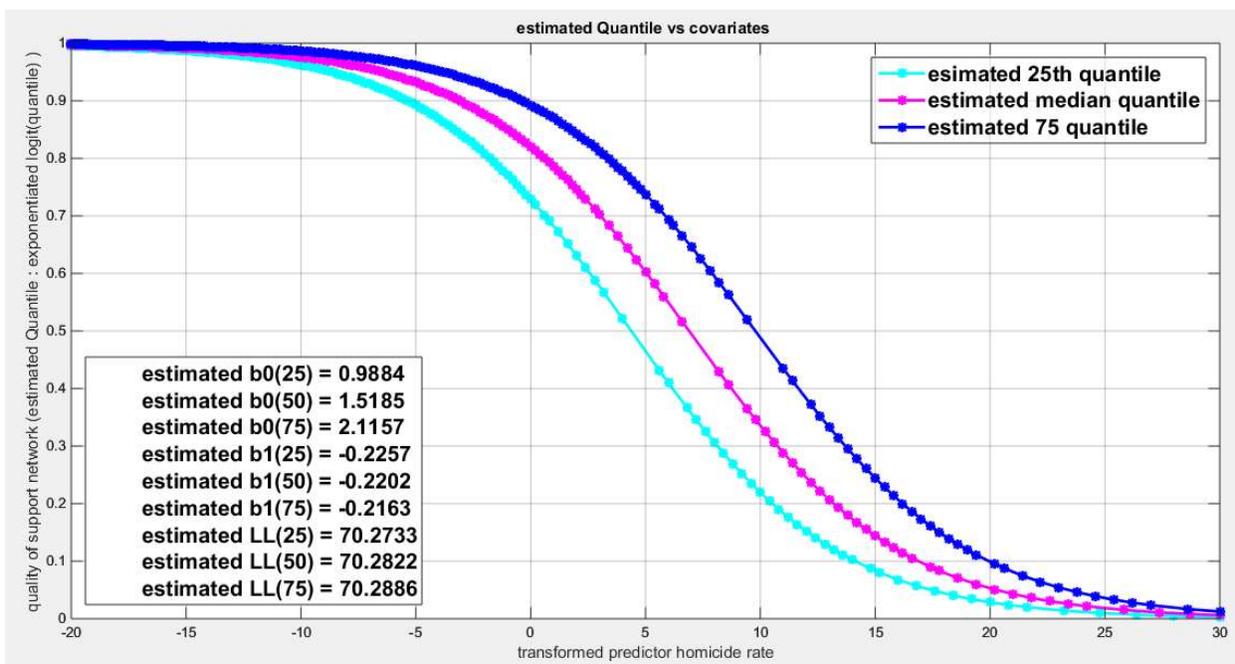

Fig. 91 shows parallel quantile curves across 25th , 50th ( median), 75th percentiles, suggesting that the predictor exerts a uniform influence on the response consistent with homoscedasticity.

The marginal correlations between the variables (the response and the predictors) are shown in Table 22. The quality of support network is negatively and shows statistical significant correlation with both the air pollution and the homicide rate. The air pollution is negatively and significantly correlated with the life expectancy. The life expectancy shows statistical and significant negative correlation with the homicide rate.

Table 22: The marginal correlation matrix Kendall tau coefficient and associated p-value:

|  | Quality of support network(Y) | Air Pollution(X1) | Life expectancy(X2) | Homicide Rate(X3) |
|---|---|---|---|---|
| Quality of support network(Y) | 1 | -0.3743 P=9.1487e-4 | 0.2079 P=0.0667 | -0.5692 P=0.0184 |
| Air Pollution (X1) | -0.3743 P=9.1487e-4 | 1 | -0.2711 P=0.0138 | 0.1207 P=0.2796 |
| Life expectancy(X2) | 0.2079 P=0.0667 | -0.2711 P=0.0138 | 1 | -0.4327 P=0.0001 |
| Homicide Rate (X3) | -0.5692 P=0.0184 | 0.1207 P=0.2796 | -0.4327 P=0.0001 | 1 |

The condition indices obtained from the standardized transformed X'X are 2.5439, 1.5928, and 1. The VIF for the air pollution is 1.2151, for life expectancy is 2.0919, for homicide rate is 1.8597. So as the largest condition index is 2.5439 less than 10 and the VIF values are less than 5 so there is no evidence of significant multi-collinearity between the predictors.

The signs of the coefficients of the marginal correlations match those signs of the conditional correlations coefficients when regressing the quality of support network response variable on each predictor, one at a time. Also in multiple regression analysis, the signs of the coefficients of the air pollution and homicide rate are consistently and negatively dependent with the quality of support network while this consistency is not observed for the sign of the regression coefficient of the life expectancy which flips according to the X removed from the full model. Life expectancy may be a proxy for the highly developed country where people can survive for long time due to the welfare in these countries but the loneliness the elderly suffered may drive them to reduce reporting the social support when in crises or need; hence the regression coefficient sign flips.

The author added these three predictors in one equation and used the different link functions, then removed each one at a time and calculated the LRT to assess the significance of this particular predictor while controlling for other predictors. The results are summarized in the following Tables 23-24. The description of these tables as regards the rows and the columns are the same as the description of the tables for the previous response variables.

Table 23: the coefficients of the parametric median regression analysis with removal of different predictors and the associated standard error

| | intercept | X1 Coeff. | X2 Coeff. | X3 Coeff. | Preserve sign | LRT & P value | LL |
|---|---|---|---|---|---|---|---|
| Full Model SE | -1.1348 14.6108 | -0.7188 0.2885 | -4.3763 3.4593 | -0.2672 0.1270 | No | 13.696 0.0033 | 73.6732 |
| Rx1 SE | 0.7805 13.684 | | -2.2194 3.2062 | -0.2764 0.1219 | No | 6.3292 0.0119 | 70.5086 |
| Rx2 SE | 0.4661 0.6348 | -0.6319 0.2744 | | -0.1641 0.0926 | Yes | 1.4767 0.2243 | 72.9349 |
| Rx3 SE | 1.1724 11.9447 | -0.7383 0.2790 | 0.9942 2.7688 | | yes | 4.6426 0.0312 | 71.3519 |
| Reduced model | 2.4568 | | | | | | 66.8252 |

Table 24: AIC, CAIC, BIC, and HQIC

| | AIC | CAIC | BIC | HQIC | R squared |
|---|---|---|---|---|---|
| Full model | -139.3464 | -138.2353 | -132.4922 | -136.8505 | 0.2840 |
| Rx1 | -135.0177 | -134.3686 | -129.8762 | -133.1453 | 0.1430 |
| Rx2 | -139.8698 | -139.2211 | -134.7290 | -137.9978 | 0.0354 |
| Rx3 | -136.7039 | -136.0552 | -131.5632 | -134.8319 | 0.1071 |
| Reduced model | -131.6505 | -131.5479 | -129.9369 | -131.0265 | |

When X1 was removed from the full model; the LL dropped significantly from 73.6732 to 70.5086 as signified by LRT value of 6.3292 with significant p value equals to 0.0119, denoting the significance of the effect of the air pollution on the quality of support network. Likely, when the homicide rate predictor was removed the same consequence was observed as shown from the table; reduction of LL from 73.6732 to 71.3519 affecting the LRT to be 4.6426 which is statistically significant (p-value=0.0312). However, when the life expectancy was removed from the full model, the LL reduction from 73.6732 to 72.9349 was minimal implicating small LRT of 1.4767 and its associated insignificant p-value of 0.2243. In other words, the life expectancy is insignificantly important in the full model and can be safely removed with no drawback. Removing X1 from the full model reverses the sign of the life expectancy in contrast to removal of X2 which preserves the signs of the remaining variables, and these preserved signs were also noted on X3 removal from the full model. But these 2 variables are significant and cannot be removed and the presence of both X1 and X3 without X2 gives LL of 72.9349 which is higher than of model with X1 alone ( LL is 71.2863) and is higher than of the model with X3 alone ( LL is 70.2822) .Also when removing X2 and keeping X1 and X3 together gives an AIC equals to -139.8698 which is also less than keeping X1 alone ( AIC is -138.5726) and is also lower than keeping X3 alone ( AIC is -136.5645). So the best model that can ascertain the predictors that affect the quality of support network are the air pollution and the homicide rate. And both predictors exert negative effect on this response variable. The more the air pollution is, the less the support the person can get. Likewise, the more the homicide rate is, the less the support the individual can have on crisis.

## Fourth Response Variable: Feeling Safe Walking Alone

Feeling safe walking alone was regressed on two predictors; one at a time then all in one full model. Figure 92 shows the scatter plot that detects the relationship between the response variable and each predictor (transformed). The author displayed figures for the logit model, figures that illustrate the estimated curve, plot between the predictor and the residuals, and the QQ plot for the empirical residuals against the theoretical residuals.

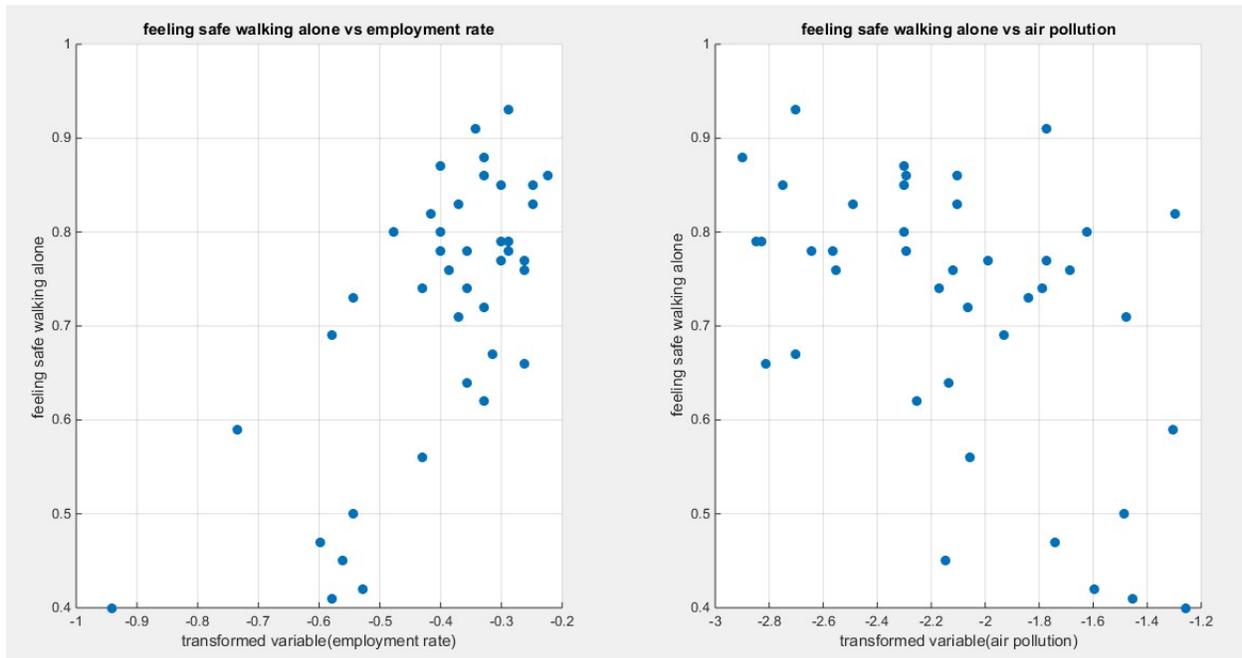

Fig. 92 shows the scatter plot of the response variable and each of the predictor. The relationship is nonlinear relationship except the relationship between the water quality and the employment rate.

The predictors are the employment rate (X1) and the air pollution (X2). Both variables are transformed using the division by 100 then taking the log of the result. The relationship can be modeled using the parametric median quantile regression. The relationship between the response variable and each of the predictor is nonlinear. Tables 25-26 show the results obtained from regressing the feeling safe walking alone on each predictor using different link functions and comparing the statistical indices as regards the estimated coefficients , the Likelihood Ratio Test (LRT ) and its p value, AIC, CAIC, BIC, HQIC and the LL.

Table 25: regressing feeling safe walking alone on the employment rate

| | Logit link function | | Log-log complementary | | Log-log median | |
|---|---|---|---|---|---|---|
| B0 | 2.5148 | | 1.0998 | | -2.4163 | |
| B1 | 3.3982 | | 1.8957 | | -2.7691 | |
| LL | -31.3853 | | -31.3853 | | 31.2662 | |
| Wald stat. of b0 | 5.9094 (p < 0.025) | | 4.9135 (p < 0.025) | | 6.9349 (p < 0.025) | |
| Wald stat. of b1 | 3.2037 (p < 0.025) | | 3.1599 (p < 0.025) | | 3.3565 (p <0.025) | |
| AIC | -58.7706 | | -58.8158 | | -58.5324 | |
| CAIC | -58.4548 | | -58.5001 | | -58.2166 | |
| BIC | -55.3434 | | -55.3887 | | -55.1052 | |
| HQIC | -57.5226 | | -57.5679 | | -57.2844 | |
| LRT | 13.6359 (p=2.2190e-4) | | 13.6812(p=2.1662e-4) | | 13.3977 (p=2.5193e-4) | |
| R-squared | 0.2829 | | 0.2837 | | 0.2788 | |
| P-value for randomized quantile residuals | 0.0630 at 5% significance level | | 0.0848 at 5% significance level | | 0.0442 at 1% significance level | |
| p-value for Cox-Snell residuals | 0.0630 at 5% significance level | | 0.0848 at 5% significance level | | 0.0442 at 1% significance level | |
| Variance-covariance matrix | 0.1811 | 0.4301 | 0.0501 | 0.1283 | 0.1214 | 0.2721 |
| | 0.4301 | 1.1251 | 0.1283 | 0.3599 | 0.2721 | 0.6806 |
| QR vs. predictor(tau,p) | (-0.0174,0.8837) | | (-0.0174,0.8837) | | (-0.0149 , 0.9015 ) | |
| CS vs. predictor(tau,p) | (-0.0174,0.8837) | | (-0.0174,0.8837) | | (-0.0149 , 0.9015 ) | |

Table 25 shows that the predictor is significant as likelihood ratio test (LRT) is high around 13. The R squared is also high for this predictor around 0.28 across the different link functions. The AIC, CAIC, BIC, HQIC and the LL are more or less equal across the different models. The LL is around 31 across the link functions. The residuals plotted against the predictors show no specific trend and they are randomly scattered. The QQ plot shows that the randomized quantile residuals are not perfectly aligned with the diagonal all through its course in similarity with the Cox Snell residuals that do not show this perfect alignment at the lower tail and the center. The estimated curve between the estimated median and the transformed predictor is increasing reflecting that the more the employment rate is, the more the percentage expressing feeling safe walking alone is. The figure for the clog-log shows the same pattern. The log-log figure has the same pattern. The difference is mainly manifested in the slope of the estimated curve of the predictor against the estimated median. To assess the assumption of constant variance in the median parametric regression model, residual-based diagnostic tests were conducted using both randomized quantile (RQ) and Cox-Snell (CS) residuals. For each type of residual, an auxiliary regression of the squared residuals on the corresponding predictor was estimated by ordinary least squares, and the null hypothesis of homoscedasticity ($H_0$: constant variance) was tested. The results indicted no significant relationship between the squared residuals and the predictor variable (CS: p=0.538, R-squared=0.00979; RQ: p=0.408, R-squared=0.0176), suggesting that the variance of the residuals remained approximately constant across the range of the predictor. Furthermore, the magnitude of the CS residuals were within a reasonable range (four values between 2.1136 and 3.0478), which supports the absence of heteroscedasticity. These findings provide evidence that the fitted median regression model satisfies the homoscedasticity assumption. These results are for the logit model. Figures 93-99 show the previous results.

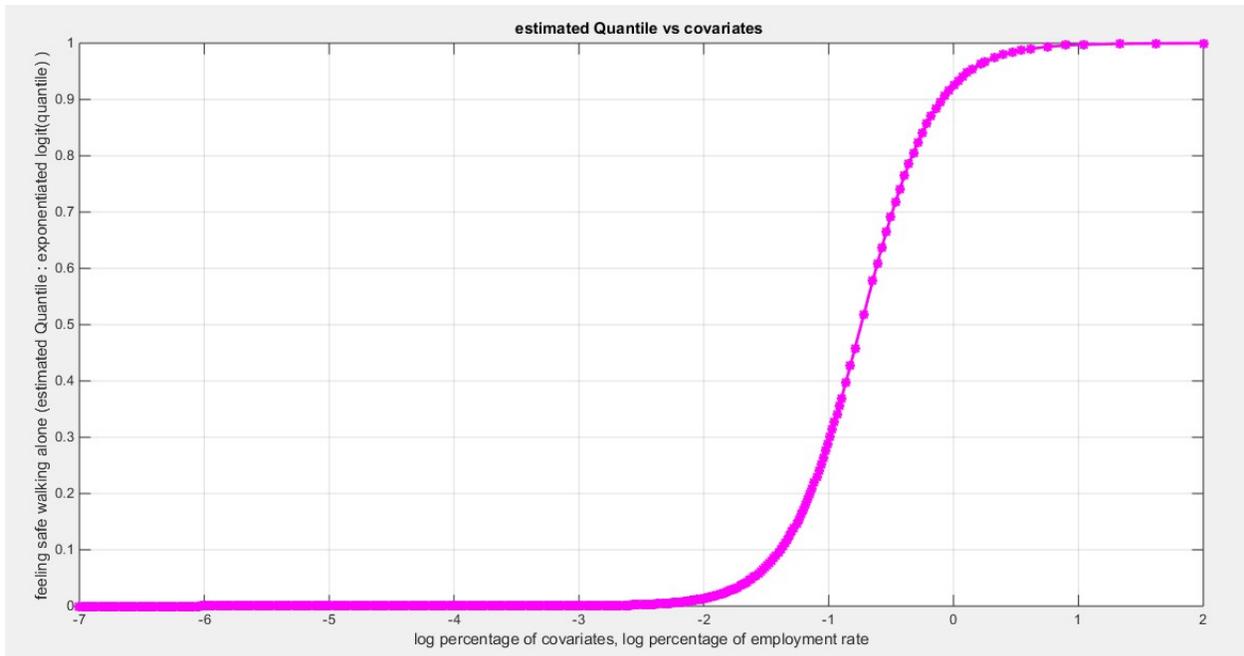

Fig. 93 shows the estimated curve plotting the transformed predictor against the estimated median (for the logit link).

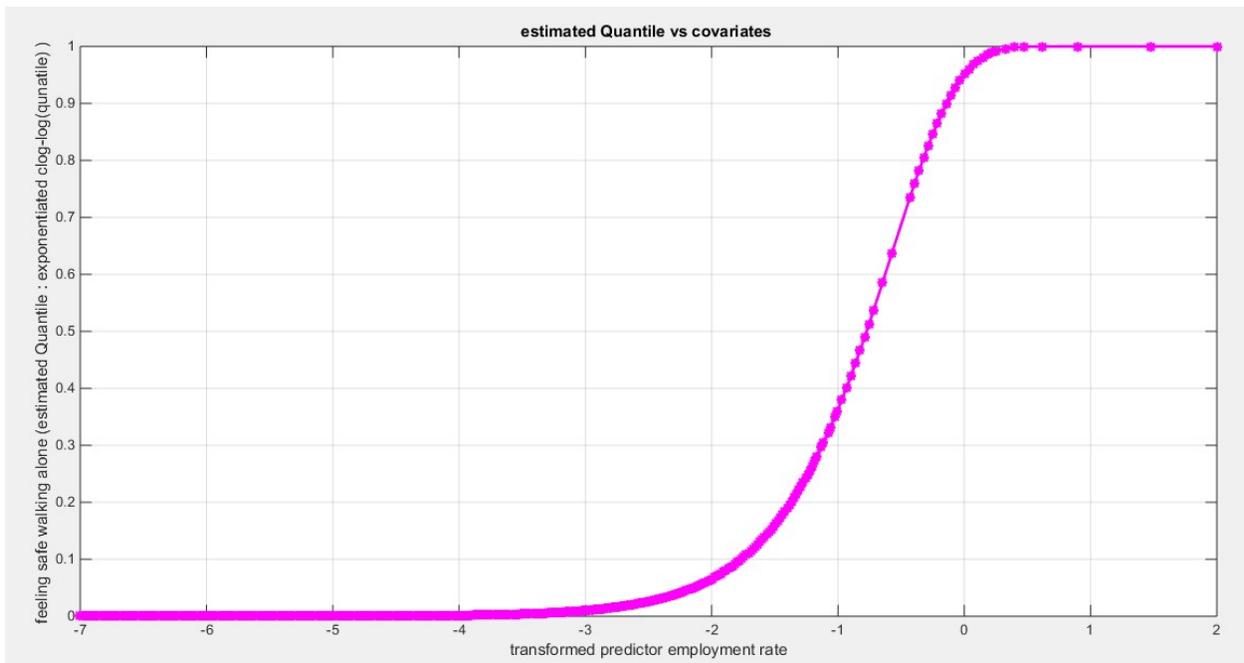

Fig. 94 shows the estimated curve plotting the transformed predictor against the estimated median (for the clog-log link).

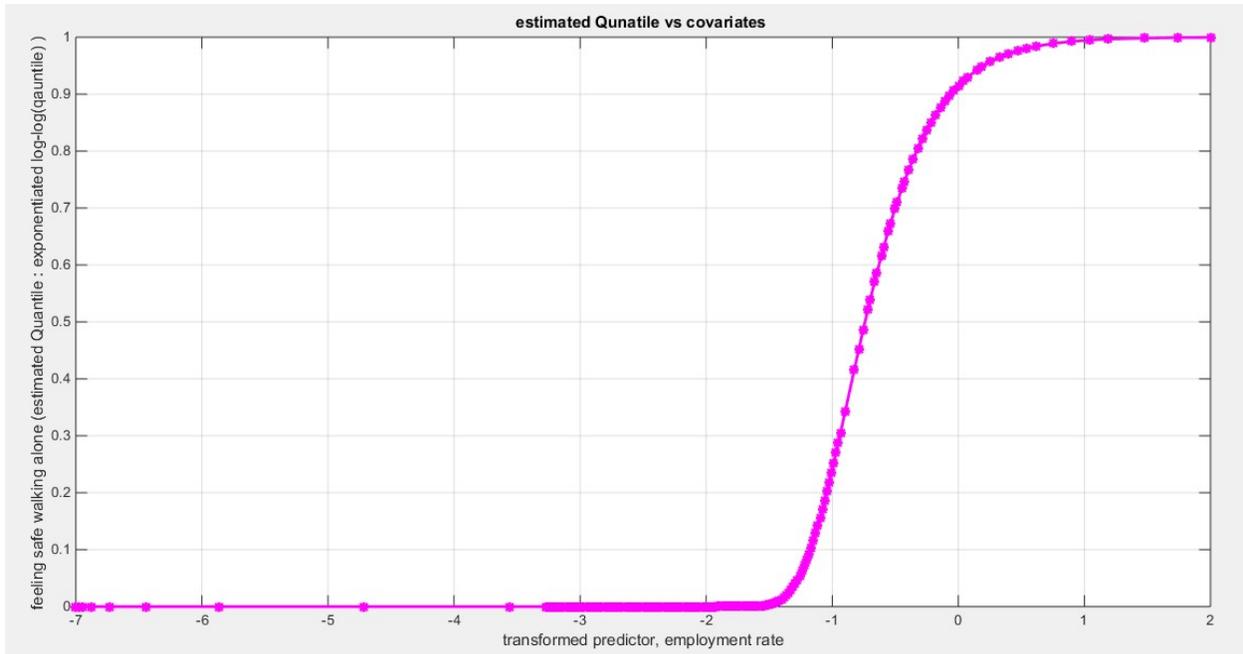

Fig. 95 shows the estimated curve plotting the transformed predictor against the estimated median (for the log-log link).

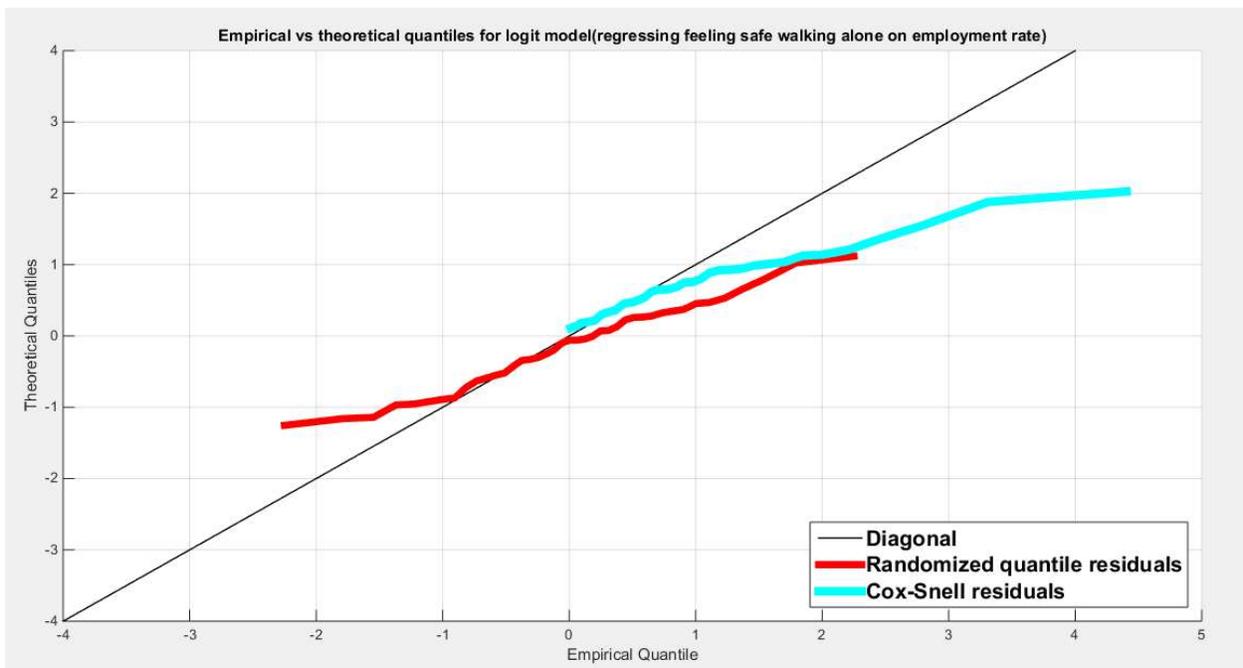

Fig. 96 shows the QQ plot of the empirical quantiles and the theoretical quantiles for both types of residuals.

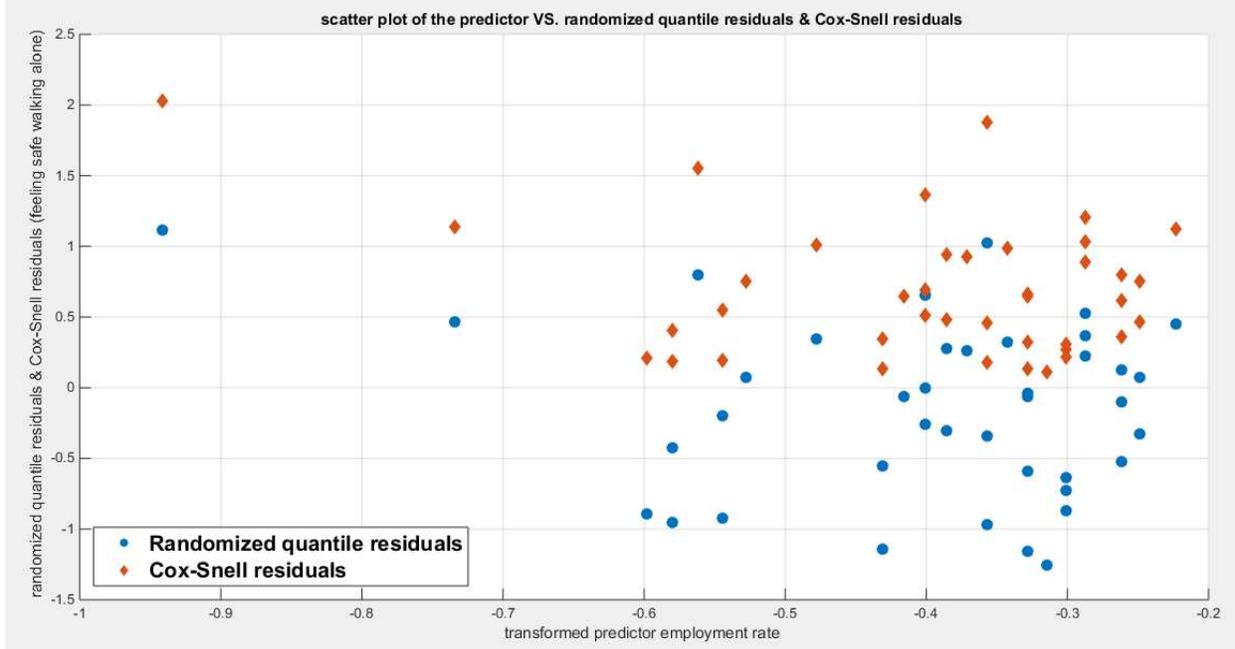

Fig. 97 shows the scatter plot of residuals of both types against transformed predictors.

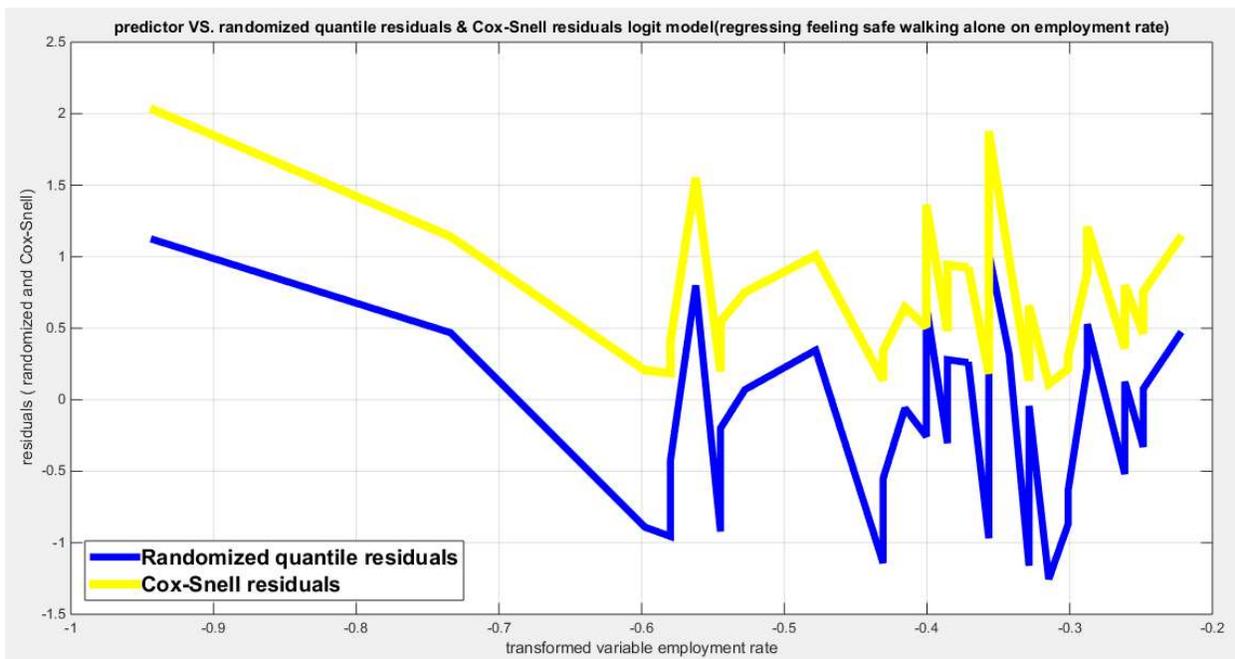

Fig. 98 shows the plot of residuals of both types against transformed predictors

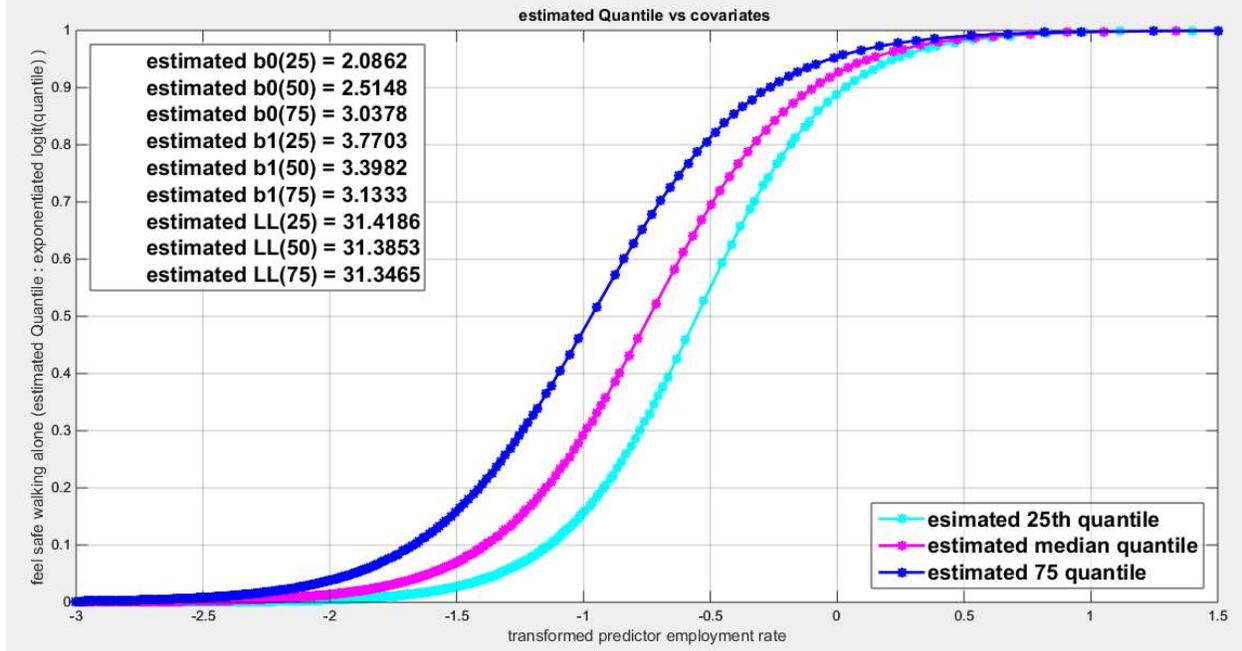

Fig. 99 shows parallel quantile curves across 25th , 50th ( median), 75th percentiles, suggesting that the predictor exerts a uniform influence on the response consistent with homoscedasticity.

Table 26: Regressing  feeling safe  walking  on air pollution

| | Logit link function | | Log-log complementary | | Log-log median | |
|---|---|---|---|---|---|---|
| B0 | -0.4385 | | -0.4842 | | 0.0944 | |
| B1 | -0.7413 | | -0.3869 | | 0.6492 | |
| LL | -28.3374 | | -28.2522 | | -28.3682 | |
| Wald stat. of b0 | 0.7358 ( p >0.025) | | 1.4844 (p>0.025) | | 0.1857 ( p >0.025) | |
| Wald stat. of b1 | 2.7362 (p <0.025) | | 2.7155 (p<0.025) | | 2.7607 (p<0.025) | |
| AIC | -52.6748 | | -52.5045 | | -52.7363 | |
| CAIC | -52.3590 | | -52.1887 | | -47.0321 | |
| BIC | -49.2476 | | -49.0773 | | -49.3092 | |
| HQIC | -51.4268 | | -51.2565 | | -51.4883 | |
| LRT | 7.5401 (p-val.=0.006) | | 7.3698 (p-val.=0.006) | | 7.6016 (p-val.=0.0058) | |
| R-squared | 0.1680 | | 0.1645 | | 0.1692 | |
| P-value for randomized quantile residuals | 0.2505 | | 0.2830 | | 0.2378 | |
| p-value for Cox-snell residuals | 0.2505 | | 0.2830 | | 0.2378 | |
| Variance-covariance matrix | 0.3552 | 0.1577 | 0.1064 | 0.0454 | 0.2585 | 0.1166 |
| | 0.1577 | 0.0734 | 0.0454 | 0.0203 | 0.1166 | 0.0553 |
| QR vs. predictor(tau,p) | (0.0159, 0.8927) | | (0.0159, 0.8927) | | (0.0159, 0.8927) | |
| CS vs. predictor(tau,p) | (0.0159, 0.8927) | | (0.0159, 0.8927) | | (0.0159, 0.8927) | |

Table 26 shows that the predictor is significant as likelihood ratio test (LRT) is high around 7; but it is less than that of the employment rate. The R squared is also high for this predictor around 0.16 across the different link functions. The AIC, CAIC, BIC, HQIC and LL are more or less equal across the different models. The LL is around 28 across the link functions. The residuals plotted against the predictors show no specific trend and they are randomly scattered. The QQ plot shows that the randomized quantile residuals are not perfectly aligned with the diagonal all through its course in similarity with the Cox Snell residuals that do not show this perfect alignment at the lower tail and the center. The estimated curve between the estimated median and the transformed predictor is decreasing reflecting that the more the air pollution is, the less the percentage expressing feeling safe walking alone is. The figure for the clog-log shows the same pattern. The log-log figure has the same pattern. The difference is mainly manifested in the slope of the estimated curve. To assess the assumption of the constant variance in the median parametric regression model, residual-based diagnostic tests were conducted using both randomized quantile (RQ) and Cox-Snell (CS) residuals. For each type of residual, an auxiliary regression of the squared residuals on the corresponding predictor was estimated by ordinary least squares, and the null hypothesis of homoscedasticity ($H_0$: constant variance) was tested. The results indicted no significant relationship between the squared residuals and the predictor variable (CS: p=0.816, R-squared=0.0014; RQ: p=0.873, R-squared=0.000667), suggesting that the variance of the residuals remained approximately constant across the range of the predictor. Furthermore, the magnitude of the CS residuals were within a reasonable range (one value 2.0259), which supports the absence of heteroscedasticity. These findings provide evidence that the fitted median regression model satisfies the homoscedasticity assumption. These results are for the logit model. Figures 100-106 show the previous results.

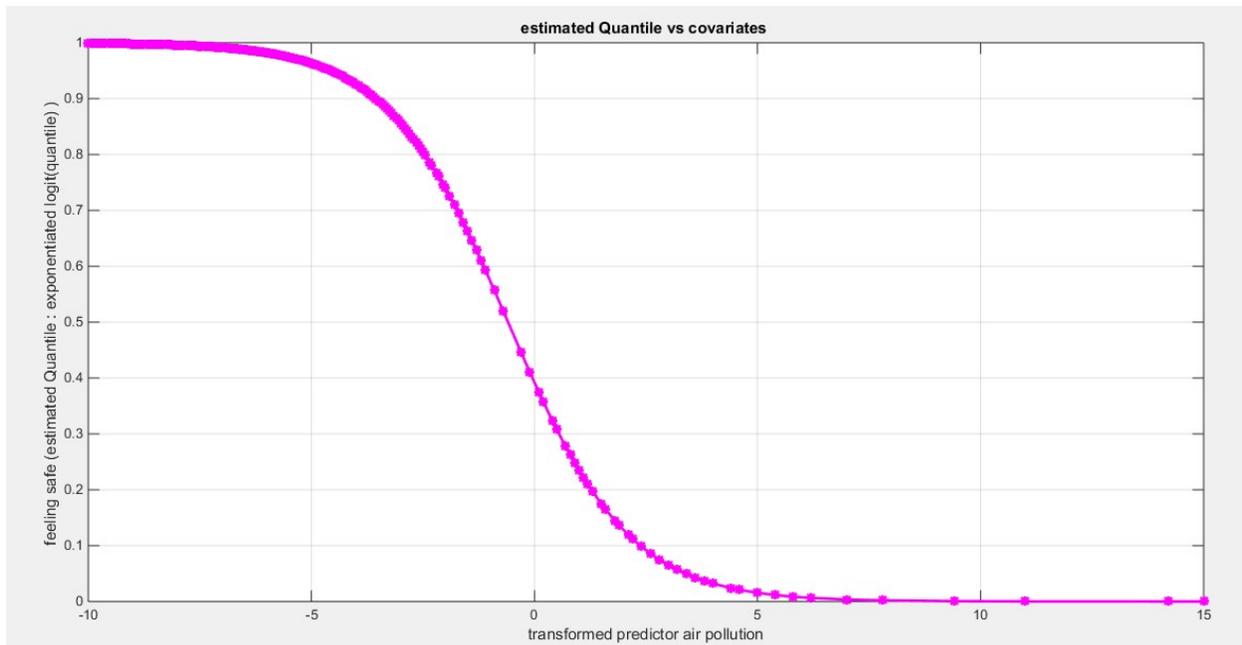

Fig. 100 shows the estimated curve plotting the transformed predictor against the estimated median (for the logit link).

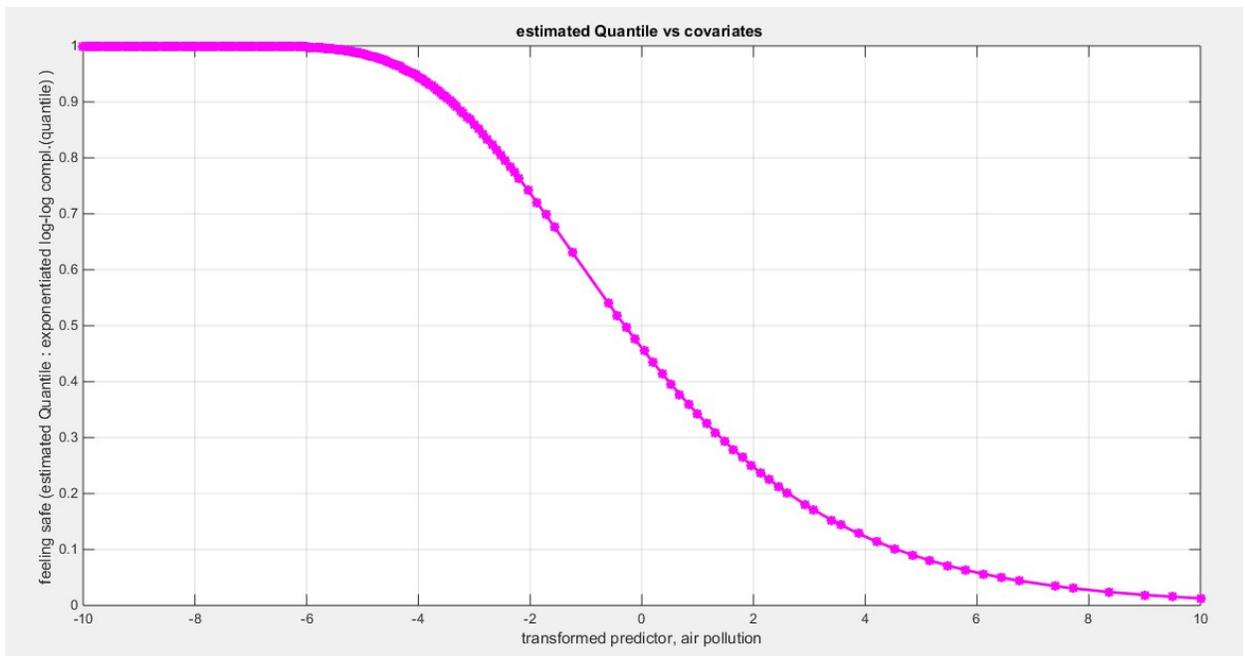

Fig. 101 shows the estimated curve plotting the transformed predictor against the estimated median (for the clog-log link).

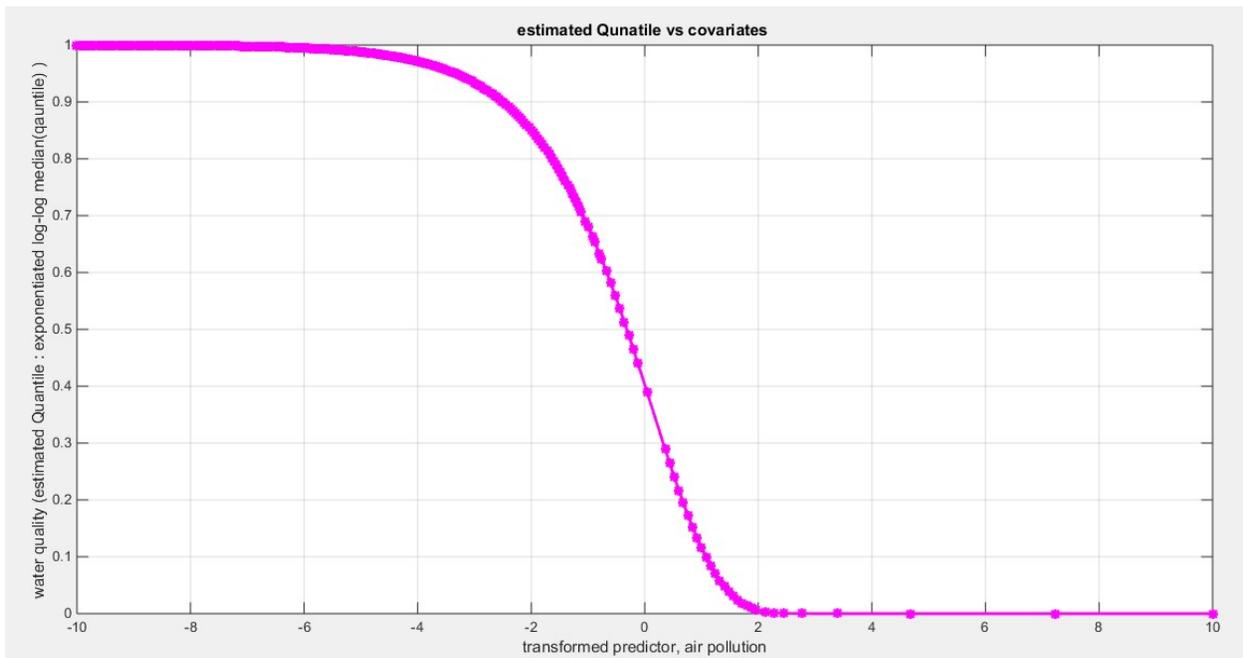

Fig. 102 shows the estimated curve plotting the transformed predictor against the estimated median (for the log-log link).

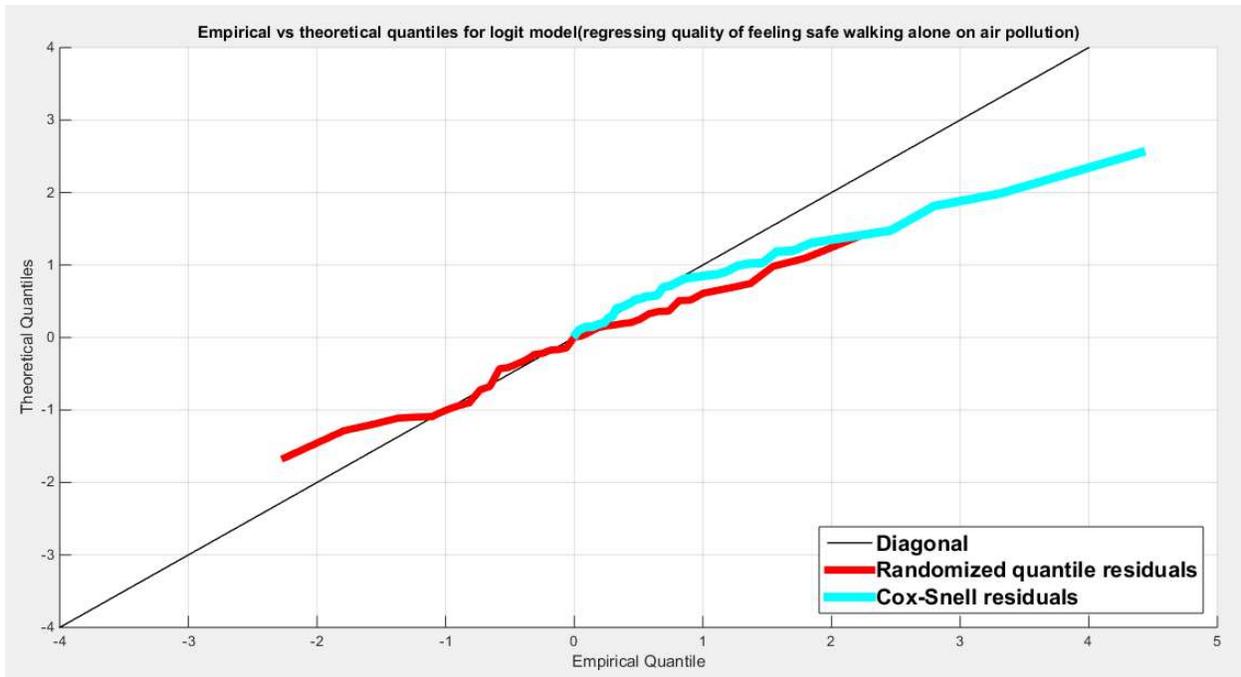

Fig. 103 shows the QQ plot of the empirical quantiles and the theoretical quantiles for both types of residuals.

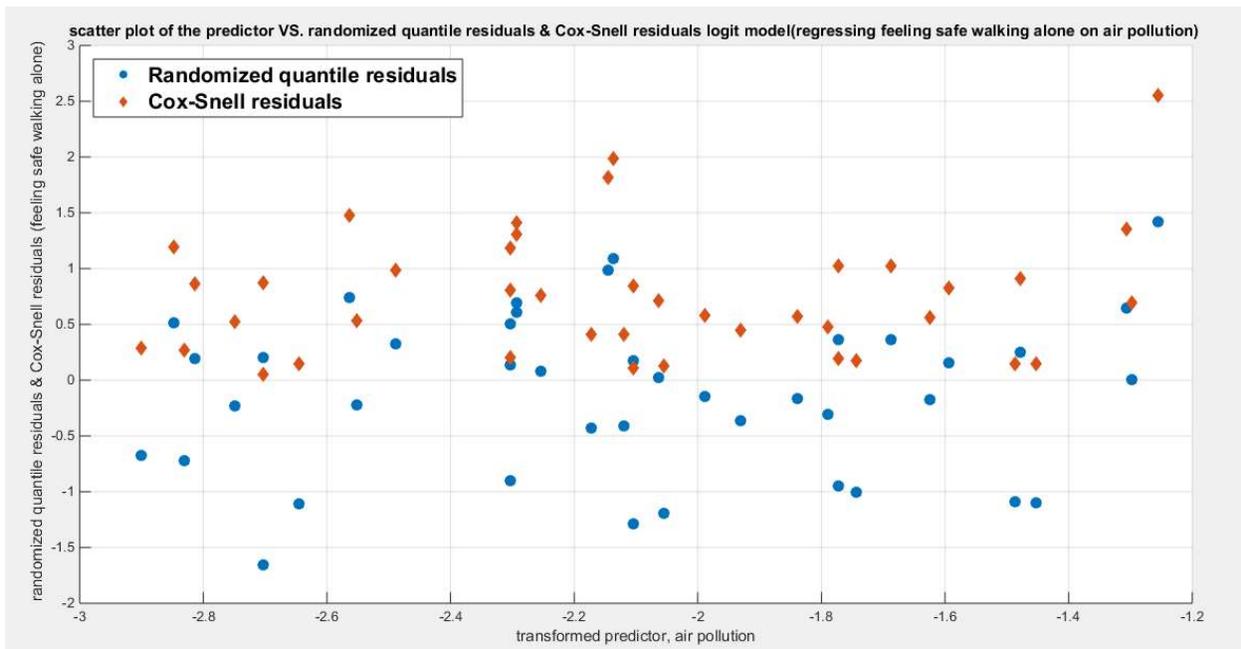

Fig. 104 shows the scatter plot of residuals of both types against transformed predictors.

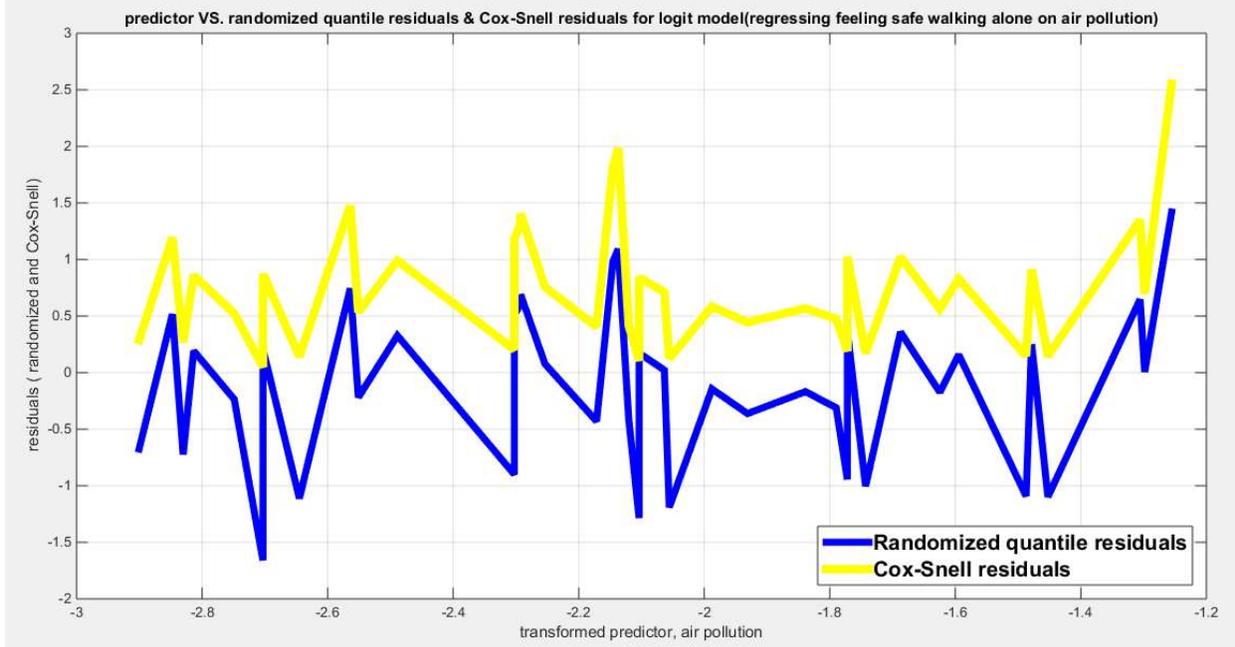

Fig. 105 shows the plot of residuals of both types against transformed predictors

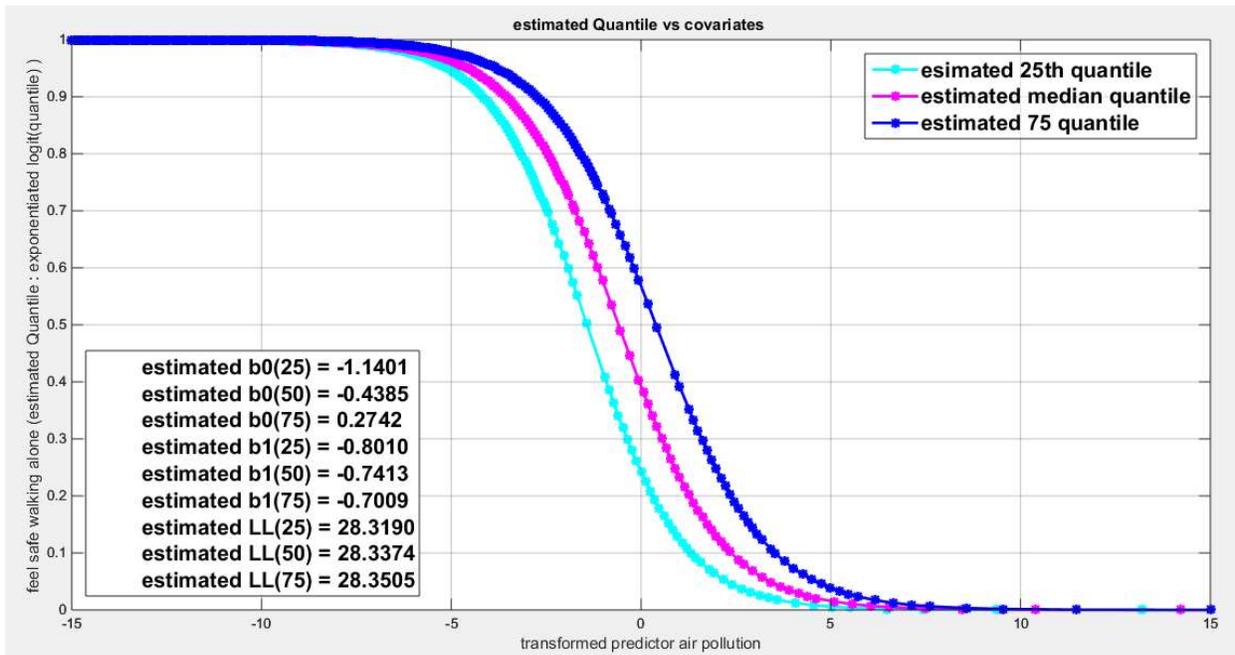

Fig. 106 shows parallel quantile curves across 25th , 50th ( median), 75th percentiles, suggesting that the predictor exerts a uniform influence on the response consistent with homoscedasticity.

The marginal correlations between the variables (the response and the predictors) are shown in Table 27. The response variable, feeling safe walking alone, shows a positive and a statistically significant correlation with the employment rate, while it displays a negative and a statistically significant correlation with the air pollution. The employment rate has a statistically significant and a negative correlation with the air pollution.

Table 27: The marginal correlation matrix Kendall tau coefficient and associated p-value:

| | Feeling safe walking alone (Y) | Employment rate (X1) | Air Pollution (X2) |
|---|---|---|---|
| **Feeling safe walking alone(Y)** | 1 | 0.4038 P=2.8625e-4 | -0.3333 P=0.0025 |
| **Employment rate (X1)** | 0.4038 P=2.8625e-4 | 1 | -0.4135 P=1.9355e-4 |
| **Air Pollution (X2)** | -0.3333 P=0.0025 | -0.4135 P=1.9355e-4 | 1 |

The condition indices obtained from the standardized transformed X'X are 2.0437, and 1. The VIF for the employment rate is 1.6040 and for the air pollution is 1.6040. So as the largest condition index is 2.0437 is less than 10 and the VIF values are less than 5 so there is no evidence of significant multi-collinearity between the predictors.

The signs of the coefficients of the marginal correlations match those signs of the conditional correlations coefficients when regressing the feeling safe walking alone response variable on each predictor, one at a time. Also in multiple regression analysis the employment rate is positively dependent with the response variable while the air pollution is negatively dependent with feeling safe walking alone consistent with simple regression.

The author added these two predictors in one equation and used the different link functions, then removed each one at a time and calculated the LRT to assess the significance of this particular predictor while controlling for the another predictor. The results are summarized in Table 28-29. The description of these tables a regards the rows and columns are similar to the descriptions of the tables of the previous variables.

Table 28: the coefficients of parametric median regression analysis with removal of different predictors and the associated standard error

| | intercept | X1 Coeff. | X2 Coeff. | Preserve sign | LRT & P value | LL |
|---|---|---|---|---|---|---|
| Full Model SE | 2.1946 11.9447 | 3.1597 0.2790 | -0.1071 2.7688 | Yes | 13.7226 0.001 | 31.4286 |
| Rx1 SE. | -0.4385 11.3807 | | -0.7413 2.5953 | Yes | 6.1825 0.019 | 28.3374 |
| Rx2 SE | 2.5148 0.5648 | 3.3982 0.2604 | | yes | 0.0867 0.7684 | 31.3853 |
| Reduced model | 1.0804 | | | | | 24.5673 |

Table 29: AIC, CAIC,BIC, and HQIC.

| | AIC | CAIC | BIC | HQIC | R squared |
|---|---|---|---|---|---|
| Full model | -56.8573 | -56.2086 | -51.7166 | -54.9853 | 0.2844 |
| Rx1 | -52.6748 | -52.3590 | -49.2476 | -51.4268 | 0.1400 |
| Rx2 | -58.7706 | -58.4548 | -55.3434 | -57.5226 | 0.0021 |
| Reduced model | -47.1347 | -47.0321 | -45.4211 | -46.5107 | |

Tables 28-29 show that when removing the employment rate (X1), LL decreases from 31.4286 to 28..3374 and the LRT is 6.1825 which is statistically significant (p=0.001); while removing X2 minimally decreases the LL from 31.4286 to 31.3853 and the LRT is 0.0867 which is statistically insignificant (p= 0.7684) denoting that the employment rate is more significant and important than the air pollution. Furthermore; the LL of the full model is minimally reduced when removing X2 since X1 is remaining and the LL of regressing the response on X1 is 31.3853. The AIC of the model involving X1 (AIC= -58.7706) is less than that of the full model with two predictors (AIC=-56.8573) and this is due to the penalty of the number of the parameters paid when number of parameters increases. So removing X2 is advised and regressing feeling safe walking alone on the employment rate is the best model.

# Section Four:

# Conclusions

The regression analysis is a fundamental statistical tool used to model the relationship between the response variable and the and one or more of the explanatory variables. The classical regression, focuses on modeling the mean response . It requires some assumptions to be fulfilled like normality, homoscedasticity and lack of outliers. When any of these assumptions is violated especially if the response distribution is skewed, heavy tailed or heteroscedastic and in the presence of outliers, the classical regression model may lead to misleading estimates and inference, in other words; model misspecification. The parametric quantile regression generalizes this framework by modeling the conditional quantiles, allowing for a richer understanding of the response distribution. It produces more reliable and interpretable results particularly for asymmetric and heavy-tailed data. In both simple and multiple settings, it provides a comprehensive view of how predictors influence different regions of the response distribution, rather than only its average behavior. It is robust to outliers so it is valuable for skewed or heavy-tailed data. It allows the response distribution to follow non-normal distribution so it flexibly provides a complete description of the response distribution at different regions and how the predictors affect its low, median and high quantiles differently. It naturally handles the heteroscedasticity violation (non-constant variance) across the predictors by allowing qauntile specific slopes. In this study, and for the above mentioned properties, the author harnessed the power of parametric quantile regression to elegantly model some of the bounded response data recorded in the OECD data platform adding depth and precision to this statistical technique. In simple parametric quantile regression, the quantile regression lines for the 25th , 59th and 75th percentiles were approximately parallel indicating that the effect of the predictor on the response variable was consistent across the conditional distributionof the response. This suggests a location shift of the effect of the distribution rather than a scale change, implying that the predictor influences the central tendency of the response without affecting its variability.

In the simple regression model, the educational attainment response variable is positively correlated with both the employment rate and the life satisfaction while it is negatively correlated with the air pollution and the homicide rate. Meanwhile, in the multiple regression models, only the employment rate appears to be the only significant predictor when controlling for other predictors. This may point to some of these predictors are acting as proxy for other unobserved variables since the coefficients of these controlled variables flipped their signs in the full or nested models. Air pollution can be considered a proxy to industrial urbanization, life satisfaction can be anticipated as a proxy to the lack of drive or motivation to education, and homicide rate can act as a proxy to urbanization and lack of faith and religion, high rates of gang violence, high rates of organized crime, and high rates of gun ownership than the rural area.

Simple regression model revealed that the water quality response variable is positively correlated with the employment rate, the life expectancy, and the life satisfaction but it is negatively correlated with the air pollution and the homicide rate. And this relationship is statistically significant. However, in multiple regression models, the only significant predictors are the life satisfaction and the employment rate. Even the employment rate can be removed and hence the life satisfaction is the most significant predictor influencing the quality of water. The air pollution can be treated as a proxy for industrial urbanization that may cause flipping of the sign of the air pollution coefficient from negative to positive. Also this is applicable to the homicide rate that can act as a proxy to the urbanization and the associated high rate of the gun ownership, high rate of the drug trafficking, gang violence, and organized crime which derives the sign of the homicide coefficient to flip.

Regressing the quality of support network on any of the air pollution, the life expectancy, or the homicide rate yields that the air pollution and the homicide rate are the only significant predictors in the simple regression models. Life expectancy can be speculated as a proxy for loneliness that can flip the sign of the life expectancy coefficient from positive to negative in multiple regression models when the three variables are added together.

The feeling safe walking alone at night is positively correlated with the employment rate and negatively correlated with the air pollution. Both predictors are statistically significant in the simple regression models. But in the multiple regression models, the employment rate surpasses the air pollution in significance and importance.

The entire nested models and the full model discussed in this paper, passed the model adequacy tests as regards the diagnostic tests of the randomized residuals and the Cox-Snell residuals. The models are well specified as regards the chosen parametric PDF of the MBUR distribution and any of the link functions like the logit, clog-log and the log-log functions. The author tested for the presence of the multicollinearity between predictors, and there were no evidence of significant multicollinearity problem. There was also no evidence of heteroscedasticity, although the parametric quantile regression accounts for this problem and can naturally manage it.

## Future work

The proxy variable effect in parametric quantile regression needs further studies to calculate their consequences on the estimation and inference process. The omitted variable bias requires more detailed research in the area of parametric quantile regression. The residual on residual studies and plot mandate more in depth investigations to reveal its estimations in this type of regression.

Appendix A: the response variables and the predictors.

| | Air poll-ution | Homi-cide rate | Life expect-ancy | Life satis-faction | Employ-ment rate | Education attainment | Water quality | Feeling safe walking alone | Quality of support network |
|---|---|---|---|---|---|---|---|---|---|
| Australia | 6.7 | 0.9 | 83 | 7.1 | 73 | 0.84 | 0.92 | 0.67 | 0.93 |
| austria | 12.2 | 0.5 | 82 | 7.2 | 72 | 0.86 | 0.92 | 0.86 | 0.92 |
| Belgium | 12.8 | 1.1 | 82.1 | 6.8 | 65 | 0.80 | 0.79 | 0.56 | 0.90 |
| Canada | 7.1 | 1.2 | 82.1 | 7 | 70 | 0.92 | 0.90 | 0.78 | 0.93 |
| Chile | 23.4 | 2.4 | 80.6 | 6.2 | 56 | 0.67 | 0.62 | 0.41 | 0.88 |
| Colombia | 22.6 | 23.1 | 76.7 | 5.7 | 58 | 0.59 | 0.82 | 0.50 | 0.80 |
| Costa rica | 17.5 | 10 | 80.5 | 6.3 | 55 | 0.43 | 0.87 | 0.47 | 0.82 |
| Czechia | 17 | 0.7 | 79.3 | 6.9 | 74 | 0.94 | 0.89 | 0.77 | 0.96 |
| Denmark | 10 | 0.5 | 81.5 | 7.5 | 74 | 0.82 | 0.93 | 0.85 | 0.95 |
| Estonia | 5.9 | 1.9 | 78.8 | 6.5 | 74 | 0.91 | 0.86 | 0.79 | 0.95 |
| Finland | 5.5 | 1.2 | 82.1 | 7.9 | 72 | 0.91 | 0.97 | 0.88 | 0.96 |
| France | 11.4 | 0.4 | 82.9 | 6.7 | 65 | 0.81 | 0.78 | 0.74 | 0.94 |
| Germany | 12 | 0.4 | 81.4 | 7.3 | 77 | 0.86 | 0.91 | 0.76 | 0.90 |
| Greece | 14.5 | 1 | 81.7 | 5.8 | 56 | 0.76 | 0.67 | 0.69 | 0.78 |
| Hungary | 16.7 | 0.9 | 76.4 | 6 | 70 | 0.86 | 0.81 | 0.74 | 0.94 |
| Iceland | 6.4 | 0.3 | 83.2 | 7.6 | 78 | 0.76 | 0.97 | 0.85 | 0.98 |
| Ireland | 7.8 | 0.5 | 82.8 | 7 | 68 | 0.85 | 0.80 | 0.76 | 0.96 |
| Isreal | 19.7 | 1.5 | 82.9 | 7.2 | 67 | 0.88 | 0.77 | 0.80 | 0.95 |
| Italy | 15.9 | 0.5 | 83.6 | 6.5 | 58 | 0.63 | 0.77 | 0.73 | 0.89 |
| Japan | 13.7 | 0.2 | 84.4 | 6.1 | 77 | | 0.87 | 0.77 | 0.89 |
| Korea | 27.3 | 0.8 | 83.3 | 5.8 | 66 | 0.89 | 0.82 | 0.82 | 0.80 |
| Latvia | 12.7 | 3.7 | 75.5 | 6.2 | 72 | 0.89 | 0.83 | 0.72 | 0.92 |
| Lithuania | 10.5 | 2.5 | 76.4 | 6.4 | 72 | 0.94 | 0.83 | 0.62 | 0.89 |
| Luxem-bourg | 10 | 0.2 | 82.7 | 7.4 | 67 | 0.74 | 0.85 | 0.87 | 0.91 |
| Mexico | 20.3 | 26.8 | 75.1 | 6 | 59 | 0.42 | 0.75 | 0.42 | 0.77 |
| Nether-lands | 12.2 | 0.6 | 82.2 | 7.5 | 78 | 0.81 | 0.91 | 0.83 | 0.94 |
| New Zealand | 6 | 1.3 | 82.1 | 7.3 | 77 | 0.81 | 0.85 | 0.66 | 0.95 |
| Norway | 6.7 | 0.6 | 83 | 7.3 | 75 | 0.82 | 0.98 | 0.93 | 0.96 |
| Poland | 22.8 | 0.5 | 78 | 6.1 | 69 | 0.93 | 0.82 | 0.71 | 0.94 |
| Portugal | 8.3 | 0.7 | 81.8 | 5.8 | 69 | 0.55 | 0.89 | 0.83 | 0.87 |
| Slovak Republic | 18.5 | 0.8 | 77.8 | 6.5 | 68 | 0.92 | 0.81 | 0.76 | 0.95 |
| Slovania | 17 | 0.4 | 81.6 | 6.5 | 71 | 0.90 | 0.93 | 0.91 | 0.95 |
| Spain | 10 | 0.7 | 83.9 | 6.5 | 62 | 0.63 | 0.76 | 0.80 | 0.93 |
| Sweden | 5.8 | 1.1 | 83.2 | 7.3 | 75 | 0.84 | 0.97 | 0.79 | 0.94 |
| Switzerland | 10.1 | 0.3 | 84 | 7.5 | 80 | 0.89 | 0.96 | 0.86 | 0.94 |
| Turkiye | 27.1 | 1 | 78.6 | 4.9 | 48 | 0.42 | 0.62 | 0.59 | 0.85 |
| United Kingdom | 10.1 | 0.2 | 81.3 | 6.8 | 75 | 0.82 | 0.82 | 0.78 | 0.93 |
| United States | 7.7 | 6 | 78.9 | 7 | 67 | 0.92 | 0.88 | 0.78 | 0.94 |
| Brazil | 11.7 | 19 | 75.9 | 6.1 | 57 | 0.57 | 0.70 | 0.45 | 0.83 |
| Russia | 11.8 | 4.8 | 73.2 | 5.5 | 70 | 0.95 | 0.62 | 0.64 | 0.89 |
| South africa | 28.5 | 13.7 | 64.2 | 4.9 | 39 | 0.48 | 0.72 | 0.40 | 0.89 |

# Brief description of each of the above indicators

Air pollution: expressed in OECD database as the percentage of the population exposed to unsafe levels of PM2.5 which is the key pollutant harmful to human health. PM2.5 refers to particulate matter smaller than 2.5 micrometers, which can penetrate deep into the lung and even enter the bloodstream. Higher concentrations indicate poor air quality.

Homicide rate: is expressed as the number on intentional homicides per 100,000 inhabitants in a year. It reflects level of extreme violence in a society and it is often used as a standard for safety and security.

Life expectancy: is typically expressed as a single number representing the expected lifespan at birth. It indicates effective healthcare system, low infant mortality, healthier life style as regards nutrition and low pollution, and good living conditions like clean water and low pollution.

Life satisfaction: is presented as an average score for a country derived from survey responses often on a scale from 0 to 10, reflecting how individuals perceive and evaluate the overall quality of their life. The 0 represents the least satisfaction and the 10 represents the most satisfaction. High scores reflect the good economic conditions, the good social support, and the good health.

Employment rate: is expressed as percentage of working-age population that is currently employed. It is calculated as the number of employed individuals divided by the total working-age population, multiplied by 100. High rates indicate strong economic conditions and economic prosperity.

Educational attainment: is presented as the percentage of a given population who has completed a specific level of education mainly the tertiary education.

Water quality: is the percentage of the people who report being satisfied with the quality of their water. This reflects the proportion of people who have access to clean water supply.

Feeling safe walking alone at night: is expressed as the percentage of individuals who report feeling safe when walking alone in their local area after dark. It reflects the social cohesion, trust in the public safety measures, and the level of the crime in the society.

Quality of support network: is represented as the percentage of the people who report having someone they can account on for support when they need. It reflects the strong social cohesion, mental and emotional well-being, greater life satisfaction, and resilience during tough times.

**Declarations:**
**Ethics approval and consent to participate**
Not applicable.
**Consent for publication**
Not applicable
**Availability of data and material**
Not applicable. Data sharing not applicable to this article as no datasets were generated or analyzed during the current study.
**Competing interests**
The author declares no competing interests of any type.
**Funding**
No funding resource. No funding roles in the design of the study and collection, analysis, and interpretation of data and in writing the manuscript are declared
**Authors' contribution**

AI ( Attia Iman ) carried the conceptualization by formulating the goals, aims of the research article, formal analysis by applying the statistical, mathematical and computational techniques to synthesize and analyze the hypothetical data, carried the methodology by creating the model, software programming and implementation, supervision, writing, drafting, editing, preparation, and creation of the presenting work.

**Acknowledgement**

Not applicable